\documentclass[english,reprint, aip,jcp,superscriptaddress,showpacs,floatfix]{revtex4-1}
\usepackage{amsmath}
\usepackage{graphicx, subfigure}
\usepackage{makeidx}

\newcommand{\bra}[1]{\left \langle #1 \right \vert}
\newcommand{\ket}[1]{\left \vert #1 \right \rangle}
\newcommand{\braket}[2]{\langle #1 \vert  #2 \rangle}

\begin{document}

\title{Quantum Dynamics in Open Quantum-Classical Systems}

\author{Raymond Kapral}
\affiliation{Chemical Physics Theory Group, Department of Chemistry, University of Toronto, Toronto, ON, M5S 3H6 Canada}

\begin{abstract}
   Often quantum systems are not isolated and interactions with their environments must be taken into account. In such open quantum systems these environmental interactions can lead to decoherence and dissipation, which have a marked influence on the properties of the quantum system. In many instances the environment is well-approximated by classical mechanics, so that one is led to consider the dynamics of open quantum-classical systems. Since a full quantum dynamical description of large many-body systems is not currently feasible, mixed quantum-classical methods can provide accurate and computationally tractable ways to follow the dynamics of both the system and its environment. This review focuses on quantum-classical Liouville dynamics, one of several quantum-classical descriptions, and discusses the problems that arise when one attempts to combine quantum and classical mechanics, coherence and decoherence in quantum-classical systems, nonadiabatic dynamics, surface-hopping and mean-field theories and their relation to quantum-classical Liouville dynamics, as well as methods for simulating the dynamics.

\end{abstract}

\maketitle



\section{Introduction} \label{sec:intro}

It is difficult to follow the dynamics of quantum processes that occur in large and complex systems. Yet, often the quantum phenomena we wish to understand and study take place in such systems. Both naturally-occurring and man-made systems provide examples: excitation energy transfer from light harvesting antenna molecules to the reaction center in photosynthetic bacteria and plants, electronic energy transfer processes in the semiconductor materials used in solar cells, proton transfer processes in some molecular machines that operate in the cell, and the interactions of the qbits in quantum computers with their environment. Although the systems in which these processes take place are complicated and large, it is often the properties that pertain to only a small part of the entire system that are of interest; for example, the electrons or protons that are transferred in a biomolecule. This subsystem of the entire system can then be viewed as an open quantum system that interacts with its environment. In open quantum systems the dynamics of the environment can influence the behavior of the quantum subsystem in significant ways. In particular, it can lead to decoherence and dissipation which can play central roles in the rates and mechanisms of physical processes. This partition of the entire system into two parts has motivated the standard system-bath picture where one of these subsystems (henceforth called the {\em subsystem}) is of primary interest while the remainder of the degrees of freedom constitute the environment or {\em bath}.

Most system-bath descriptions focus on the dynamics of the subsystem density matrix, which is obtained by tracing over the bath degrees of freedom: $\hat{\rho}_s={\rm Tr}_b \hat{\rho}$. If such a program were carried out fully an exact equation of motion for $\hat{\rho}_s$ could be derived and no information about the bath would be lost in this process. Of course, for problems of most interest where the bath is very large with complicated interactions this is not feasible and would defeat the motivation behind the system-bath partition. Consequently, the influence of the bath on the dynamics of the subsystem is embodied in dissipative and other coupling terms in the subsystem evolution equation.

\begin{figure}[hc]
\begin{center}
\includegraphics[width=0.85\columnwidth]{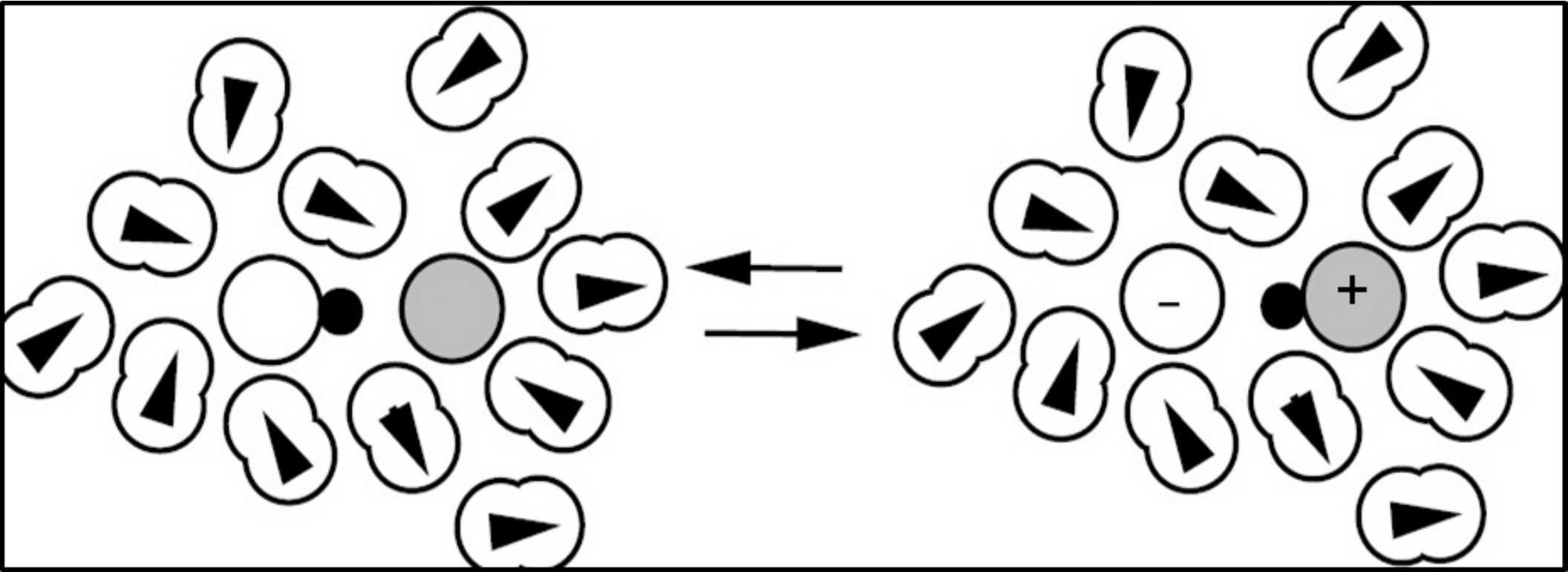}
\caption{Schematic representation showing the local solvent structure around the phenol-triethylamine complex. The covalent form $\mbox{PhO-H}\cdots\mbox{NR}_{3}$ of the phenol-amine complex (left) is unfavorably solvated by the polar solvent molecules. This induces a proton transfer giving rise to the ionic form $\mbox{PhO}^{-}\cdots\mbox{H-}\mbox{NR}_{3}^{+}$ (right). Subsequent solvent dynamics can lead to solvent polarization that favors the covalent form and the reverse proton transfer.}
\label{fig:ptransfer}
\end{center}
\end{figure}

There are many instances where more detailed information about the bath dynamics and its coupling to the subsystem is important. Examples are provided by proton and electron transfer processes in condensed phases or biological systems. As a specific example, consider the proton transfer reaction in a phenol-amine complex, $\mbox{PhO-H}\cdots\mbox{NR}_{3}\rightleftharpoons\mbox{PhO}^{-}\cdots\mbox{H-}\mbox{NR}_{3}^{+}$, when the complex is solvated by polar molecules (see Fig.~\ref{fig:ptransfer}). The proton transfer events are strongly correlated with local solvent  collective polarization changes. Subtle changes in the orientations of neighboring solvent molecules can induce proton transfers within the complex, which, in turn, influence the polarization of the solvent. The treatment of the dynamics in such cases requires detailed information about the dynamics of the environment and its coupling to the quantum process. It is difficult to capture such subtle effects without fully accounting for dynamics of individual solvent molecules in the bath.

When investigating the dynamics of a quantum system it is often useful and appropriate to take into account the characteristics of the different degrees of freedom that comprise the system. The fact that electronic and nuclear motions occur on very different time scales, as a result of the disparity in their masses, forms the basis for the Born-Oppenheimer approximation where the nuclear-configuration-dependent electronic energy is used as the potential energy for the evolution of the nuclear degrees of freedom. This distinction between electronic and nuclear degrees of freedom is an example of the more general partition of a quantum system into subsystems with different characteristics.

Since the scale separation in the Born-Oppenheimer approximation is approximate, it can break down and its breakdown leads to coupling among many electronic energy surfaces. When this occurs, the evolution is no longer described by adiabatic dynamics on a single potential energy surface and nonadiabatic effects become important. Nonadiabatic dynamics plays an essential role in the description of many physical phenomena, such as photochemical processes where transitions among various electronic states occur as a result of avoided crossings of adiabatic states or conical intersections between potential energy surfaces.

In the examples presented above the molecules comprising the bath are often much more massive than those in the subsystem ($M \gg m$). This fact motivates the construction of a quantum-classical description where the bath, in the absence of interactions with the quantum subsystem, is described by classical mechanics. Mixed quantum-classical methods provide a means to investigate quantum dynamics in large complex systems, since fully quantum treatments of the dynamics of such systems are not feasible. The study of such open quantum-classical systems is the main topic of this review. Since quantum and classical mechanics do not easily mix, one must consider the properties of schemes that combine these two types of mechanics. One such scheme, quantum-classical Liouville dynamics, will be discussed in detail and its features will be compared to other quantum-classical and full quantum methods.

\section{Open Quantum Systems}\label{sec:open}

Since the quantum systems we study are rarely isolated and interact with the environments within which they reside, the investigation of the dynamics of such open quantum systems is a worthy endeavor. The full description of the time evolution of a composite quantum system comprising a subsystem and bath is given by the quantum Liouville equation,
\begin{equation}
\frac{\partial}{\partial t} \hat{\rho}(t) = -\frac{i}{\hbar}
[\hat{H}, \hat{\rho}(t)], \label{eq:qle}
\end{equation}
where $\hat{\rho}(t)$ is the density matrix at time $t$, $\hat{H}$ is the total Hamiltonian, and the square brackets denote the commutator.

Introducing some of the notation that will be used in this paper, we denote by $\hat{q}=\{\hat{q_i}\},\;i=1,...,n $ the coordinate operators for the $n$ subsystem degrees of freedom with mass $m$, while the remaining $N$ bath degrees of freedom with mass $M$ have coordinate operators $\hat{Q}=\{\hat{Q_i}\},\;i=1,...,N$. (The formalism is easily generalized to situations where the masses $m$ and $M$ depend on the particle index.) The total Hamiltonian takes the form
\begin{equation}
\hat{H} = \frac{\hat{P}^2}{2M} + \frac{\hat{p}^2}{2m} +
\hat{V}(\hat{q}, \hat{Q}), \label{eq:hamiltonian}
\end{equation}
where the momentum operators for the subsystem and bath are $\hat{p}$ and $\hat{P}$, respectively. It is convenient to write the potential energy operator, $\hat{V}(\hat{q}, \hat{Q})$ as a sum of subsystem, bath and coupling contributions: $\hat{V}(\hat{q}, \hat{Q})=\hat{V}_s(\hat{q}) + \hat{V}_b(\hat{Q}) + \hat{V}_c(\hat{q}, \hat{Q})$. In this case the Hamiltonian operator can be written as a sum of contributions,
\begin{equation}
\hat{H} = \hat{h}_s + \hat{H}_b +\hat{V}_c,
\label{eq:H-breakup}
\end{equation}
where $\hat{h}_s=\frac{\hat{p}^2}{2m} +\hat{V}_s(\hat{q})$ is the quantum subsystem Hamiltonian, $\hat{H}_b=\frac{\hat{P}^2}{2M} +\hat{V}_b(\hat{Q})$ is the quantum bath Hamiltonian and $\hat{V}_c$ is the coupling between these two subsystems.

Most often in considering the dynamics of such open quantum systems one traces over the bath since it is the dynamics of the subsystem that is of interest. As noted in the Introduction, a considerable research effort has been devoted to the construction of equations of motion for the reduced density matrix, $\hat{\rho}_s(t)= {\rm Tr}_b \hat{\rho}(t)$. The Redfield equation~\cite{redfield65} describes the dynamics of a subsystem weakly coupled to a bath with suitably fast bath relaxation time scales, since a Born-Markov approximation is made in its derivation. In a basis of eigenstates of $\hat{h}_s$, $\hat{h}_s |\lambda \rangle =\epsilon_\lambda |\lambda \rangle$, it has the form,
\begin{equation}
\frac{\partial }{\partial t} \rho_s^{\lambda \lambda'}(t)=-i \omega_{\lambda \lambda'} \rho_s^{\lambda \lambda'}(t) +R_{\lambda \lambda';\nu \nu'}
\rho_s^{\nu \nu'}(t),
\end{equation}
where the summation convention has been used. This convention will be used throughout the paper when confusion is unlikely.  Here $\omega_{\lambda \lambda'} = (\epsilon_\lambda -\epsilon_{\lambda'})/\hbar$, while the second term on the right accounts for dissipative effects due to the bath. Remaining within the Born-Markov approximation, the general form of the equation of motion for a reduced density matrix that guarantees its positivity is given by the Lindblad equation~\cite{lindblad76},
\begin{eqnarray}
\frac{\partial }{\partial t} \hat{\rho}_s(t)&=&-\frac{i}{\hbar}[\hat{h}_s,  \hat{\rho}_s(t)] \\
&&+\frac{1}{2} \sum_j \left( [\hat{L}_j\hat{\rho}_s(t),\hat{L}_j^\dagger]+ [\hat{L}_j,\hat{\rho}_s(t)\hat{L}_j^\dagger] \right),\nonumber
\end{eqnarray}
where the $\hat{L}_j$ are operators that account for interactions with the bath. In addition to these equations, a number of other expressions for the evolution of the reduced density matrix have been derived. These include various master equations and generalized quantum master equations.
There is a large literature dealing with open quantum systems, which is described and surveyed in books on this topic.~\cite{0book-davies,weiss99, breuer06} In such reduced descriptions information about the bath is contained in parameters that enter in the operators that describe the coupling between the subsystem and bath. Also, quantum-classical versions of the Redfield~\cite{toutounji05} and Lindblad~\cite{toutounji01} equations have been derived.

There are many applications, such as those mentioned in the Introduction, where a more detailed treatment of the bath dynamics and its interactions with the subsystem is required, even though one's primary interest is in the dynamics of the subsystem. If, as we suppose here, the systems we wish to study are large and may involve complex molecular constituents, a full quantum mechanical treatment is beyond the scope of existing computational power and algorithms. Currently, the only viable way to simulate the dynamics of such systems is by using mixed quantum-classical schemes. Quantum-classical methods in a variety of forms and derived in a variety of ways have been used to simulate the dynamics.~\cite{herman94,tully98,martinez98,kapral06_2,tully12} Mean field and surface-hopping methods are widely employed and will be discussed in some detail below. Mixed quantum-classical dynamics~\cite{agostini13} based the on the exact time-dependent potential energy surfaces derived from the exact decomposition of electronic and nuclear motions~\cite{abedi10} has been constructed.
In addition, semiclassical path integral formulations of quantum mechanics~\cite{sun97,sun98,makri98,miller09,makri12} and ring polymer dynamics methods~\cite{ring-polymer13} have been developed to approximate the dynamics of open quantum systems.

In the next section the specific version of mixed quantum-classical dynamics that is the subject of this review, quantum-classical Liouville dynamics, will be described. The passage from quantum to classical dynamics is itself a difficult problem with an extensive literature, and decoherence is often invoked to effect this passage.~\cite{0book-joos03,zurek91} Considerations based on decoherence can also be used motivate the use of mixed quantum-classical descriptions.~\cite{shiokawa02} Mean-field and surface-hopping methods suffer from difficulties related to the treatment of coherence and decoherence, and these methods will be discussed in the context of the quantum-classical Liouville equation, which is derived and discussed in the next two sections. Some applications of the theory to specific systems will be presented in order to test the accuracy of this equation description and the algorithms used to simulate its dynamics.

\section{Quantum-Classical Liouville Dynamics}\label{sec:qcl}

The first step in constructing a quantum-classical Liouville description is to introduce a phase space representation of the bath degrees of freedom in preparation for the passage to the classical bath limit. This is conveniently done by introducing a partial Wigner transform~\cite{wigner32} over the bath degrees of freedom defined by
\begin{equation}
\hat{\rho}_W (R, P) = \frac{1}{(2 \pi \hbar)^N}\int dZ \; e^{i P \cdot
Z/\hbar} \langle R - \frac{Z}{2} | \hat{\rho} | R + \frac{Z}{2}
\rangle, \label{eq:wigner1}
\end{equation}
with an analogous expression for the partial Wigner transform of an operator $\hat{A}_W(R,P)$ in which the prefactor $(2 \pi \hbar)^{-N}$ is absent. We let $X=(R,P)$ to simplify the notation. The quantum Liouville equation then takes the form,
\begin{eqnarray}
\frac{\partial }{\partial t} \hat{\rho}_W (X,t)& = &
-\frac{i}{\hbar} \left(\hat{H}_W
e^{\hbar \Lambda /2i}\hat{\rho}_W(t)
-  \hat{\rho}_W(t) e^{\hbar \Lambda
/2i}\hat{H}_W \right). \label{eq:qle_wigner}
\nonumber \\
\end{eqnarray}
To obtain this equation the formula for the Wigner transform of a product of operators~\cite{imre67},
\begin{equation}
( \hat{A} \hat{B} )_W (X) = \hat{A}_W(X) e^{\hbar
\Lambda /2i} \hat{B}_W (X), \label{eq:wigner2}
\end{equation}
was used. Here the operator $\Lambda = \overleftarrow{\nabla}_P \cdot \overrightarrow{\nabla}_R - \overleftarrow{\nabla}_R \cdot \overrightarrow{\nabla}_P$, where the arrows denote the directions in which the derivatives act, is the negative of the Poisson bracket operator,
\begin{eqnarray}
\hat{A}_W \Lambda \hat{B}_W&=& -\left(\nabla_R \hat{A}_W \cdot \nabla_P \hat{B}_W - \nabla_P \hat{A}_W \cdot \nabla_R \hat{B}_W \right)\nonumber \\ &\equiv &-\{\hat{A}_W. \hat{B}_W \}
\end{eqnarray}
The partial Wigner transform of the total Hamiltonian is,
\begin{equation}
\hat{H}_W(X) = \frac{P^2}{2M} + \frac{\hat{p}^2}{2m} + \hat{V} (\hat{q}, R)\equiv  \frac{P^2}{2M} + \hat{h}(\hat{q},R). \label{eq:hamiltonian_wigner}
\end{equation}
We have dropped the subscript W on the potential energy operator to simplify the notation; when the argument contains $R$ the partial Wigner transform is implied.

\subsection*{Derivation of the QCLE}
The quantum-classical Liouville equation (QCLE) can be derived by formally expanding the exponential operators on the right side of Eq.~(\ref{eq:qle_wigner}) to ${\mathcal O}(\hbar)$.~\cite{alek81,geras82} The truncation of the series expansion can be justified for systems where the masses of particles in the environment are much greater than those of the subsystem, $ M \gg m$.~\cite{kapral99} Scaling similar to that in the microscopic derivation of the Langevin equation for Brownian motion from the classical Liouville equation~\cite{mazur70} can be used for this purpose, and we may write the equations in terms of the reduced bath momenta, $\tilde{P}=\mu P$ where $\mu = (m/M)^{1/2}$. In this variable the kinetic energies of the light and heavy particle systems are comparable so that $P$ is of order $M^{1/2}$. To see this more explicitly we introduce scaled units where energy is expressed in the unit $\epsilon_0$, time in $t_0=\hbar/\epsilon_0$ and length in units of $\lambda_m=(\hbar^2 /m \epsilon_0)^{1/2}$. Using these length and time units, the scaling factor for the momentum is $p_m=(m\lambda_m/t_0)=(m \epsilon_0)^{1/2}$. Thus, in terms of the scaled variables $R'=R/\lambda_m$ and $P'=\tilde{P}/p_m$ we have
\begin{equation}
e^{\hbar \Lambda /2i} =e^{\mu \Lambda' /2i}= 1+\mu \Lambda' /2i +{\mathcal O}(\mu^2),
\end{equation}
where the prime on $\Lambda$ indicates that it is expressed in the primed variables. Note that for a system characterized by a temperature $T$ the small parameter $\mu$ can be written as the ratio of the thermal de Broglie wavelengths $\lambda_M=(\hbar^2/M k_BT)^{1/2}$ and $\lambda_m=(\hbar^2/m k_BT)^{1/2}$ of the heavy bath and light subsystem particles, respectively, $\lambda_M/\lambda_m=\mu$, and truncation of the dynamics to terms of ${\mathcal O}(\mu)$ effectively averages out the quantum bath oscillations on the longer quantum length scale of the light subsystem.

Inserting the expression for the exponential Poisson bracket operator, valid to ${\mathcal O}(\mu)$,  into the scaled version of Eq.~(\ref{eq:qle_wigner}) and returning to unscaled variables we obtain the quantum-classical Liouville equation~\cite{kapral99},
\begin{eqnarray} \label{eq:dmatabs}
&&\frac{\partial}{\partial t} \hat{\rho}_W(X,t)=-i \hat{\mathcal
L}\hat{\rho}_W(t)= -\frac{i}{\hbar}
[\hat{H}_W, \hat{\rho}_W(t)] \\
&& \qquad \quad
+\frac{1}{2} \Big( \{\hat{H}_W,\hat{\rho}_W(t)\}
- \{\hat{\rho}_W(t), \hat{H}_W\} \Big).\nonumber
\end{eqnarray}
Additional discussion of this equation can be found in the literature~\cite{kapral06_2,kapral99,alek81,geras82,boucher88,zhang88,donoso98,horenko02,shi04a,thorndyke05,burghardt11}.
Comparison of the second and third equalities in this equation defines the QCL operator $i\hat{{\mathcal L}}$, and given this definition the formal solution of the QCLE is
\begin{equation}
\hat{\rho}_W(X,t)=e^{-i \hat{\mathcal L}t} \hat{\rho}_W(X) , \label{eq:dmatabs-soln}
\end{equation}
where we let $\hat{\rho}_W(X)=\hat{\rho}_W(X,0)$ here and in the following to simplify the notation. The QCLE (\ref{eq:dmatabs}) may also be written in the form~\cite{nielsen01},
\begin{equation}
\frac{\partial}{\partial t} \hat{\rho}_W(X,t)=-\frac{i}{\hbar}\Big(
\stackrel{\rightarrow}{{\mathcal H}}_{\Lambda}\hat{\rho}_W(t)-\hat{\rho}_W(t)
\stackrel{\leftarrow}{{\mathcal H}}_{\Lambda}\Big), \label{eq:dmatabs-fb}
\end{equation}
which resembles the quantum Liouville equation but the quantum Hamiltonian operator is replaced by the forward and backward operators,
\begin{eqnarray}\label{eq:sah}
{\stackrel{\rightarrow}{{\cal H}}_{\Lambda}}=\hat{H}_W\left(1+ \frac{\hbar \Lambda}{2i}\right), \quad
{\stackrel{\leftarrow}{{\cal H}}_{\Lambda}}= \left(1+ \frac{\hbar \Lambda}{2i}\right)\hat{H}_W.
\end{eqnarray}
This form of the evolution equation has been used to discuss the statistical mechanical properties of QCL dynamics~\cite{nielsen01}, and will be used later to derive approximate solutions to the QCLE.

In applications it is often more convenient to evolve an operator rather than the density matrix and we may easily write the evolution equations for operators. Starting from the Heisenberg equation of motion for an operator $\hat{B}$,
\begin{equation}
\frac{d }{d t}\hat{B}(t)=\frac{i}{\hbar} [\hat{H},\hat{B}(t)],
\end{equation}
one can carry out an analogous calculation to find the QCLE for the partial Wigner transform of this operator:
\begin{equation}
\frac{d}{d t} \hat{B}_W(X,t)=i \hat{\mathcal
L}\hat{B}_W(t), \label{eq:dmatabs-op}
\end{equation}
whose formal solution can be written as
\begin{equation}
\hat{B}_W(X,t)=e^{i \hat{\mathcal L}t} \hat{B}_W(X). \label{eq:dmatabs-op-soln}
\end{equation}

\subsection*{QCLE from linearization}\label{sec:Equation}
The QCLE, when expressed in the adiabatic or subsystem bases, has been derived from linearization of the path integral expression for the density matrix by Shi and Geva~\cite{shi04a}. It can also be derived in a basis-free form by linearization~\cite{bonella10} and it is instructive to sketch this derivation here to see how the QCLE can be obtained from a perspective that differs from that discussed in the previous subsection.

The time evolution of the quantum density operator from time $t$ to a short later time $t+\Delta t$ is given by
\begin{equation}\label{eq:Heisenberg2}
\hat{\rho}(t+\Delta t)=e^{-{i\over\hbar}{\hat H}\Delta t}\hat{\rho}(t)e^{{i\over\hbar}{\hat H} \Delta t} .
\end{equation}
Writing the Hamiltonian in the form $\hat{H}= \hat{P}^2/2M +\hat{h}({\hat q},{\hat Q})$, for this short time interval, a Trotter factorization of the propagators can be made:
\begin{equation}
e^{\pm{i\over\hbar}{\hat H} \Delta t}\approx
e^{\pm{i\over\hbar}{{\hat P}^2\over 2M}\Delta t}
e^{\pm{i\over\hbar} {\hat h}({\hat Q})\Delta t} +{\mathcal O}(\Delta t^2).
\end{equation}
For simplicity, we have suppressed the $\hat{q}$ dependence in $\hat{h}$ but kept the $\hat{Q}$ dependence since it is required in the derivation. Working in the $\{Q\}$ representation for the bath, inserting resolutions of the identity, and evaluating the contributions coming from the kinetic energy operators that appear in the resulting expression, we obtain
\begin{eqnarray}
&&\langle Q | \hat{\rho}(t+\Delta t) | Q' \rangle = \int dQ_0 dP_0
dQ_0' dP_0' \;e^{{i\over\hbar}P_0\cdot (Q-Q_0)}  \nonumber \\
&& \qquad \qquad \times e^{-{i\over\hbar}{P^2_0\over 2M}\Delta t}  e^{-{i\over\hbar} {\hat h}(Q_0)\Delta t} \langle Q_0 | \hat{\rho}(t) | Q_0' \rangle \nonumber \\
&& \qquad \qquad \times
e^{-{i\over\hbar}P_0'\cdot (Q'-Q_0')} e^{{i\over\hbar} {P'^2_0\over
2M}\Delta t}  e^{{i\over\hbar} {\hat h}(Q_0')\Delta t}.
\end{eqnarray}
Next, we make the change of variables ${\bar R} =(Q+Q')/2$ and $Z = Q-Q'$, along with similar variable changes
for the momenta $\bar{P}=P+P'$ and $\Delta P = (P-P')/2$. In the new variables, the density matrix element is
\begin{eqnarray} \label{eq:dent+dt}
&&\langle {\bar R} + {Z \over 2} | \hat{\rho}(t+\Delta t) |
{\bar R} - {Z \over 2}\rangle =  \\
&& \qquad \int d{\bar R}_0 d{\bar P}_0
dZ_0 d\Delta P_0 \; e^{{i\over\hbar}{\bar P}_0\cdot (Z-Z_0)}
\nonumber \\
&&\qquad  \times e^{{i\over\hbar}\Delta P_0 \cdot ({\bar R}-{\bar R}_0)} e^{-{i\over\hbar} {{\bar P}_0\over M}\Delta P_0 \Delta t}
e^{-{i\over\hbar} {\hat h}({\bar R}_0+{Z_0\over 2})\Delta t} \nonumber \\
&& \qquad  \times \langle {\bar R}_0 + {Z_0\over
2} | \hat{\rho}(t) | {\bar R}_0 - {Z_0\over 2} \rangle e^{{i\over\hbar} {\hat h}({\bar R}_0-{Z_0\over 2})\Delta t} .\nonumber
\end{eqnarray}
We may now make use of the definition of the partial Wigner transform (see Eq.~(\ref{eq:wigner1})) in the expression for the matrix element of the density operator in Eq.~(\ref{eq:dent+dt}) to derive an equation of motion for ${\hat \rho}_W({\bar R},{\bar P},t)$. To do this we first expand the exponentials that depend on $\Delta t$ to first order in this parameter; e.g., $e^{-{i\over\hbar}{{\bar P}_0\over M}\cdot \Delta P_0 \Delta t}\approx 1 - {i\over\hbar} {{\bar P}_0\over M}\cdot \Delta P_0 \Delta t$. We may then use this expansion to compute the finite difference expression $(\langle {\bar R} + {Z\over 2} | \hat{\rho}(t+\Delta t) | {\bar R} - {Z\over 2}\rangle -\langle {\bar R} + {Z\over 2} | \hat{\rho}(t) |{\bar R} - {Z\over 2}\rangle )/\Delta t$. Finally we multiply the equation by $e^{-{i\over\hbar}{\bar P}\cdot Z}$, integrate the result over $Z$ and take the limit $\Delta t \to 0$. The result of these operations is
\begin{eqnarray}\label{eq:MixedEv1}
&&\frac{\partial}{\partial t}{\hat \rho}_W({\bar X},t) = -  {{\bar P} \over M} \cdot \nabla_{\bar R} {\hat \rho}_W({\bar X},t)\\
&&  +{i\over\hbar} \int dZ e^{-{i\over \hbar} {\bar P}\cdot Z} {\hat h}({\bar R}+{Z\over 2})
\langle {\bar R} + {Z\over 2} | \hat{\rho}(t) | {\bar R} - {Z\over 2} \rangle \nonumber \\ \nonumber
&& - {i\over\hbar} \int dZ e^{-{i\over \hbar} {\bar P}\cdot Z}
\langle {\bar R} + {Z\over 2} | \hat{\rho}(t) | {\bar R} - {Z\over 2} \rangle {\hat h}({\bar R} -{Z \over 2}), \nonumber
\end{eqnarray}
where $\bar{X}=(\bar{R},\bar{P})$. This integro-differential equation describes the full quantum evolution of the density matrix element; however, it is not a closed equation for ${\hat \rho}_W(t)$ because of the dependence of ${\hat h}({\bar R}\pm{Z \over 2})$ on $Z$. If we make use of the expansion of this operator to linear order in $Z$, ${\hat h}({\bar R}\pm{Z \over 2}) \approx {\hat h}({\bar R}) \pm \frac{Z}{2} \cdot \nabla_{\bar{R}} {\hat h}({\bar R})$ when performing the integrals in the right side of Eq.~(\ref{eq:MixedEv1}), we obtain the QCLE in Eq.~(\ref{eq:dmatabs}). The linearization approximation can be justified for systems where $M \gg m$.~\cite{bonella10} The same scaled variables introduced above in the first derivation may also be used to re-express Eq.~(\ref{eq:MixedEv1}) in scaled form. In this scaled form one may show that the expansion in $Z$ is equivalent to an expansion in the mass ratio parameter $\mu$.

\subsection*{QCLE in a dissipative environment}

At times it may be convenient to further partition the bath into two subsets of degrees of freedom, $X=(X_0,X_a)$, where the $X_0$ variables are directly coupled to the quantum subsystem and the remainder of the (usually large number of) degrees of freedom denoted by $X_a$ only participate in the subsystem dynamics indirectly through their coupling to $X_0$. In such a case we can project these $X_a$ degrees of freedom out of the QCLE to derive a dissipative evolution equation for the quantum subsystem and the directly coupled $X_0$ variables~\cite{kapral01}. For example, such a description could be useful in studies of proton or electron transfer in biomolecules where remote portions of the biomolecule and solvent need not be treated in detail but, nevertheless, these remote degrees of freedom do provide a source of decoherence and dissipation on the relevant degrees of freedom.

For a system of this type the partially Wigner transformed total Hamiltonian of the system is,
\begin{eqnarray}
   \hat{H}_W(X) & = & \frac{P_a^2}{2M} + \frac{P_0^2}{2M} + \frac{\hat{p}^2}{2m} + \hat{V}(\hat{q},R_0, R_a) \nonumber \\
   & \equiv & \frac{P_a^2}{2M} + \frac{P_0^2}{2M} + \hat{h}(R) \;.
\end{eqnarray}
The potential energy operator, $\hat{V}(\hat{q}, R_0, R_a)=\hat{V}(\hat{q},R_0)+ V_a(R_a) + V_{0a}(R_0,R_a)$, includes all of the coupling contributions discussed above, namely, the potential energy operator $\hat{V}(\hat{q},R_0)$ for the quantum subsystem and directly coupled degrees of freedom, the potential energy of the outer bath $V_a$ and the coupling between the two bath subsystems, $V_{0a}$.

An evolution equation for the reduced density matrix of the quantum subsystem and directly coupled $X_0$ degrees of freedom,
\begin{equation}
\hat{\rho}_W(X_0,t)= \int dX_a\; \hat{\rho}_{W}(X_0,X_a,t),
\end{equation}
can be obtained by using projection operator methods.~\cite{nakajima58,0chap-zwanzig61} The result of this calculation is a dissipative QCLE, which takes the form~\cite{kapral01},
\begin{eqnarray}
&&\frac{\partial }{\partial t}\hat{\rho}_{W}(X_0t) =
-i\hat{{\mathcal L}} \hat{\rho}_W(t) -
{\mathcal F}\cdot \frac{\partial }{\partial P_0} \hat{\rho}_W(t)
\nonumber \\
&&
\qquad + \zeta(R_0):
\frac{\partial }{\partial P_0} \left(\frac{P_0}{M} +
k_BT\frac{\partial }{\partial P_0}\right)
\hat{\rho}_W(t),
\label{eq:mqcfp}
\end{eqnarray}
where $i\hat{{\mathcal L}}$ is the QCL operator introduced earlier but now only for the quantum subsystem and $X_0$ bath degrees of freedom.
The effects of the less relevant $X_a$ bath degrees of freedom are accounted for by the mean force ${\mathcal F}$ defined by ${\mathcal F}(R_0)=
-\langle \partial V_{0a}/\partial R_0 \rangle_0 \equiv \langle F_{0a} \rangle_0$, where the average is over a canonical equilibrium distribution involving the Hamiltonian $H_0=\frac{P_a^2}{2M}+ V_a(R_a) + V_{0a}(R_0,R_a)$. The Fokker-Planck-like operator in Eq.~(\ref{eq:mqcfp}) depends on the fixed particle friction tensor, $\zeta(R_0)$, defined by
\begin{equation}
\zeta(R_0)=\int_0^\infty dt \; \langle \delta F_{0a}(t) \delta F_{0a} \rangle_0/k_BT,
\end{equation}
where $\delta F_{0a}= F_{0a}-{\mathcal F}$,  and its time evolution is given by the classical dynamics of the $X_a$ degrees of freedom in the field of the fixed $R_0$ coordinates. The quantum-classical limit of the multi-state Fokker-Planck equation introduced by Tanimura and Mukamel~\cite{mukamel94} is similar to the dissipative QCLE~(\ref{eq:mqcfp}) when expressed in the subsystem basis.

\section{Some properties of the QCLE}\label{sec:properties}

The QCLE specifies the time evolution the density matrix of the entire system comprising the subsystem and bath and conserves the energy of the system. If the coupling potential $\hat{V}_c(\hat{q},R)$ in the Hamiltonian is zero, the density matrix factors into a product of subsystem and bath density matrices, $\hat{\rho}(X,t)=\hat{\rho}_s(t) \rho_b(X,t)$. In this limit the subsystem density matrix satisfies the quantum Liouville equation,
\begin{equation}
\frac{\partial }{\partial t} \hat{\rho}_s (t)= -\frac{i}{\hbar} [\hat{h}_s,\hat{\rho}_s(t)],
\label{eq:sub}
\end{equation}
and bath phase space density satisfies the classical Liouville equation,
\begin{equation}
\frac{\partial }{\partial t} {\rho}_b (X,t)= \{H_b(X),\rho_b(X,t)  \}.
\label{eq:bath}
\end{equation}

While the bath evolves by classical mechanics when it is not coupled to the quantum subsystem, its evolution is no longer classical when coupling is present. As we shall see in more detail below, not only does the bath serve to account for the effects of decoherence and dissipation in the subsystem, it is also responsible for the creation of coherence. Conversely, the subsystem can interact with the bath to modify its dynamics. This leads to a very complicated evolution, but one which incorporates many of the features that are essential for the description of physical systems.

Often, when considering the dynamics of a quantum system coupled to a bath, the bath is modeled by a collection of harmonic oscillators which are bilinearly coupled to the quantum subsystem. In this case we may write the coupling potential as $\hat{V}_{c}(\hat{q}, R)=\hat{C}(\hat{q})\cdot R$. The partially Wigner transformed Hamiltonian then takes the form,
\begin{eqnarray}
\hat{H}_W(X) &=&  \frac{\hat{p}^2}{2m} + \hat{V}_s(\hat{q})+ \frac{P^2}{2M} + {V}_h(X)+\hat{C}(\hat{q})\cdot R \nonumber \\
&\equiv& \hat{h}_s +H_h(X)+\hat{C}(\hat{q})\cdot R \label{eq:harmonic_wigner}
\end{eqnarray}
where $H_h$ is the harmonic oscillator bath Hamiltonian. When the Hamiltonian has this form one may show easily that $(\hat{H}_W \Lambda^2\hat{\rho}_W(t)-  \hat{\rho}_W(t) \Lambda^2 \hat{H}_W )=0$. Consequently, when the exponential Poisson bracket operators in Eq.~(\ref{eq:qle_wigner}) are expanded in a power series, the series truncates at linear order and we obtain the QCLE in the form given in Eq.~(\ref{eq:dmatabs-fb}); thus, the QCLE is exact for general quantum subsystems which are bilinearly coupled to harmonic baths. For more general Hamiltonian operators the series does not truncate and QCL dynamics is an approximation to full quantum dynamics.

Quantum and classical mechanics do not like to mix. The coupling between the smooth classical phase space evolution of the bath and the quantum subsystem dynamics with quantum fluctuations on small scales presents challenges for any quantum-classical description. The QCLE, being an approximation to full quantum dynamics, is not without defects. One of its features that requires consideration is its lack of a Lie algebraic structure. The quantum commutator bracket $(i/\hbar)[\hat{A},\hat{B}]$ and Poisson bracket $\{A,B\}$ for quantum and classical mechanics,  respectively, are bilinear, skew symmetric, and satisfy the Jacobi identity, so that these brackets have Lie algebraic structures. The quantum-classical bracket, $(\hat{A}_W,\hat{B}_W)_{QC}=(i/\hbar)[\hat{A}_W,\hat{B}_W]-(\{\hat{A}_W,\hat{B}_W\}-\{\hat{B}_W,\hat{A}_W\})/2$, which is the combination of the commutator and the Poisson bracket terms does not have such a Lie algebraic structure. While this bracket is bilinear and skew symmetric, it does not exactly satisfy the Jacobi identity. Instead, the Jacobi identity is satisfied only to order $\hbar$ (or $\mu$ if scaled variables are considered):
$(\hat{A}_W,(\hat{B}_W,\hat{C}_W))+(\hat{C}_W,(\hat{A}_W,\hat{B}_W)) +(\hat{B}_W,(\hat{C}_W,\hat{A}_W))={\cal O}(\hbar)$.
The lack of a Lie algebraic structure, its implications for the dynamics, and the construction of the statistical mechanics of quantum-classical systems were discussed earlier~\cite{nielsen01,0chap-kapral02} where full details may be found. For example, the standard linear response derivations of quantum transport properties have to be modified, and in quantum-classical dynamics the evolution of a product of operators is not the product of the evolved operators; this is true only to order $\mu$. These feature are not unique to QCL dynamics and almost all mixed quantum-classical methods used in simulations suffer from such defects, although they are rarely discussed. Mixed quantum-classical dynamics and its algebraic structure continue to attract the attention of researchers.~\cite{salcedo96,salcedo12,prezhdo97b,sergi05,prezhdo06,salcedo07,agostini07,hall_reginatto05,hall08}

One way to bypass some of the difficulties in the formulation of the statistical mechanics of quantum-classical systems that are associated with a lack of a Lie algebraic structure is to derive expressions for average values and transport property using full quantum statistical mechanics. Then, starting with these exact quantum expressions, one may approximate the quantum dynamics by quantum-classical dynamics.~\cite{sergi04,kim05,kim05a,hsieh14} In this framework the expectation value of an observable $\hat{B}_W(X)$ is given by,
\begin{equation}
\overline{B(t)}={\rm Tr}_s \int dX \; \hat{B}_W(X,t) \hat{\rho}_W(X),
\end{equation}
where $\hat{\rho}_W(X)$ is the partial Wigner transform of the initial quantum density operator and the evolution of $\hat{B}_W(X,t)$ is given by the QCLE. Similarly, the expressions for transport coefficients involve time integrals of correlation functions $C_{AB}(t)$ of the form,
\begin{equation}
\label{eq:real_time_corr}
C_{AB}(t) = \frac{1}{Z_Q}\int dX\; \left[ \left(e^{-\beta\hat{H}}\hat{A}\right)_W(X)
 \hat{B}_W(X,t) \right],
\end{equation}
where $\big(e^{-\beta\hat{H}}A\big)_W(X)$ is the partial Wigner transform of the product of the quantum canonical density operator and the operator $\hat{A}$, and the time evolution of $\hat{B}_W(X,t)$ is again given by the QCLE. Such formulations preserve the full quantum equilibrium structure which, while difficult to compute, is computationally much more tractable than full quantum dynamics.~\cite{poulsen03,ananth10} The importance of quantum versus classical equilibrium sampling on reactive-flux correlation functions, whose time integrals are reaction rate coefficients, has been investigated in the context of quantum-classical Liouville dynamics.~\cite{kim05a} In this review we shall focus on dynamics but, when applications are considered, the above equations that contain the quantum initial or equilibrium density matrices will be used.

\section{Surface hopping, coherence and decoherence} \label{sec:surf-hop}
Surface-hopping methods are commonly used to simulate the nonadiabatic dynamics of quantum-classical systems. In such schemes the bath phase space variables follow Newtonian trajectories on single adiabatic surfaces. Nonadiabatic effects are taken into account by hops between different adiabatic surfaces that are governed by probabilistic rules.

\begin{figure}[htbp]
\begin{center}
\includegraphics[width=0.8\columnwidth]{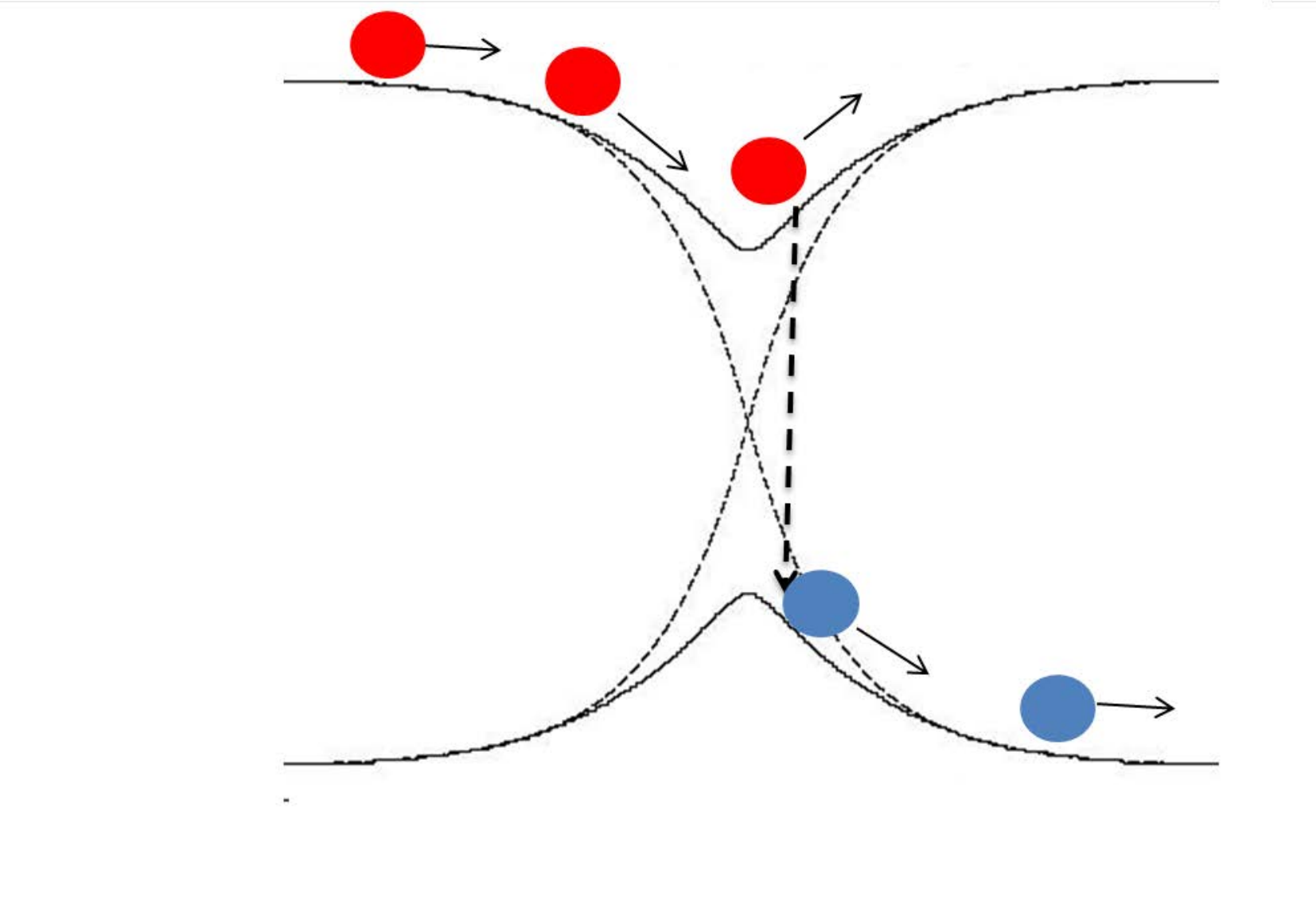}
\caption{Schematic representation of the evolution of a wave packet in a two-level system with a simple avoided crossing. The diabatic (crossing curves) and adiabatic (avoided crossing curves) are shown. Following the nonadiabatic transition from the upper to lower adiabatic surfaces, the system continues to evolve on the lower surface until the next nonadiabatic transition.}
\label{fig:surf-hop}
\end{center}
\end{figure}

One of the most widely used schemes is Tully's fewest-switches surface hopping.~\cite{tully90,tully91,tully98} In this method one assumes that the electronic wave function $|\psi(R(t),t)\rangle$ depends on the time-dependent nuclear positions $R(t)$, whose evolution is governed by a stochastic algorithm. More specifically, choosing to work in a basis of the instantaneous adiabatic eigenfunctions of the Hamiltonian $\hat{h}(R(t))$, $\hat{h}(R(t)) |\alpha;R(t) \rangle =E_\alpha(R(t)) |\alpha;R(t) \rangle$, we may expand the wave function as $|\psi(R(t),t)\rangle= \sum_\alpha c_\alpha(t) |\alpha;R(t) \rangle$. An expression for the time evolution of the subsystem density matrix $\hat{\rho}_s(t)=|\psi(R(t),t)\rangle \langle \psi(R(t),t) |$ can be obtained by substitution into the Schr\"odinger equation. The equations of motion for its matrix elements, $\rho_s^{\alpha \alpha'}(t)= c_\alpha(t) c^*_{\alpha'}(t)$ are given by,
\begin{eqnarray} \label{eq:sub-den-SH}
&&\frac{d \rho_s^{\alpha \alpha'}(t)}{dt}= -i \omega_{\alpha \alpha'}(R(t))\rho_s^{\alpha \alpha'}(t)\\
 && \qquad  -\frac{P(t)}{M} \cdot  d_{\alpha \beta}(R(t)) \rho_s^{\beta \alpha'}(t) -\frac{P(t)}{M} \cdot d^*_{\alpha' \beta}(R(t)) \rho_s^{\alpha \beta}(t).\nonumber
\end{eqnarray}
In this equation $d_{\alpha \beta}$ is the nonadiabatic coupling matrix element, $d_{\alpha \beta}= \langle \alpha; R | \frac{\partial}{\partial R}|\beta;R \rangle$. From this expression the rate of change of the population in state $\alpha$ may be written as
\begin{equation}
\dot{\rho}_s^{\alpha \alpha} =-\frac{2P}{M} \cdot  \Re(d_{\alpha \alpha''} \rho_s^{\alpha'' \alpha}),
\end{equation}
where, for simplicity, we have suppressed the time dependence in the variables, and $\Re$ stands for the real part. This rate has contributions from transitions to and from all other states $\alpha''$. Consider a single specific state $\beta$. Then transitions into $\alpha$ from $\beta$ and out of $\alpha$ to $\beta$ will determine the rate of change of the $\alpha$ population due to transitions involving this $\beta$ state. In fewest-switches surface hopping the transitions $\beta \to \alpha$ are dropped and the transition rate for $\alpha \to \beta$, $r_{\alpha \to \beta}$,  is adjusted to give the correct weighting of populations:
\begin{equation}
r_{\alpha \to \beta}=-\frac{2P}{M} \cdot \frac{\Re(d_{\alpha \beta} \rho_s^{\beta \alpha})}{\rho_s^{\alpha \alpha}}.
\end{equation}
This transition rate is used to construct surface-hopping trajectories that specify the evolution of the phase space variables $(R(t),P(t))$ as follows: When the system is in state $\alpha$, the coordinates evolve by Newtonian trajectories on the $\alpha$ adiabatic surface. Transitions to other states $\beta$ occur with probabilities per unit time, $p_{\alpha \to \beta}= r_{\alpha \to \beta} \Theta(r_{\alpha \to \beta})$. Since the rates may take negative values, the Heaviside function $\Theta(x)$ sets the probability zero for negative values of the rate. If the transition to state $\beta$ occurs, the momentum of the system is adjusted to conserve energy and the system then propagates on the $\beta$ adiabatic surface. The momentum adjustment is taken to occur along the direction of the nonadiabatic coupling vector and is given by $P \to P+\Delta P^{FS}_{\alpha \beta}$, with \begin{eqnarray} \label{eq:tully-mom-jump}
\Delta P^{FS}_{\alpha \beta}&=& \hat{d}_{\alpha \beta}\Big({\rm sgn} (P\cdot \hat{d}_{\alpha \beta}) \sqrt{(P \cdot
\hat{d}_{\alpha \beta})^2 + 2\Delta E_{\alpha \beta} M}\nonumber \\
&&-(P \cdot \hat{d}_{\alpha \beta})\Big),
\end{eqnarray}
The form that the stochastic evolution takes can be seen from an examination of Fig.~\ref{fig:surf-hop}, which schematically shows the evolution of a wave packet that starts on the upper adiabatic surface of a two level system with a simple avoided crossing. (This is Tully's simple avoided crossing model.~\cite{tully90}) When the system enters the region of strong nonadiabatic coupling near the avoided crossing, nonadiabatic transitions to the lower state are likely, a surface hop occurs and the system then continues to evolve on the lower surface after momentum adjustment.
For upward transitions it may happen that there is insufficient energy in the environment to insure energy conservation. In this case the transition rule needs to be modified, usually by setting the transition probability to zero. This scheme is very easy to simulate and captures much of the essential physics of the nonadiabatic dynamics.

Fewest-switches surface hopping does suffer from some defects associated with the fact that decoherence is not properly treated. The transition probability depends on the off-diagonal elements of the density matrix but no mechanism for their decay is included in the model. As a result, the fewest-switches surface hopping model overestimates coherence effects and retains memory which can influence the probabilities of subsequent hops. Several methods have been proposed to incorporate the effects of decoherence in mixed quantum-classical theories and, in particular, in surface-hopping schemes.~\cite{neria93,hammesschiffer94,bittner95,bittner97,schwartz96,bedard05,subotnik11,subotniki11a,subotnik11b,subotnik11c,subotnik11d,prezhdo12} In many of these methods a term of the form, $- \gamma \rho_s^{\alpha \alpha'}(t)$, is appended to the equation of motion for the off-diagonal elements of the subsystem density matrix to account for the decay of coherence. The decoherence rate $\gamma$ is estimated using perturbation theory or from physical considerations involving the overlap of nuclear wave functions. In the remainder of this section we discuss how the QCLE accounts for decoherence and comment on its links to surface-hopping methods.

\subsection*{QCL dynamics in the adiabatic basis and decoherence}
Since surface-hopping methods are often formulated in the adiabatic basis, it is instructive to discuss the dynamical picture that emerges when the QCLE is expressed in this basis. Adopting an Eulerian description, the adiabatic energies, $E_\alpha (R)$, and the adiabatic states, $ | \alpha; R \rangle$, depend parametrically on the coordinates of the bath.  We may then take matrix elements of Eq.~(\ref{eq:dmatabs}),
\begin{equation}
\frac{\partial}{\partial t} \langle \alpha; R |\hat{\rho}_W(X,t) |\alpha'; R\rangle=-i \langle \alpha; R | \hat{\mathcal
L}\hat{\rho}_W(t)| \alpha'; R\rangle, \label{eq:dmat-aid-abs}
\end{equation}
to find an evolution equation for the density matrix elements, $ \langle \alpha; R | \hat{\rho}_W (X, t) |\alpha';R \rangle
=\rho_W^{\alpha \alpha'} (X, t)$. Evaluation of the matrix elements on the right side of this equation yields an expression for the QCL superoperator~\cite{kapral99},
\begin{eqnarray}\label{eq:L-off}
   i {\mathcal L}_{\alpha \alpha', \beta \beta'} &=&
  (i\omega_{\alpha \alpha'} + iL_{\alpha \alpha'}) \delta_{\alpha \beta}
\delta_{\alpha' \beta'} - {\mathcal J} _{\alpha \alpha', \beta \beta'}\nonumber \\
&\equiv &i {\mathcal L}^{(0)}_{\alpha \alpha', \beta \beta'}  - {\mathcal J} _{\alpha \alpha', \beta \beta'}.
\end{eqnarray}
Here the frequency $\omega_{\alpha \alpha'}(R)=(E_{\alpha}-E_{\alpha'})/\hbar \equiv \Delta E_{\alpha\alpha'}(R)/\hbar$ (now in the adiabatic basis), and $iL_{\alpha \alpha'}$ is the classical Liouville operator
\begin{equation}\label{eq:adiabatic_superoperator}
iL_{\alpha \alpha'}=\frac{P}{M}\cdot{\partial \over \partial R} +
\frac{1}{2} \left( F_{\alpha} + F_{\alpha'} \right)
\cdot{\partial  \over \partial P}\;,
\end{equation}
and involves the Hellmann-Feynman forces, $F_{\alpha}= -{\partial E_\alpha(R)}/{\partial R}$. The superoperator, ${\mathcal J}$, whose matrix elements are
\begin{eqnarray}
   {\mathcal J}_{\alpha \alpha',\beta \beta'} & = & - d_{\alpha \beta} \cdot \left(\frac{P}{M}
+\frac{1}{2} \Delta E_{\alpha\beta}{\partial  \over \partial P}\right)\delta_{\alpha' \beta'} \nonumber \\
   & - & d^*_{\alpha' \beta'}\cdot \left(\frac{P}{M}+\frac{1}{2} \Delta E_{\alpha' \beta'}
{\partial  \over \partial P}\right) \delta_{\alpha \beta},
\label{eq:jdef}
\end{eqnarray}
couples the dynamics on the individual and mean adiabatic surfaces so that the evolution is no longer described by Newtonian dynamics.

The resulting QCLE in the adiabatic representation reads,
\begin{equation}
   \frac{\partial}{\partial t} \rho^{\alpha \alpha'}_W (X, t) = - i  {\cal L}_{\alpha \alpha', \beta \beta'} \rho^{\beta \beta'}_W (X, t).
\label{eq:adiabatic_qcle}
\end{equation}
To simplify we shall often use a formal notation and write Eq.~(\ref{eq:adiabatic_qcle}) as
\begin{equation}
   \frac{\partial}{\partial t} \rho_W (X, t) = - i  {\cal L} \rho_W (X, t),
\label{eq:adiabatic_qcle-formal}
\end{equation}
where $\rho_W$ and ${\mathcal L}$ (without ``hats") are understood to be a matrix and superoperator, respectively, in the adiabatic basis.

Insight into the nature of QCL dynamics can be obtained as follows. If the operator ${\mathcal J} _{\alpha \alpha', \beta \beta'}$ is dropped the resulting equation of motion for the diagonal elements of the density matrix is
\begin{equation}
\left(\frac{\partial}{\partial t}  +iL_{\alpha} \right)\rho^{\alpha \alpha}_W (X, t)=  0,
\end{equation}
which implies that the phase space density is constant along trajectories on the $\alpha$ adiabatic surface,
\begin{equation}
\rho^{\alpha \alpha}_W (X, t)=e^{-iL_{\alpha}(t-t_0)}\rho^{\alpha \alpha}_W (X, t_0)=\rho^{\alpha \alpha}_W (X(t_0), t_0),
\end{equation}
 where
\begin{equation}
\dot{R}(t)=\frac{P(t)}{M}, \quad \dot{P}(t)=-\frac{\partial }{\partial R(t)} E_\alpha(R(t)),
\end{equation}
with the notation $X(t)=X$.
The off-diagonal density matrix elements satisfy
\begin{equation}
   \left(\frac{\partial}{\partial t}+ iL_{\alpha \alpha'}\right) \rho^{\alpha \alpha'}_W (X, t) = - i\omega_{\alpha \alpha'}(R)  \rho^{\alpha \alpha'}_W (X, t),
\end{equation}
whose solution is
\begin{eqnarray}
 \rho^{\alpha \alpha'}_W (X, t)&=& e^{-i(L_{\alpha \alpha'}+\omega_{\alpha \alpha'})(t-t_0)}\rho^{\alpha \alpha'}_W (X, t_0)\nonumber \\
 &=&{\mathcal W}_{\alpha \alpha'}(t,t_0)\rho^{\alpha \alpha'}_W (X(t_0), t_0),
\end{eqnarray}
where ${\mathcal W}_{\alpha \alpha'}(t,t_0)=e^{-i\int_{t_0}^t dt' \; \omega_{\alpha \alpha'}(R(t'))}$ and the evolution of the phase space coordinates of the bath is given by
\begin{equation}
\dot{R}(t)=\frac{P(t)}{M}, \quad \dot{P}(t)=-\frac{1}{2}\frac{\partial }{\partial R(t)}\left( E_\alpha(R(t))+ E_{\alpha'}(R(t))\right),
\end{equation}
The off-diagonal elements accumulate a phase in the course of their evolution on the mean of the two $\alpha$ and $\alpha'$ adiabatic surfaces.

The momentum derivative terms in ${\mathcal J}$ are responsible for the energy transfers that occur to and from the bath when the subsystem density matrix changes its quantum state. Consequently the subsystem and bath interact with each other and the dynamics of both the subsystem and bath are modified in the course of the evolution. Further, we can see from the structure of the QCLE that there are continuous changes to the subsystem quantum state and bath momenta during the evolution, as opposed to the jumps that appear in surface-hopping schemes. Nonetheless, links to surface-hopping methods can be made.

Subotnik, Ouyang and Landry~\cite{subotnik11c} established a connection between fewest-switches surface-hopping and the QCLE. They investigated what must be done to the equations describing fewest-switches surface hopping in order to obtain the QCL dynamics. Since there are continuous bath momentum changes in QCL dynamics and discontinuous changes in fewest-switches surface hopping, there are limitations on the nuclear momenta. An important element in their analysis is the fact that terms of the form, $-\gamma^{(\alpha)}_{\alpha \alpha'} \rho^{\alpha \alpha'}(t)$, that account for decoherence must be added to the fewest-switches approach.  The specific form of the decoherence rate in their analysis is
\begin{equation}\label{eq:subotnik-decoh}
\gamma_{\alpha \alpha'}^{(\alpha)} \approx \frac{1}{2}(F_{\alpha'}-F_{\alpha})\cdot \frac{1}{\rho^{\alpha \alpha'}}\frac{\partial \rho^{\alpha \alpha'}}{\partial P}
\end{equation}
The superscript $(\alpha)$ indicates that evolution is on the $\alpha$ adiabatic surface and all quantities on the right are taken to evolve on this surface. An analogous expression can be written for $\gamma_{\alpha \alpha'}^{(\alpha')}$.

Recall that surface-hopping schemes assume that the dynamics occurs on single adiabatic surfaces between hops.  Given this fact, we can understand the need for such a term by viewing QCL dynamics in a frame of reference corresponding to motion along single adiabatic surfaces. To see this consider the equation of motion for an off-diagonal element of the density matrix as given by the QCLE. From Eqs.~(\ref{eq:L-off})-(\ref{eq:adiabatic_qcle}) we have
\begin{eqnarray}
\frac{\partial}{\partial t} \rho^{\alpha \alpha'}_W (X, t) &=& -(i\omega_{\alpha \alpha'} + iL_{\alpha \alpha'}) \rho^{\alpha \alpha'}_W (t) \nonumber \\&&+ {\mathcal J} _{\alpha \alpha', \beta \beta'}\rho^{\beta \beta'}_W (t).
\end{eqnarray}
Defining the material derivative for the flow on the $\alpha$ adiabatic surface as
\begin{equation}
\frac{d^\alpha}{dt}=\frac{\partial}{\partial t}+ iL_{\alpha},
\end{equation}
we obtain
\begin{eqnarray}
 \frac{d^\alpha}{dt}\rho^{\alpha \alpha'}_W (X, t) &=& \Big(-i\omega_{\alpha \alpha'} -
\frac{1}{2} \left( F_{\alpha'} - F_{\alpha} \right)
\cdot{\partial  \over \partial P}\Big) \rho^{\alpha \alpha'}_W (t) \nonumber \\
 &&+ {\mathcal J} _{\alpha \alpha', \beta \beta'}\rho^{\beta \beta'}_W (t).
\end{eqnarray}
We see that the second term on the right side of this equation is just the decoherence factor that appears in Eq.~(\ref{eq:subotnik-decoh}). The fact that decoherence depends on the difference between the forces is a common factor in many of the models for decoherence mentioned above. The decoherence contribution is difficult to compute in its current form because of the bath momentum derivative and it is usually approximated in applications.~\cite{subotnik11c}

\subsection*{Surface-hopping solution of the QCLE}

As discussed above, the dynamics prescribed by the QCLE is not in the form of surface hopping since quantum state and bath momentum changes as embodied in the ${\mathcal J}$ superoperator occur continuously throughout the evolution. The effects of ${\mathcal J}$  can be seen by considering the formal solution of Eq.~(\ref{eq:adiabatic_qcle}),
\begin{equation}
\rho^{\alpha \alpha'}_W(X,t)=\left(e^{-i{\mathcal L} t} \right)_{\alpha \alpha', \alpha_N \alpha_N'}\rho^{\alpha_N \alpha_N'}_W(X,0).
\label{eq:formal}
\end{equation}
The time interval $t$ can be divided into $N$ segments of lengths $\Delta t_j$ so that for the $jth$ segment
$t_j-t_{j-1}=\Delta t_j= \Delta t$. Without approximation we may then write
\begin{equation}
\left( e^{-i{\cal L}t}\right)_{\alpha_0 \alpha_0' \alpha_N \alpha_N'}= \prod_{j=1}^N \Big( e^{-i {\mathcal L} (t_j-t_{j-1})}
\Big)_{\alpha_{j-1}\alpha_{j-1}', \alpha_{j}\alpha_{j}'},
\label{eq:slice0}
\end{equation}
where $\alpha_0=\alpha$ and $\alpha_0'=\alpha'$. In each short time segment we can write
\begin{eqnarray}\label{eq:short-time}
&&\Big( e^{-i {\mathcal L} \Delta t}
\Big)_{\alpha_{j-1}\alpha_{j-1}', \alpha_{j}\alpha_{j}'}
\approx {\mathcal W}_{\alpha_{j-1}\alpha_{j-1}'}(\Delta t)e^{-iL_{\alpha_{j-1}\alpha_{j-1}'}\Delta t} \nonumber \\
&&\quad \qquad \times \left(\delta_{\alpha_{j-1} \alpha_{j}} \delta_{\alpha_{j-1}' \alpha_{j}'} + \Delta t {\mathcal J}_{\alpha_{j-1}\alpha_{j-1}', \alpha_{j}\alpha_{j}'} \right).
\end{eqnarray}
If this expression for the short time evolution is substituted in to Eqs.~(\ref{eq:formal}) and (\ref{eq:slice0}), the resulting form for the density matrix is represented as a sum of contributions involving increasing numbers of nonadiabatic transitions governed by the ${\mathcal J}$ operators. The first term in the series is just ordinary adiabatic dynamics if a diagonal density matrix element is considered; for an off-diagonal element the dynamics takes place on the mean of two surfaces and incorporates a phase factor as discussed earlier. The higher order terms in the series involve nonadiabatic transitions between such adiabatic evolution segments.

More specifically, the operator ${\mathcal J}$ contains terms which can be written as follows:
\begin{eqnarray}\label{eq:Jform}
&&d_{\alpha \beta} \cdot \left(\frac{P}{M} +\frac{1}{2}E_{\alpha\beta}{\partial  \over \partial P}\right) \nonumber \\
&& \qquad = \frac{P}{M} \cdot d_{\alpha \beta}
 \left(1+ \frac{1}{2} \frac{M \Delta E_{\alpha \beta}}{(P \cdot \hat{d}_{\alpha \beta})}  \frac{\partial}{\partial (P \cdot \hat{d}_{\alpha \beta})} \right)\nonumber \\
 && \qquad = \frac{P}{M} \cdot d_{\alpha \beta}
 \left(1+ M \Delta E_{\alpha \beta} \frac{\partial}{\partial {\mathcal Y}_{\alpha \beta}} \right),
\end{eqnarray}
where ${\mathcal Y}_{\alpha \beta}=(P \cdot \hat{d}_{\alpha \beta})^2$.
The second equality shows that the momentum changes in the bath occur along the direction of the nonadiabatic coupling matrix element while the third equality shows that the momentum changes can be expressed in terms of an $R$-dependent prefactor ($\Delta E_{\alpha \beta}(R)$) multiplying a derivative with respect to the square of the momentum along $\hat{d}_{\alpha \beta}$.

If the momentum derivative is approximated by finite differences, a branching tree of trajectories will be generated; each branch corresponding to the increment in the bath momentum in the finite difference form the derivative~\cite{nielsen00b}. The number of trajectories will then grow exponentially and the dynamics cannot be propagated for long times and large nonadiabatic coupling. Such a branching tree of trajectories can be avoided and a surface-hopping description can be obtained by making the momentum-jump approximation described below.~\cite{kapral99,0chap-kapral02,kapral06_2}

An expression for the ${\mathcal J}$ operator for small $M \Delta E_{\alpha \beta}$ can be obtained by approximating the factor in parentheses in the last line of Eq.~(\ref{eq:Jform}) as
\begin{equation}
 \left(1+ M \Delta E_{\alpha \beta} \frac{\partial}{\partial {\mathcal Y}_{\alpha \beta}} \right) \approx e^{M \Delta E_{\alpha \beta} \frac{\partial}{\partial {\mathcal Y}_{\alpha \beta}}} \equiv j_{\alpha \beta}.
\end{equation}
The operator $j_{\alpha \beta}$ acts as a momentum translation operator on any function $f(P)$. If we decompose the momentum into its components parallel and perpendicular to the direction of the nonadiabatic coupling matrix element $\hat{d}_{\alpha \beta}$ we have
\begin{eqnarray}
P&=&(P \cdot \hat{d}_{\alpha \beta})\hat{d}_{\alpha \beta}+(P \cdot \hat{d}^\perp_{\alpha \beta})\hat{d}^\perp_{\alpha \beta}\nonumber \\
&=&{\rm sgn} (P\cdot
\hat{d}_{\alpha \beta}) \sqrt{{\mathcal Y}_{\alpha \beta}}\hat{d}_{\alpha \beta}+(P \cdot \hat{d}^\perp_{\alpha \beta})\hat{d}^\perp_{\alpha \beta}.
\end{eqnarray}
Then $j_{\alpha \beta} f(P)=f(P+\Delta P_{\alpha \beta})$, where
\begin{eqnarray}
\Delta P_{\alpha \beta}&=& \hat{d}_{\alpha \beta}\Big({\rm sgn} (P\cdot \hat{d}_{\alpha \beta}) \sqrt{(P \cdot
\hat{d}_{\alpha \beta})^2 + \Delta E_{\alpha \beta} M}\nonumber \\
&&-(P \cdot \hat{d}_{\alpha \beta})\Big),
\end{eqnarray}
and the momentum along the direction of the nonadiabatic coupling matrix element is changed by the action of this operator. Note that this expression for the momentum adjustment is very similar to that in Eq.~(\ref{eq:tully-mom-jump}) for the fewest-switches surface hopping algorithm, the only difference being a factor of two multiplying $\Delta E_{\alpha \beta}$. This factor arises because in fewest-switches surface hopping transitions occur between single adiabatic states corresponding to populations; instead, in QCL dynamics transitions change only one index of the density matrix and correspond to changes from, say, a diagonal density matrix element to an off-diagonal element. It then takes two (or more generally an even number) of quantum transitions to effect a population change; hence, two of these half changes are needed to adjust the momentum in a population change.

In this {\em momentum-jump} approximation the operator ${\mathcal J}$ is given by
\begin{equation}
   {\mathcal J}_{\alpha \alpha',\beta \beta'} \approx - \frac{P}{M} \cdot d_{\alpha \beta} j_{\alpha\beta}\delta_{\alpha' \beta'}
   -\frac{P}{M} \cdot d^*_{\alpha' \beta'}j_{\alpha' \beta'}\delta_{\alpha \beta}.
\label{eq:jumpJ}
\end{equation}

If the momentum-jump expression for ${\mathcal J}$ is used in Eq.~(\ref{eq:short-time}) and the terms in the series in Eq.~(\ref{eq:slice0}) are evaluated by Monte Carlo sampling, a solution in terms of surface-hopping trajectories can be obtained.~\cite{mackernan02,sergi03b,mackernan08} To see this in more detail it is convenient to introduce some notation for the pairs of quantum indices that appear in the expressions given above. We define an index $s$ as $s= \alpha n + \alpha'$ with the pair $(\alpha \alpha')$, where $0 \leq \alpha, \alpha' < n$ for an $n$-state quantum subsystem.~\cite{mackernan02} Then Eqs.~(\ref{eq:slice0}) and (\ref{eq:short-time}) can be written more compactly as
\begin{eqnarray}
&&\left( e^{i{\cal L}t}\right)_{s_0 s_N}\approx \sum_{s_1 s_2 \dots
s_{N-1}} \prod_{j=1}^N {\mathcal W}_{s_{j-1}}(t_j-t_{j-1})  \nonumber \\
&& \qquad \times e^{-iL_{s_{j-1}}(t_j-t_{j-1})}\Big(\delta_{s_{j} s_{j-1}} + \Delta t J_{s_{j-1} s_{j}}\Big) .
\label{eq:alg}
\end{eqnarray}
To propagate the dynamics through one time interval, the positions and momenta and the phase factor are evaluated at time $\Delta t$ by applying $e^{-iL_{s_{j-1}} \Delta t}$. Then, given $s_{j-1}$,  $s_{j}$ is chosen uniformly from the set of allowed final states and a weight associated with the number of final states is applied. Once the final state is chosen, the non-adiabatic coupling matrix element $d_{s_{{j-1}},s_j}$ at the updated position can be computed. Next, a probability, $\pi$, for a nonadiabatic transition is defined as,
\begin{equation}
\pi = \frac{|\frac{P}{M} \cdot d_{s_{j-1} s_j}| \Delta t }{\Big( 1 +
|\frac{P}{M} \cdot d_{s_{j-1} s_j}| \Delta t \Big)},
\end{equation}
and is used to determine if a transition occurs. If no transition occurs by the Monte Carlo sampling, then a weight $1/(1 - \pi)$ is included to account for this failure. If a transition does occur, a weight $1/\pi$ is applied and the bath momenta are adjusted by the momentum-jump operator as discussed above.

If the transition is from an excited state to a lower state the excess energy can always be deposited into the bath. However, if the transition is from a state of lower energy to higher energy, the energy needed for this transition will have to be removed from the bath. As in fewest-switches surface hopping, it may happen that bath degrees of freedom do not have sufficient energy for this process to take place. Then the argument of the square root in the expression for $\Delta P_{\alpha \beta}$ will be negative and the expression cannot be used. In such a circumstance the transition is not allowed and the evolution continues on the current adiabatic surface.

These features are a consequence of the making the momentum-jump approximation. In the exact QCL dynamics, as noted above, there are continuous bath momentum changes along the trajectory from the ${\mathcal J}$ terms. The energy of the ensemble is conserved but there is no requirement that individual trajectories in a trajectory picture conserve energy.

\begin{figure}[htbp]
\begin{center}
\includegraphics[width=0.8\columnwidth]{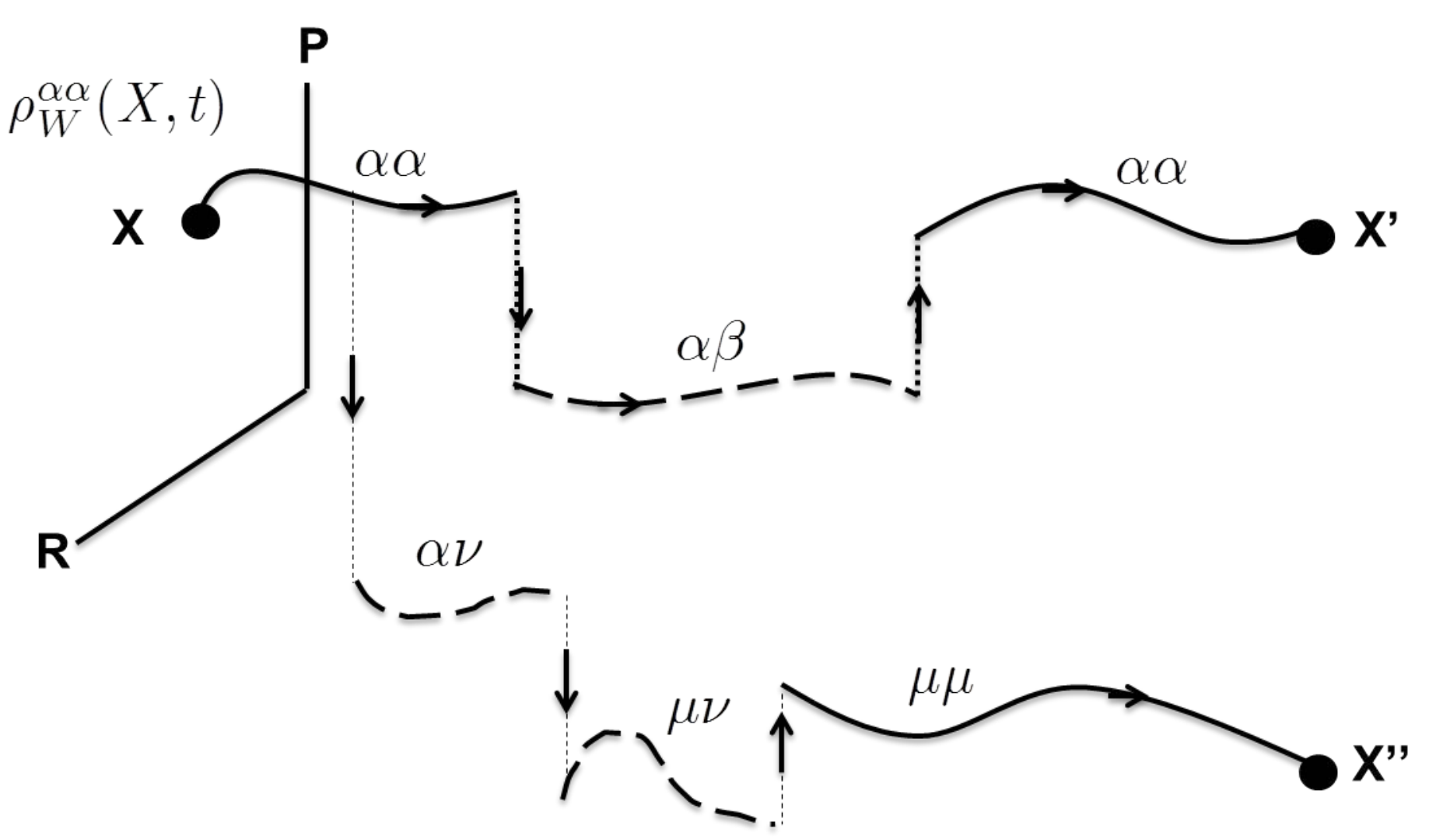}
\caption{Schematic representation of stochastic trajectories that contribute to the population in state $\alpha$ at time $t$. The solid lines indicate propagation on single adiabatic surfaces while the dashed lines indicate propagation on the mean of two adiabatic surfaces accompanied by a phase factor. The vertical dotted lines indicate nonadiabatic surface-hopping transitions accompanied by momentum shifts.}
\label{fig:QCLE-traj}
\end{center}
\end{figure}

Figure~\ref{fig:QCLE-traj} is an illustration of two of the possible trajectories that contribute to the surface-hopping solution of the QCLE for the diagonal ($\alpha \alpha$) density matrix element at phase point $X$ at time $t$.
Following the upper trajectory backward in time, evolution on the $\alpha$ adiabatic surface proceeds until, at some time, a nonadiabatic transition to an off-diagonal state $\alpha \beta$ occurs. Quantum coherence is created by this nonadiabatic event; the system evolves on the mean of the $\alpha$ and $\beta$ adiabatic surfaces and carries a phase factor ${\mathcal W}_{\alpha \beta}$. Proceeding along the trajectory, another nonadiabatic transition occurs and this transition takes the system back to the $\alpha$ population state and, after evolution on this surface, it ends at phase point $X'$. In the second transition the quantum coherence that was created in the first transition is destroyed. The other sample trajectory in the figure shows that sequences of nonadiabatic transitions can lead to more complex evolution in ``off-diagonal" space before the coherence is destroyed. The ensemble of all such trajectories will contribute to the solution of the density matrix. Thus, we see that decoherence is automatically taken into account in this description and will play an essential role in determining the dynamics.

These largely qualitative considerations form the basis for the sequential-short-time-propagation~\cite{mackernan02,sergi03b} and more the refined Trotter-based surface-hopping~\cite{mackernan08} algorithms. Both of these algorithms make use of the momentum-jump approximation and are surface-hopping schemes. In addition both involve ``hops"  between adiabatic surfaces, or the means of adiabatic surfaces, based on weight functions that are designed to simulate the evolution prescribed by the QCLE. The contributions that yield the solution must then include reweighting to compensate for the chosen transition probabilities. While the dynamics is easily simulated for relatively short times, the trajectory contributions contain both weight factors and the signs of the $P\cdot d_{\alpha \beta}/M$ terms that enter the equations. As a result of the sign oscillations and the accumulation of Monte Carlo weights, instabilities in some trajectories can develop for long times. The number of trajectories needed to accurately simulate the dynamics will then grow. Filtering out the unstable trajectories can ameliorate this problem but at the expense of introducing systematic errors. Several filtering methods have been suggested and employed in applications.~\cite{hanna05,mackernan08,sergi13} Nevertheless, simulations on  variety of systems (some described in Sec.~\ref{sec:appl}) have shown that these surface-hopping schemes for the solution of the QCLE often provide very accurate solutions. In addition, several other methods have been constructed to simulate the evolution of the QCLE~\cite{donoso98,wan00,santer01,wan02,horenko02} and the development of effective simulation methods is an active area of research.

\section{Mean-field methods and approximate solutions of the QCLE}\label{sec:approx}

\subsection*{Mean-field theory neglects correlations in the QCLE}
Mean-field methods are frequently used to study the nonadiabatic dynamics of complex systems since they provide a simple trajectory description of the dynamics that is easy to simulate. The standard mean-field description of quantum-classical systems follows from the QCLE when correlations are neglected.~\cite{geras82,cecamchap} In general, the density operator may be written as a product of subsystem and bath density functions plus a term that accounts for correlations: $\hat{\rho}_W(X,t) = \hat{\rho}_s(t) \rho_b(X,t) + \hat{\rho}_{cor}(X,t)$. Here $\rho_b(X,t)= {\rm Tr}_s \hat{\rho}_W(X,t)$ and  $\hat{\rho}_s(t) = \int dX\; \hat{\rho}_W(X,t)$. Substituting this form for $\hat{\rho}_W(X,t)$ into the QCLE and dropping all terms involving $\hat{\rho}_{cor}(X,t)$ leads to two coupled equations. The equation for the bath density is
\begin{equation}
\frac{\partial }{\partial t}\rho_b(X,t)= \{H_{{\rm
eff}}, \rho_b(X,t) \}, \label{eq:rhob}
\end{equation}
with an effective Hamiltonian given by $H_{{\rm eff}}=H_b+ {\rm Tr}_s (\hat{V}_c \hat{\rho}_s(t))$, while the subsystem density matrix satisfies
\begin{equation}
\frac{\partial \hat{\rho}_s(t)}{\partial t}= -\frac{i}{\hbar}
[\hat{h}_s+\int dX\; \hat{V}_c\rho_b(X,t) ,
\hat{\rho}_s(t)], \label{eq:seqn}
\end{equation}
Equation~(\ref{eq:rhob}) admits a solution of the form $\rho_b(X,t)=\delta(X-X(t))$ where
\begin{equation}
\dot{R}(t)=\frac{P(t)}{M},\quad \dot{P}(t)=-\frac{\partial V_{{\rm eff}}(R(t))}{\partial R(t)},
\end{equation}
with $V_{{\rm eff}}(R(t))= V_b(R(t))+{\rm Tr}_s (\hat{V}_c(R(t)) \hat{\rho}_s(t))$.
The equation for the subsystem density may then be written as,
\begin{equation}
\frac{\partial }{\partial t}\hat{\rho}_s(t)= -\frac{i}{\hbar}
[\hat{h}_s+\hat{V}_c(R(t)), \hat{\rho}_s(t)]. \label{eq:seqn2}
\end{equation}
These equations are the Ehrenfest mean-field equations of motion.~\cite{ehrenfest27,dirac30,mclachlan64}

When expressed in a basis of the instantaneous adiabatic states these equations take the form,
\begin{equation}\label{eq:phase:MF}
\dot{R}(t)=\frac{P(t)}{M},\quad \dot{P}(t)=  \Re( F_{\alpha \alpha'}(R(t)) \rho_s^{\alpha' \alpha}),
\end{equation}
where the force matrix elements are
\begin{equation}\label{eq:force-mat-el}
F_{\alpha \alpha'}(R)=F_\alpha(R)\delta_{\alpha \alpha'}+ \Delta E_{\alpha \alpha'}(R) d_{\alpha \alpha'}(R).
\end{equation}
The subsystem density matrix elements satisfy
\begin{eqnarray}\label{eq:sub-MF}
&&\frac{\partial}{\partial t} \rho_s^{\alpha \alpha'}(t)= -i \omega_{\alpha \alpha'}(R(t)) \rho_s^{\alpha \alpha'}(t)\\
&& \qquad -\frac{P(t)}{M}\cdot d_{\alpha \beta}(R(t)) \rho_s^{\beta \alpha'}(t)
-\frac{P(t)}{M}\cdot d^*_{\alpha' \beta}(R(t)) \rho_s^{\alpha \beta}(t),\nonumber
\end{eqnarray}
which has the same the same form as Eq.~(\ref{eq:sub-den-SH}) in the discussion of surface hopping; however, now the bath phase space coordinates evolve according to the mean-field equations (\ref{eq:phase:MF}).

Both the utility and difficulties of mean-field dynamics have been discussed often in the literature.~\cite{tully98,tully12,prezhdo97,subotnik10} In particular, since the classical degrees of freedom evolve subject to a potential that is the average of all subsystem quantum states, the mean-field dynamics will not be able to capture aspects of the dynamics where the potential energy surfaces differ markedly and trajectories populate the levels with very different probabilities. Because of these problems, several methods have been proposed to modify mean-field dynamics to correct these difficulties. These methods include combinations of mean-field and surface-hopping dynamics~\cite{prezhdo97,truhlar04,truhlar04b,bedard05,subotnik10,prezhdo14}, which are designed to allow the system to evolve to a single quantum state in regions where the coupling vanishes.

In the remainder of this section we consider approximate solutions to the QCLE which have a mean-field-like character but are not equivalent to the simple mean-field theory outlined above. The results we present below are derived by employing a mapping of the discrete subsystem states onto single-occupancy oscillator states, as in earlier semi-classical path integral methods~\cite{miller78,sun98,stock97,thoss99,miller01,bonella03,bonella05,dunkel08}. The mapping representation yields a continuous phase-space-like representation of the quantum degrees of freedom. We first show how the QCLE can be written in this mapping basis and then describe how approximate solutions can be constructed.

\subsection*{QCLE in the mapping basis}

The representation of the QCLE (\ref{eq:dmatabs}) in the mapping basis can be carried out by mapping either the adiabatic or subsystem quantum states onto oscillator states. We first consider a mapping representation of subsystem quantum states, while results for adiabatic states will be presented later in this section.

The subsystem basis was defined earlier by the solutions of the eigenvalue problem, $\hat{h}_s |\lambda \rangle = \epsilon_\lambda |\lambda \rangle$, where $\hat{h}_s$ is the quantum subsystem Hamiltonian defined below Eq.~(\ref{eq:H-breakup}). We may then take matrix elements of Eq.~(\ref{eq:dmatabs}),
\begin{equation}
\frac{\partial}{\partial t} \langle \lambda |\hat{\rho}_W(X,t) |\lambda' \rangle=-i \langle \lambda | \hat{\mathcal
L}\hat{\rho}_W(t)| \lambda' \rangle , \label{eq:dmat-sub-abs}
\end{equation}
to evaluate the matrix elements of the QCL operator, ${\cal L}_{\lambda \lambda', \nu \nu'}$, in this basis. We obtain~\cite{kapral99},
\begin{eqnarray}
&&i{\cal L}_{\lambda \lambda', \nu \nu'} = i (\omega_{\lambda \lambda'}
+ L_{b})\delta_{\lambda \nu} \delta_{\lambda' \nu'}
- \frac{i}{\hbar}(\delta_{\lambda \nu} V_c^{\nu' \lambda'}- V_c^{\lambda \nu} \delta_{\lambda' \nu'}) \nonumber \\
& & \qquad \qquad - \frac{1}{2} \left(\delta_{\lambda' \nu'}\frac{\partial
V_c^{\lambda \nu}}{\partial R} + \delta_{\lambda \nu}\frac{\partial V_c^{\nu' \lambda'}}{\partial R}\right)
\cdot \frac{\partial}{\partial P},
\label{eq:Lsubsystem_basis}
\end{eqnarray}
where $\omega_{\lambda \lambda'} = (\epsilon_\lambda -
\epsilon_{\lambda'})/\hbar$, $V_c^{\lambda \lambda'} = \langle \lambda | \hat{V}_c | \lambda' \rangle$, $iL_{b} =
\frac{P}{M} \cdot \frac{\partial}{\partial R} + F_b(R) \cdot \frac{\partial}{\partial P}$, and $F_b(R) = -\partial V_b/\partial
R$ is the force exerted by the bath.

The mapping basis provides another way to write the subsystem (or adiabatic) representation of QCLE. In the mapping representation~\cite{chap-schwinger65,miller78,meyer79,stock05,thoss99} the eigenfunctions of an $n$-state quantum subsystem are replaced with eigenfunctions of $n$ fictitious harmonic
oscillators with occupation numbers limited to 0 or 1: $|\lambda\rangle\rightarrow|m_{\lambda}\rangle=|0_{1}, \cdots,1_{\lambda},\cdots0_{n}\rangle$. A matrix element of the density $\hat{\rho}_W(X)$ in the subsystem basis, $\rho^{\lambda \lambda'}_W(X)$, can be written in mapping form as
\begin{equation}
\rho_{W}^{\lambda\lambda'}(X)=\langle\lambda|\hat{\rho}_{W}(X)|\lambda'\rangle=\langle m_{\lambda}|\hat{\rho}_{m}(X)|m_{\lambda'}\rangle,
\end{equation}
 where
\begin{equation}
\hat{\rho}_{m}(X)=\rho_{W}^{\lambda\lambda'}(X)\hat{a}_{\lambda}^{\dag}\hat{a}_{\lambda'},
\end{equation}
with an analogous expression for an operator. The mapping annihilation and creation operators are given by
\begin{equation}
\hat{a}_{\lambda}=\sqrt{\frac{1}{2\hbar}}(\hat{q}_{\lambda}
+i\hat{p}_{\lambda}),\quad\hat{a}_{\lambda}^{\dag}
=\sqrt{\frac{1}{2\hbar}}(\hat{q}_{\lambda}-i\hat{p}_{\lambda}).
\label{eq:creation}
\end{equation}
They satisfy the commutation relation $[ \hat{a}_\lambda, \hat{a}^{\dag}_{\lambda'}]   =  \delta_{\lambda,\lambda'}$, and act on the single-excitation mapping states to give $\hat{a}^{\dag}_\lambda \ket{0}  =  \ket{m_\lambda}$ and $\hat{a}_\lambda \ket{m_\lambda}  =  \ket{0}$, where $\ket{0} = \ket{0_1 \dots 0_{n}}$ is the ground state of the mapping basis.

Because of the equivalence of matrix elements in the subsystem and mapping bases, we can write Eq.~(\ref{eq:dmat-sub-abs}) as
\begin{equation}
\frac{\partial}{\partial t} \langle m_\lambda |\hat{\rho}_m(X,t) |m_{\lambda'} \rangle=-i \langle m_\lambda | \hat{\mathcal
L}_m\hat{\rho}_m(t)| m_{\lambda'} \rangle . \label{eq:dmat-map-abs}
\end{equation}
Here $\hat{\mathcal L}_m$ has the same form as $\hat{\mathcal L}$ in Eq.~(\ref{eq:dmatabs}) but with the Hamiltonian replaced by the corresponding mapping Hamiltonian. Provided we restrict our calculations to mapping function matrix elements, we have the following alternative formal expression for the QCLE:
\begin{equation}
\frac{\partial}{\partial t} \hat{\rho}_m(X,t) =-i  \hat{\mathcal L}_m\hat{\rho}_m(t). \label{eq:dmat-map-formal}
\end{equation}

By taking a Wigner transform of this equation in the mapping space, we can cast the equation of motion into a form where the discrete quantum degrees of freedom are described by continuous position and momentum variables.~\cite{kim-map08} This can be done by making use of an $n$-dimensional coordinate space representation of the mapping basis. More specifically, we take matrix elements of the equation with respect to $\{|r-z/2\rangle, |r+z/2\rangle \}$ and then take the Wigner transform defined as
\begin{equation}\label{eq:unproj-den}
\rho_m({\mathcal X}) = \frac{1}{(2\pi \hbar)^n} \int dz\; e^{ip\cdot z/\hbar}
\langle r-\frac{z}{2}|\hat{\rho}_{m}(X)|r+\frac{z}{2}\rangle,
\end{equation}
where ${\mathcal X}=(x,X)$ with $x=(r,p)$. To evaluate the terms in the Wigner transform of $\hat{\mathcal L}_m\hat{\rho}_m(t)$ we again make use of the rule for the Wigner transform of a product of operators in Eq.~(\ref{eq:wigner2}) to obtain,
\begin{eqnarray}
 &  & \frac{\partial}{\partial t}\rho_{m}({\mathcal X},t)=-\frac{2}{\hbar}H_{m}
 \sin(\frac{\hbar\Lambda_{m}}{2})\rho_{m}(t)\label{eq:pmweq}\\
 &  & \qquad+\frac{\partial H_{m}}{\partial R}
 \cos(\frac{\hbar\Lambda_{m}}{2})\cdot\frac{\partial \rho_{m}(t)}{\partial P}
 -\frac{P}{M}\cdot\frac{\partial \rho_{m}(t)}{\partial R},\nonumber
\end{eqnarray}
where the negative of the Poisson bracket operator on the mapping phase space coordinates is defined as $\Lambda_{m}=\overleftarrow{\nabla_{p}}\cdot\overrightarrow{\nabla_{r}} -\overleftarrow{\nabla_{r}}\cdot\overrightarrow{\nabla_{p}}$. The
Hamiltonian in the mapping basis is
\begin{equation} \label{eq:mapH}
H_{m}({\mathcal X}) = \frac{P^{2}}{2M}+V_{0}(R)
+ \frac{\bar{h}_{\lambda\lambda'}}{2\hbar} (r_{\lambda}r_{\lambda'}
 +p_{\lambda}p_{\lambda'}),
\end{equation}
where $h_{\lambda\lambda'}(R)=\langle\lambda|\hat{h}(R)|\lambda'\rangle$, $\bar{h}_{\lambda\lambda'}(R)=h_{\lambda\lambda'}(R)-({\rm Tr}\; \hat{h})\delta_{\lambda\lambda'}/n$ and $V_0(R)=V_b(R)+{\rm Tr}\; \hat{h}/n$.
Evaluating the exponential Poisson bracket operators and making use of the fact that the mapping Hamiltonian is a quadratic function of the mapping coordinates, we have,
\begin{eqnarray}
 &  & \frac{\partial}{\partial t}\rho_{m}({\mathcal X},t)=\{H_{m},\rho_{m}(t)\}_{{\mathcal X}}\label{eq:awt3}\\
 &  & \quad - \frac{\hbar}{8}\frac{\partial h_{\lambda\lambda'}}{\partial R}(\frac{\partial}{\partial r_{\lambda'}}\frac{\partial}{\partial r_{\lambda}}+\frac{\partial}{\partial p_{\lambda'}}\frac{\partial}{\partial p_{\lambda}})\cdot\frac{\partial}{\partial P}\rho_{m}(t),\nonumber
 \end{eqnarray}
where $\{A_{m},B_{m}(t)\}_{{\mathcal X}}$ is a Poisson bracket in the full mapping-bath phase space of the system. We may write Eq,~(\ref{eq:awt3}) more compactly as
\begin{equation}\label{eq:awt3-sup}
\frac{\partial}{\partial t}\rho_{m}({\mathcal X},t)=-i{\mathcal{L}}_{m}\rho_{m}(t)=(-i{\mathcal{L}}_{m}^{PB} - i{\mathcal{L}}_{m}^{\prime})\rho_{m}(t),
\end{equation}
where the QCL operator is given by the sum of two contributions: a Poisson-bracket term, $i{\mathcal{L}}_{m}^{PB}$, which gives rise to Newtonian evolution in the ${\mathcal X}$ phase space, and a second term, $i{\mathcal{L}}_{m}^{\prime}$, which involves derivatives with respect to both mapping and bath variables. This latter operator accounts for a portion of the influence of the quantum subsystem on the bath~\cite{nassimi10} and the dynamics that results when this term is included can be described by an ensemble of ``entangled"  trajectories~\cite{kelly12}, analogous to but different from the entangled trajectories that arise in the solutions of the Wigner transformed quantum Liouville equation constructed by Donoso, Zheng and Martens.~\cite{donoso01,donoso03}

\subsection*{Poisson-bracket mapping equation and its extensions} \label{sec:PBME}

A simple approximation to Eq.~(\ref{eq:awt3-sup}) is obtained by neglecting the difficult $i{\mathcal{L}}_{m}^{\prime}$ operator to obtain the Poisson-bracket mapping equation,
\begin{eqnarray}\label{eq:pbme}
\frac{\partial}{\partial t}\rho_{m}({\mathcal X},t)=-i{\mathcal{L}}_{m}^{PB} \rho_{m}(t),
\end{eqnarray}
where the explicit form of $i{\mathcal{L}}_{m}^{PB}$ is
\begin{equation} \label{eq:QCLMop}
i{\mathcal{L}}_{m}^{PB}= -\{H_{m}, \; \;\}_{{\mathcal X}}
 = \Big(\frac{\partial H_m}{\partial \mathcal{P}}\cdot \frac{\partial }{\partial \mathcal{R}} -\frac{\partial H_m}{\partial \mathcal{R}}\cdot  \frac{\partial }{\partial \mathcal{P}}
\Big),
\end{equation}
It is possible to solve this equation in characteristics leading to a solution in terms of an ensemble of independent trajectories that satisfy the Hamiltonian set of equations,
\begin{eqnarray}\label{eq:pbme-ham-eqns}
\frac{d \mathcal{\chi}_\mu}{dt}= \frac{\partial
  H_{m}}{\partial \mathcal{\pi}_\mu}, \qquad
\frac{d \mathcal{\pi}_\mu}{dt}= -\frac{\partial
  H_{m}}{\partial \mathcal{\chi}_\mu},
\end{eqnarray}
where $\mathcal{\chi}=(r,R)$ and $\mathcal{\pi}=(p,P)$. These equations have appeared earlier in mapping formulations based on semi-classical path integral formulations of the dynamics~\cite{stock05,miller01,miller07}.

The solutions of Poisson-bracket mapping equation often provide a quantitatively accurate description of the dynamics~\cite{stock05,kim-map08,nassimi10,kelly12}, but for some systems this approximation may not provide accurate results and even artifacts in the dynamics may appear~\cite{kelly12,rekik13}. Common with other approaches that use the mapping representation, these difficulties can be traced to the fact that the dynamics may take the system out of the physical space~\cite{stock05,bonella01,bonella05}.

The dynamics will be confined to the physical space provided that mapping operators act on mapping functions $|m_\lambda \rangle$. In mapping space we have the completeness relations $\hat{{\mathcal P}}=\sum_{\lambda=1}^n |m_\lambda \rangle \langle m_\lambda|=1$, where $\hat{{\mathcal P}}$ is the projector onto the complete set of mapping states. We may then consider operators projected onto the physical space to ensure that they act there. The density operator projected onto the physical space is given by
\begin{eqnarray}
\hat{\rho}^{{\mathcal P}}_m(X)=|m_\lambda\rangle {\rho}_W^{\lambda \lambda'}(X) \langle m_{\lambda'}|.
\end{eqnarray}
Taking the Wigner transform of this operator, we obtain
\begin{eqnarray}\label{eq:proj-rho-proof}
&&{\rho}_{m}^{{\mathcal P}}({\mathcal X})= \int dz\; e^{ip\cdot z/\hbar}
\langle r-\frac{z}{2}|\hat{\rho}^{{\mathcal P}}_{m}(X)|r+\frac{z}{2}\rangle = \\
&&\int dz\; e^{ip\cdot z/\hbar}
\langle r-\frac{z}{2}|m_\lambda\rangle \langle
m_{\lambda}|\hat{\rho}_{m}(X)|m_{\lambda'}\rangle \langle m_{\lambda'}|r+\frac{z}{2}\rangle \nonumber \\
&& = (2 \pi \hbar)^n g_{\lambda' \lambda}(x) \int dx' \; g_{\lambda \lambda'}(x') \rho_m(x',X)
\equiv {\mathcal P} \rho_m({\mathcal X}).\nonumber
\end{eqnarray}
The quantity $g_{\lambda \lambda'}(x)$ is defined by
\begin{equation}\label{eq:g-def}
g_{\lambda \lambda'}(x)=\frac{1}{(2\pi \hbar)^n}\int dz\; e^{ip\cdot z/\hbar}
\langle r-\frac{z}{2}|m_{\lambda'}\rangle \langle m_{\lambda}|r+\frac{z}{2} \rangle,
\end{equation}
and its explicit form is
\begin{eqnarray} \label{eq:g-expression}
&&g_{\lambda \lambda'}(x)=\phi(x)\\
&& \times \frac{2}{\hbar}
\Big(r_\lambda r_{\lambda'}+ p_{\lambda}p_{\lambda'} - i(r_\lambda p_{\lambda'} - r_{\lambda'} p_{\lambda})  - \frac{\hbar}{2}\delta_{\lambda\lambda'} \Big) , \nonumber
\end{eqnarray}
where $\phi(x)=(\pi \hbar)^{-n}\exp{(-x^2/\hbar)}$ is a normalized Gaussian function.  Here $x^2 =  r_\lambda r_\lambda
+ p_\lambda p_\lambda$ in the Einstein summation convention.

One may show that the projected density satisfies the QCLE in the mapping basis~\cite{kelly12},
\begin{equation} \label{eq:qcl-rho-proj}
\frac{\partial}{\partial t} \rho_m^{{\mathcal P}}({\mathcal X},t) = -i{\mathcal{L}}_{m}\rho_{m}^{{\mathcal P}}(t),
\end{equation}
and that the QCL operator, $i{\mathcal L}_{m}$, commutes with the projection operator:
\begin{eqnarray}
&&\int d{\mathcal X}\; B_m({\mathcal X})i {\mathcal L}_m {\mathcal P} \rho_m({\mathcal X}) \\
&& \qquad \quad = \int d{\mathcal X}\; B_m({\mathcal X}){\mathcal P} i {\mathcal L}_m \rho_m({\mathcal X}).
\nonumber
\end{eqnarray}
The same is not true for the Poisson-bracket mapping approximation; if $i{\mathcal{L}}_{m}$ is replaced by $i{\mathcal{L}}_{m}^{PB}$ in the above equation the identity is no longer satisfied.~\cite{kelly12} Therefore, unlike the evolution under the full QCL operator, the evolution prescribed by the Poisson-bracket mapping operator may take the dynamics out of the physical space. As a consequence, there are instances where this approximation fails and it is desirable to construct schemes for solving the full mapping form of the QCLE~(\ref{eq:awt3}).

\vspace{0.1in}\noindent
{\em Extensions of the Poisson-bracket mapping solution}: In order to break the mean-field character of the Poisson-bracket mapping solution and ensure that the solutions remain in the physical space, the operator $i{\mathcal{L}}_{m}^\prime$ that accounts for additional correlations between the quantum subsystem and the bath must not be neglected. Several approaches have been proposed to do this. In circumstances when the Poisson-bracket mapping solution fails, it is still often very accurate at short times. This feature has been exploited by Kelly and Markland~\cite{kelly-markland13} who combined the Poisson-bracket mapping solution with an exact generalized master equation derived using projection operator methods. Using this approach, accurate solutions could be obtained for long times. The utility of the method was verified through calculations on a model for condensed phase charge transfer where both fewest-switches surface hopping and mean-field approaches failed.

Rather than dropping the $i{\mathcal{L}}_{m}^\prime$ operator, Kim and Rhee~\cite{rhee14} constructed an approximation to this term by making use of its simpler form in the subsystem basis. Their approximation,
\begin{equation}
i{\mathcal{L}}_{m}^\prime \approx \frac{n}{2(n+4)\hbar}\frac{\partial h^{\lambda \lambda'}}{\partial R}
\Big(r_\lambda r_{\lambda'}+ p_{\lambda}p_{\lambda'} - \frac{\hbar}{2}\delta_{\lambda\lambda'} \Big),
\end{equation}
leads to a simple set of equations for the dynamics. The results of simulations on symmetric and asymmetric spin-boson models for a variety of parameters are in good agreement with exact results, even for long times. Such approximations to and use of the mapping form of the QCLE could prove to be very useful in applications to complex systems.

\subsection*{Forward-Backward Trajectory Solution} \label{sec:theory}
The formal solution of the quantum Liouville equation~(\ref{eq:qle}) can be written as
\begin{equation}\label{eq:formal-full-Q}
\hat{\rho}(t)= e^{-i\hat{H}t/\hbar}\hat{\rho}(0)e^{i\hat{H}t/\hbar},
\end{equation}
which is the starting point for the derivation of a number of forward-backward evolution methods for the solution of quantum and mixed quantum-classical dynamics.~\cite{sun99,thompson99,wang00,thoss01,bonella03,bonella05,miller07,bukhman09,huo11,huo12_jcp,makri12,makri12}

A more accurate approximate solution to the QCLE may be constructed by starting with the formal solution of Eq.~(\ref{eq:dmatabs-fb}):
\begin{eqnarray}\label{eq:rhoformalsol}
\hat{\rho}_W(X,t) =  {\mathcal S} \left(
e^{-i{\stackrel{\rightarrow}{\mathcal{H}}_{\Lambda}}t/\hbar}\hat{\rho}_W(X)
e^{i{\stackrel{\leftarrow}{\mathcal{H}}_{\Lambda}}t/\hbar} \right)
\end{eqnarray}
which is the analog of Eq.~(\ref{eq:formal-full-Q}). Here the ${\mathcal S}$ operator specifies the order in which the forward and backward evolution operators act on $\hat{\rho}_W(X)$.~\cite{nielsen01,hsieh12,hsieh13} We may write the formal solution of the QCLE for an operator (\ref{eq:dmatabs-op-soln}) in a similar form,
\begin{eqnarray}\label{eq:bformalsol}
\hat{B}_W(X,t)=  {\mathcal S}\left(
e^{i{\stackrel{\rightarrow}{{\cal H}}_{\Lambda}}t/\hbar}\hat{B}_W(X)
e^{-i{\stackrel{\leftarrow}{{\cal H}}_{\Lambda}}t/\hbar} \right),
\end{eqnarray}
and, since this expression will be used in the applications discussed in Sec.~\ref{sec:appl}, we shall construct the approximate solution for the time evolution of this operator. We shall see that the approximate solution to the QCLE given below~\cite{hsieh12,hsieh13}, which utilizes this starting point, bears a close connection to linearized forward-backward propagation schemes.

Consider the matrix elements of $\hat{B}_W(X,t)$ in the subsystem basis,
\begin{equation}\label{eq:bsol-mat-ele0}
{B}^{\lambda \lambda'}_W(X,t)=  \langle \lambda|{\mathcal S}\left(
e^{i{\stackrel{\rightarrow}{{\cal H}}_{\Lambda}}t/\hbar}\hat{B}_W(X)
e^{-i{\stackrel{\leftarrow}{{\cal H}}_{\Lambda}}t/\hbar} \right)|\lambda' \rangle.
\end{equation}
We may now follow the procedure given in the previous section and convert to a representation in mapping states:
\begin{equation}\label{eq:bsol-mat-ele}
{B}^{\lambda \lambda'}_W(X,t)=  \langle m_\lambda|{\mathcal S}\left(
e^{i{\stackrel{\rightarrow}{{\cal H}^m_{\Lambda}}}t/\hbar}\hat{B}_m(X)
e^{-i{\stackrel{\leftarrow}{{\cal H}^m_{\Lambda}}}t/\hbar} \right)|m_{\lambda'} \rangle,
\end{equation}
where ${\stackrel{\rightarrow}{{\cal H}^m_{\Lambda}}}=\hat{H}_m(1+ \hbar \Lambda/2i)$,
with an analogous definition for ${\stackrel{\leftarrow}{{\cal H}^m_{\Lambda}}}$.
The formal solution of QCLE for an operator in the mapping basis then reads
\begin{equation}\label{eq:bformalsol-map}
\hat{B}_m(X,t)={\mathcal S} \left(
e^{i{\stackrel{\rightarrow}{{\cal H}^m_{\Lambda}}}t/\hbar}\hat{B}_m(X)
e^{-i{\stackrel{\leftarrow}{{\cal H}^m_{\Lambda}}}t/\hbar} \right).
\end{equation}

Instead of going directly to a coordinate representation of the mapping equation as for the Poisson-bracket mapping equation, we introduce a coherent state basis $\ket{z}$ in the mapping space,
\begin{equation}
\label{eq:coherent_state}
\hat{a}_{\lambda} \ket{z}= z_\lambda \ket{z}, \quad \bra{z}\hat{a}_{\lambda}^\dagger= z_\lambda^* \bra{z},
\end{equation}
where $\ket{z}=\ket{z_1,\dots,z_n}$ and the eigenvalues are $z_\lambda= (q_\lambda + i  p_\lambda )/\sqrt{\hbar}$. The mean coordinates and momenta of the harmonic oscillators entering the coherent state $\ket{z}$ are $q=(q_1, \dots, q_{n})$ and $p=(p_1, \dots, p_{n})$, respectively. The coherent states form an overcomplete set and the inner product of two states is
\begin{equation}
\bra{z}z' \rangle = e^{-(|z-z'|^2) -i (z\cdot z^{\prime *}-z^{*}\cdot z^{\prime})}.
\end{equation}
The resolution of identity in the coherent state basis is given by
\begin{equation}\label{eq:coh_resol}
1= \int \frac{d^2z}{\pi^{n}} \ket{z}\bra{z},
\end{equation}
where $d^2z=d(\Re(z)) d(\Im(z))=dq dp/(2\hbar)^n$.

The forward and backward evolution operators in Eq.~(\ref{eq:bformalsol-map}) may now be written as a concatenation of $M$ short-time
evolution segments with $\Delta t_i = \tau$ and $M\tau = t$.  In each short-time interval $\Delta t_i$, we introduce two sets of coherent
states, $\ket{z_i}$ and $\ket{z^{\prime}_i}$ using the resolution of the identity (\ref{eq:coh_resol}) in order to compute the forward and backward time evolution operators, respectively. Evaluating the resulting expression (details can be found in Refs.~[\onlinecite{hsieh12}] and [\onlinecite{hsieh13}]), the matrix elements of Eq.~(\ref{eq:bformalsol-map}) can be written as
\begin{eqnarray}\label{eq:mqc_soln-mat2}
&&{B}^{\lambda \lambda'}_W (X,t) =\int \prod_{i=1}^M \frac{d^2z_i}{\pi^{n}} \frac{d^2z'_i}{\pi^{n}}
\langle m_{\lambda} \ket{z_1} \bra{z^\prime_1} m_{\lambda'}\rangle \nonumber \\
&& \qquad \quad
e^{i{\mathcal L}_e(X,z_1,z_1') \Delta t_1} \Big(
\braket{z_1(\Delta t_1)}{z_2}\nonumber \\
&& \qquad \quad
    e^{i{\mathcal L}_e(X,z_2,z_2') \Delta t_2} \Big( \braket{z_2(\Delta t_2}{z_3} \dots \ket{z_M}\nonumber \\
&& \qquad \quad
e^{i{\mathcal L}_e(X,z_M,z_M') \Delta t_M} \Big( \braket{z_M(\Delta t_M}{m_{\mu}} \nonumber \\
&& \qquad \quad {B}^{\mu \mu'}_W(X) \braket{m_{\mu'}}{z^{\prime}_M(\Delta t_M)}\Big)  \\
&& \qquad \quad \bra{z^\prime_M}  \dots
\ket{z^\prime_2(\Delta t_2)} \Big)
\braket{z^\prime_2}{z^\prime_1(\Delta t_1)}  \Big).\nonumber
\end{eqnarray}
The evolution operators in this equation should be evaluated sequentially, from smallest to largest times, by taking the bath phase space propagators in expressions such as $e^{i{\mathcal L}_e(X,z_i,z_i') \Delta t_i} ( \cdots )$ to act on all quantities in the parentheses, including other propagators at later times. The bath phase space propagator is given by
\begin{equation}
i {\mathcal L}_{e}(X,z,z')=\frac{P}{M}  \cdot \frac{\partial}{\partial R} -\frac{\partial }{\partial R}V_e(X,z,z')  \cdot \frac{\partial}{\partial P},
\end{equation}
where
\begin{eqnarray}\label{eq:hamil_pot}
V_e(X,z,z') & = &  V_{{\rm fb}}(R) +\frac{1}{2}\left(h^{\lambda\lambda^{\prime}}z^*_\lambda z_{\lambda^{\prime}} + h^{\lambda\lambda^{\prime}}z^{\prime *}_{\lambda}z^{\prime}_{\lambda^{\prime}}\right), \nonumber \\
& \equiv & V_{{\rm fb}}(R) + \frac{1}{2} \left( h_{cl}(R,z) + h_{cl}(R,z') \right),
\end{eqnarray}
with $V_{{\rm fb}}(R)= V_b(R) -{\rm Tr}\; \hat{h}$. In obtaining the expression for ${B}^{\lambda \lambda'}_W (X,t)$ in Eq.~(\ref{eq:mqc_soln-mat2}), we used the exact form for the coherent state evolution under a quadratic Hamiltonian $\hat{h}_m(R)$: $ e^{-i\hat{h}_m \frac{\Delta t_i}{\hbar}}\ket{z}= \ket{z(\Delta t_i)}$, where the trajectory evolution of $z_\lambda$ is governed by
\begin{equation}
\frac{dz_{\lambda}}{dt} =-\frac{i}{\hbar}\frac{\partial h_{cl}(R,z)}{\partial{z}^*_{\lambda}},
\end{equation}
with $ h_{cl}(R,z)=h^{\lambda\lambda^{\prime}}z^*_\lambda z_{\lambda^{\prime}}$. The dynamics of the quantum subsystem comprises discontinuous segments of coherent state trajectories, since the coherent state variables $z_i$ and $z_{i+1}$ are independent of each other. While this feature complicates the simulation of the dynamics, in the limit of sufficiently small time steps this formulation will yield an exact solution of the QCLE.

However, to obtain a simple tractable solution involving a continuous trajectory evolution, we make the approximation that the inner products, $\langle z_i(t_i) \ket{z_{i+1}}$, are orthogonal: $\langle z_i(t_i) \ket{z_{i+1}} \approx \pi^{n}\delta(z_{i+1}-z_i(t_i))$.

\begin{figure}[htbp]
\begin{center}
\includegraphics[width=\columnwidth]{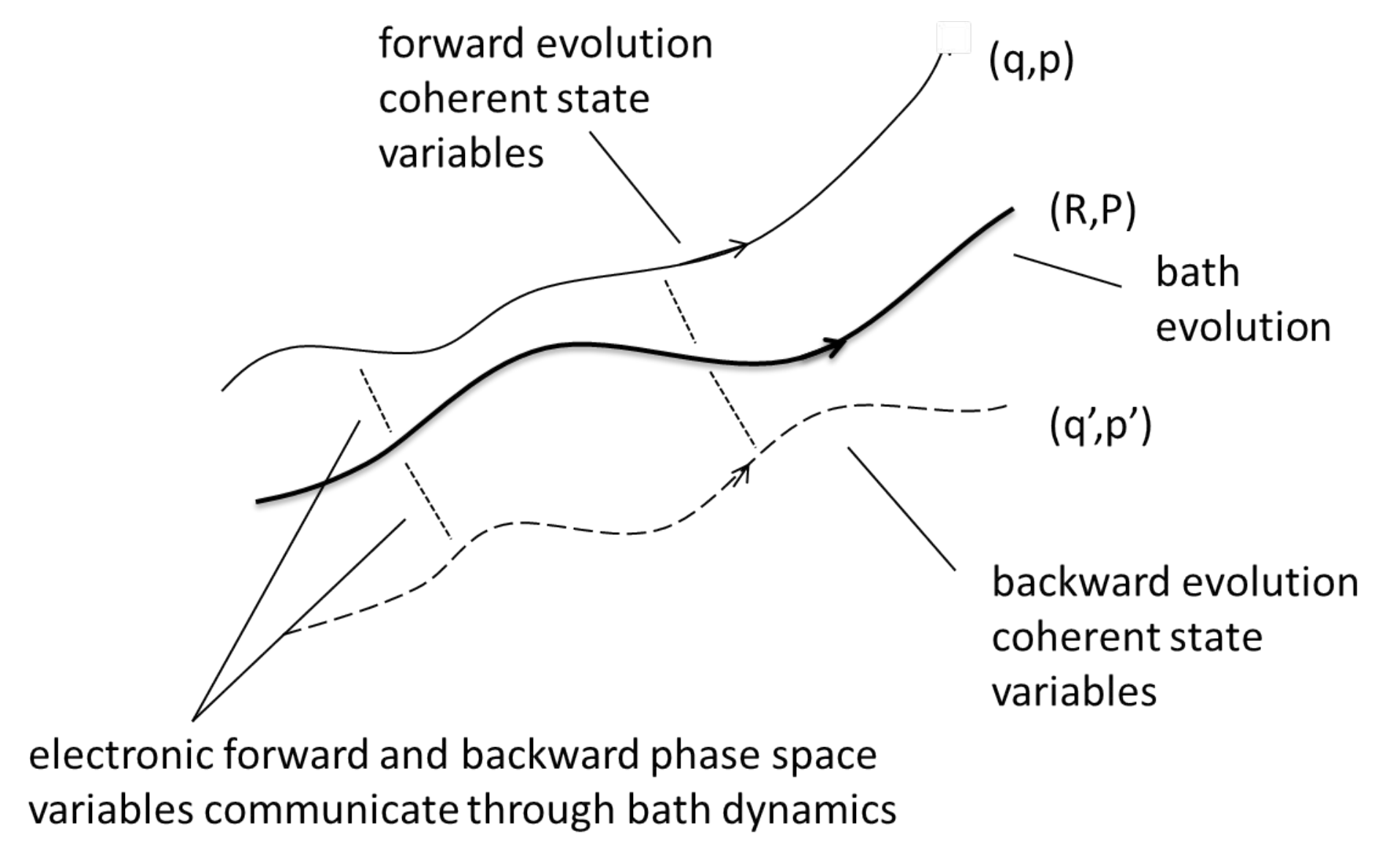}
\caption{Schematic representation of a trajectory contribution to the forward-backward trajectory solution. The evolution of the bath phase space variables $(R,P)$ (central dark trajectory) couples to the evolution of the $(q,p)$ and $(q',p')$ coherent state phase space variables.}
\label{fig:FBTS-schem}
\end{center}
\end{figure}

With this approximation, Eq.~(\ref{eq:mqc_soln-mat2}) reduces to the compact expression,
\begin{eqnarray}\label{eq:mqc_soln-mat6}
&&{B}^{\lambda \lambda'}_W (X,t) =\int dx dx^{\prime} \phi(x) \phi(x')
\frac{1}{\hbar}  (q_{\lambda}+ip_{\lambda})(q'_{\lambda^\prime}-ip'_{\lambda^\prime}) \nonumber \\
&& \quad \times {B}^{\mu \mu'}_W(X(t))\frac{1}{\hbar} (q_{\mu}(t)-ip_{\mu}(t))(q'_{\mu^\prime}(t)+ip'_{\mu^\prime}(t)),
\end{eqnarray}
where $x=(q, p)$ gives the real and imaginary parts of $z$, $dx =dq dp$, and $\phi(x)=\left(\hbar\right)^{-n}e^{-\sum_\nu (q_\nu^2+p_\nu^2)/\hbar}$ is the normalized Gaussian distribution function. In the extended phase space of $(X,x,x')=(\chi,\pi)$, with $\chi = (R,q,q')$, and $\pi = (P, p, p')$, the trajectories follow Hamiltonian dynamics,
\begin{eqnarray}\label{eq:fbts}
\frac{d \mathcal{\chi}_\mu}{dt}= \frac{\partial
  H_{e}(\chi,\pi)}{\partial \pi_\mu}, \qquad
\frac{d \mathcal{\pi}_\mu}{dt}= -\frac{\partial
  H_{e}(\chi,\pi)}{\partial \chi_\mu},
\end{eqnarray}
where
\begin{eqnarray}\label{eq:fb-eff-H}
&&H_{e}(\chi,\pi)=P^2/2M + V_{{\rm fb}}(R) \\
&&\qquad + \frac{1}{2\hbar}h_{\lambda\lambda'}(R)(q_\lambda q_{\lambda'}+ p_\lambda p_{\lambda'}+q'_\lambda q'_{\lambda'}+ p'_\lambda p'_{\lambda'}).\nonumber
\end{eqnarray}
This is the forward-backward trajectory solution. It is very simple to simulate since it only involves propagating Newtonian trajectories in an extended phase space whose dimension is four times the number of quantum states plus the number of bath degrees of freedom. A schematic representation of the nature of the trajectories is given in Fig.~\ref{fig:FBTS-schem}. The forward and backward quantum mapping coherent state phase space variables couple to the evolution of the bath variables but are not directly coupled to each other.
Efficient schemes have been constructed to evolve the dynamics.~\cite{kelly12} As we shall see below, the forward-backward trajectory solution often yields an excellent approximation to the dynamics but does fail in some circumstances. The set of evolution equations~(\ref{eq:fbts}) for the forward-backward trajectory solution are similar to those that arise in the partial linearized density matrix method~\cite{huo11}; in fact, they are identical if the system Hamiltonian is traceless. To understand this difference, note that the quantity $V_{{\rm fb}}(R)=V_b(R)-{\rm Tr}\; \hat{h}$ appears in the evolution equations for the forward-backward trajectory solution. The trace term arose from the need to use an anti-normal order for the product of the annihilation and creation operators when evaluating the short-time propagator. If this trace term is not present, the solution will not satisfy the differential form of the QCLE, and the derivation will depend on how one chooses to write the Hamiltonian operator (for example, as a sum of trace and traceless parts).

\noindent
{\it Adiabatic basis}: In some instances it is more convenient to carry out calculations in the adiabatic basis, since the adiabatic states can be obtained from quantum electronic structure calculations. The forward-backward trajectory solution can be formulated in this basis as follows.~\cite{hsieh13b} We may define adiabatic versions of the forward and backward mixed quantum-classical Hamiltonians, $\stackrel{\rightarrow}{{\mathcal H^a_\Lambda}}=\ket{\alpha;R}\stackrel{\rightarrow}{{\mathcal H^a}}_{\alpha\alpha'}\bra{\alpha';R}$, with
\begin{eqnarray}
\label{eq:map1}
&& \stackrel{\rightarrow}{{\mathcal H^a}}_{\alpha\alpha'}
\equiv
\left(\frac{P^2}{2M} + V_0(R) +
  E_\alpha(R)\right)\delta_{\alpha\alpha'} +  \\
&& \qquad
\frac{\hbar}{2i}\left[
\frac{P}{M}\cdot \frac{\stackrel{\rightarrow}{\partial}}{\partial R} \delta_{\alpha\alpha'} +
2\frac{P}{M}\cdot d_{\alpha\alpha'} + F_W^{\alpha\alpha'}\cdot
\frac{\stackrel{\rightarrow}{\partial}}{\partial P} \right], \nonumber
\end{eqnarray}
and $\stackrel{\leftarrow}{{\mathcal H^a_\Lambda}}=\ket{\alpha;R}\stackrel{\leftarrow}{{\mathcal H^a}}_{\alpha\alpha'}\bra{\alpha';R}$, with a definition for $\stackrel{\leftarrow}{{\mathcal H^a}}_{\alpha\alpha'}$ similar to that given above for the right-acting operator.
Given these definitions, the adiabatic matrix elements of the operator equation,
\begin{equation}
\frac{\partial}{\partial t} \hat{B}_W(X,t)
= \frac{i}{\hbar}\left( \stackrel{\rightarrow}{{\mathcal H^a_\Lambda}}\hat{B}_W(X,t) - \hat{B}_W(X,t) \stackrel{\leftarrow}{{\mathcal H^a_\Lambda}}\right),
\end{equation}
just reproduce the QCLE~(\ref{eq:adiabatic_qcle}) in the adiabatic basis. Following a strategy like that for calculations in the subsystem basis, the mapping transformation,
\begin{eqnarray}\label{eq:admapdef}
&&\ket{\alpha;R}\stackrel{\rightarrow}{\mathcal H^a}_{\alpha\alpha^{'}}\bra{\alpha;R}
 \rightarrow  \stackrel{\rightarrow}{\mathcal H^a_{m}} \equiv
\stackrel{\rightarrow}{\mathcal H^a}_{\alpha\alpha^{'}}\hat{b}^{\dag}_\alpha
\hat{b}_{\alpha'},  \\
&&\ket{\alpha;R}B^{\alpha\alpha'}_W\bra{\alpha';R} \rightarrow
\hat B_m(X) \equiv B^{\alpha\alpha'}_W \hat{b}^{\dag}_{\alpha} \hat{b}_{\alpha'}, \nonumber
\end{eqnarray}
is introduced. The annihilation and creation operators, $\hat{b}_\alpha$ and  $\hat{b}^{\dag}_\alpha$, respectively, now act on the single excitation states corresponding to the occupancy of the adiabatic states: $\ket{0}=\hat{b}_\alpha \ket{m_\alpha}$ and $\ket{m_\alpha} = \hat{b}^{\dag}_{\alpha} \ket{0}$. The mapping matrix elements of the adiabatic mapping operators are identical to the matrix elements of operators in the adiabatic basis; for example,
\begin{equation}
\bra{\alpha;R}\stackrel{\rightarrow}{{\mathcal H^a_\Lambda}}\ket{\alpha';R}=\stackrel{\rightarrow}{{\mathcal H^a}}_{\alpha\alpha'}= \bra{m_\alpha} \stackrel{\rightarrow}{\mathcal H^a_{m}}\ket{m_{\alpha'}}.
\end{equation}

To complete the calculation, coherent states $\ket{y}$ are introduced such that
$\hat{b}_\alpha\ket{y} = y_\alpha \ket{y}$, where $y_\alpha =\frac{1}{\sqrt{\hbar}}
(\tilde{q}_\alpha+i\tilde{p}_\alpha)$ and $\tilde{x} = (\tilde{q},\tilde{p})$ and, following the steps used in the subsystem basis calculation, the expression for ${B}^{\alpha\alpha'}_W (X,t) $ has a form identical to that in Eq.~(\ref{eq:mqc_soln-mat6}). In this adiabatic formulation the evolution equations for the bath variables are
\begin{equation}\label{eq:eomad-bath}
\frac{dR}{dt}= \frac{P}{M},\quad
\frac{dP}{dt} =  -\frac{\partial V_0}{\partial R}
+F_{\alpha \alpha'} \frac{1}{2}\left(y_\alpha y^*_{\alpha'} + y^\prime_\alpha y^{\prime *}_{\alpha'}\right),
\end{equation}
where the force matrix elements are defined by Eq.~(\ref{eq:force-mat-el}). The structure of these equations is similar to that of the bath mean-field equations~(\ref{eq:phase:MF}) but the bath momenta evolve under a mean force that depends on the forward and backward coherent states $\ket{y}$ and $\ket{y'}$.

The quantum coherent state variables evolve by
\begin{eqnarray}\label{eq:eomad-sub0}
\frac{ d y_\alpha  }{ dt } & = & -i \frac{E_{\alpha}}{\hbar} y_\alpha
-\left(d_{\alpha\alpha'}(R)\cdot\frac{P}{M}\right)y_{\alpha'}, \nonumber \\
\frac{ d y'_\alpha  }{ dt } & = & -i \frac{E_{\alpha}}{\hbar} y'_\alpha
-\left(d_{\alpha\alpha'}(R)\cdot\frac{P}{M}\right)y'_{\alpha'},
\end{eqnarray}
or, written as equations for $y_\alpha y_{\alpha'}^*$, by
\begin{eqnarray}\label{eq:eomad-sub}
&&\frac{ d y_\alpha y_{\alpha'}^* }{ dt } = -i \omega_{\alpha \alpha'} y_\alpha y_{\alpha'}^*\\
&&\qquad - \frac{P}{M} \cdot d_{\alpha \beta}(R) y_{\beta}y_{\alpha'}^*
- \frac{P}{M} \cdot d^*_{\alpha '\beta}(R) y_{\alpha} y_{\beta}^*,\nonumber
\end{eqnarray}
with an analogous set of equations for the backward propagating subsystem variables. Each of these sets of equations has a form identical to the mean-field equations of motion for the subsystem density matric elements in Eq.~(\ref{eq:sub-MF}). Thus, although the forward-backward trajectory solution provides a more sophisticated treatment of the dynamics, it nevertheless has a mean-field character. This mean-field nature stems from the orthogonality approximation made on the coherent state overlap matrices. This approximation leads to a simple trajectory description (in the subsystem basis) which necessarily endows it with a mean-field character. In order to break this mean-field structure one must relax the orthogonality approximation, and we next describe how this may be done.

\subsection*{Jump forward-backward solution}

Returning to the subsystem mapping representation, Eq.~(\ref{eq:mqc_soln-mat2}) has the general structure,
\begin{eqnarray}\label{eq:mqc_map-jump}
{B}^{\lambda \lambda'}_W (X,t) &=&\sum_{\mu \mu'}\int \prod_i  \frac{d^2z_i}{\pi^{n}} \frac{d^2z'_i}{\pi^{n}}
\cdots    \\
&&  \times \langle z_i(t_i) \ket{z_{i+1}} \cdots \bra{z^\prime_{i+1}} z^\prime_i(t_1)\rangle \cdots . \nonumber
\end{eqnarray}
If the orthogonality approximation were not invoked, one would have to evaluate the coherent state integrals at each intermediate time step and compute the integrals by Monte Carlo or some other sampling method. This would give rise to an exponentially large set of trajectories, making the algorithm impracticable. However, the orthogonality approximation need not be relaxed at every time step. For example, given a total of $M$ time steps in the expression for the matrix element, we may select $L=M/J$ time steps that are $J$ steps apart for possible relaxation of the orthogonality approximation. The steps at which this occurs may be chosen randomly by using a given binary sequence $\{\kappa_1 ,\dots, \kappa_L\}$, to determine when to fully evaluate the coherent state integrals. If at the $vJ$-th time step $\kappa_v=1$, the full integral is performed (by some sampling method); otherwise, if $\kappa_v=0$ the orthogonality approximation is applied.  The matrix element $B^{\lambda\lambda'}_W(X,t)$ is given by an average over all possible binary sequences as,
\begin{eqnarray}\label{eq:mqc_soln-jump-avg}
B^{\lambda\lambda'}_W(X,t) &=& \sum_{\kappa_1,\dots \kappa_L} P_{\{\kappa\}}{B}^{\lambda \lambda'}_{\kappa_1,\dots,\kappa_{L}} (X,t),
\end{eqnarray}
where $P_{\{\kappa\}}$ denotes the discrete probability distribution of a given binary sequence of $\{\kappa_1 ,\dots, \kappa_L\}$.  This is the jump forward-backward trajectory solution. The continuous forward-backward trajectories experience discontinuous jumps in the forward and backward subsystem phase variable, and between such jumps the evolution is governed by Eq.~(\ref{eq:fbts}). This method is closely related to the iterative partial linearized density matrix method~~\cite{huo12_jcp}. Both methods make use of stochastic sampling at intermediate times. The method is also similar in spirit to schemes that combine surface-hopping and mean-field methods. Finally we note that the forward-backward trajectory solution in the adiabatic basis can be reformulated to yield other variants of the jump forward-backward solution that could prove useful in applications~\cite{hsieh13b}, but such solutions have not been fully explored.

\section{Simulations of the Dynamics} \label{sec:appl}

The validity and accuracy of quantum dynamical methods are often tested on standard simple models that are designed to include features present in more complex realistic systems. In this section we shall present results for several such models in order to test how the solutions of the QCLE compare to full quantum dynamics and, as well, to determine the utility and accuracy of some of the algorithms for the simulation of this equation. Rather than presenting an exhaustive review of work along these lines, the focus of this section will be restricted to results obtained using the Trotter-based surface-hopping method, the forward-backward trajectory solution, and its jump extension. As discussed in previous sections, the Trotter-based scheme makes use of the momentum-jump approximation to arrive at a surface-hopping picture, while the forward-backward trajectory solution imposes orthogonality of coherent states to obtain a simple trajectory picture. The jump forward-backward solution can yield a numerically exact solution of the QCLE, provided a sufficient number of ``jumps" are taken, but this quickly become computationally infeasible for some systems for long times. Calculations on a variety of models using quantum-classical Liouville dynamics have been carried out using the Poisson-bracket mapping approximation~\cite{mackernan08,kim-map08,nassimi10,kelly12} as well as number of other computational schemes~\cite{donoso98,santer01,wan00,wan02,horenko02}, and this literature can be consulted for details. The general conclusion from these studies is that the QCLE solutions agree very well with exact quantum results for a wide variety of systems; however, approximations that are made in some simulation algorithms may fail in some circumstances. Special emphasis will be given here to models that challenge the simulation algorithms.

\subsection*{Spin-Boson and FMO Models}
We begin with a discussion of two models for which the QCLE provides an exact description of full quantum dynamics: the spin-boson and Fenna-Matthews-Olson models. Both models describe systems where an $n$-level quantum system is bilinearly coupled to a harmonic bath.

Spin-boson models have been studied often since they provide a simple description for a wide range of physical phenomena and are some of the first systems used to gauge the efficacy of quantum dynamics algorithms~\cite{weiss99}. Although all three QCLE simulation methods described earlier have been used to simulate this model~\cite{mackernan02a,mackernan08,kim-map08,bonella09}, here we give the results using the forward-backward trajectory solution and its extension including jumps. The partially Wigner transformed Hamiltonian for the spin-boson model is,
\begin{eqnarray}\label{eq:sbmodel}
\hat{H}_W(X) & = & \sum_{i=1}^{N_b} \left( \frac{P_i^2}{2M_i} + \frac{1}{2}M_i\omega_i^2R_i^2 - c_i R_i\hat{\sigma}_z \right) \nonumber \\
& & + \epsilon \hat{\sigma}_z - \Omega \hat{\sigma}_x,
\end{eqnarray}
where $M_i$ and $\omega_i$ are the mass and frequency of bath oscillator $i$, respectively, $c_i$ controls the bilinear coupling strength between the oscillator $i$ and the two-level quantum subsystem, $\Omega$ is the coupling strength between the two quantum levels, $\epsilon$ is the bias, and $\hat{\sigma}_{z(x)}$ is a Pauli matrix.  The bilinear coupling is characterized an ohmic spectral density, $J(\omega) = \pi\sum_i c_i^2/(2M_i\omega_i)\delta(\omega-\omega_i)$, where
$c_i = (\xi\Delta M_j)^{1/2}\omega_i$, $\omega_i = -\omega_c \ln(1-i\Delta\omega/\omega_c)$, and $\Delta\omega = \omega_c(1-e^{-\omega_{max}/\omega_c})/N_B$ with $\omega_c$ the cut-off frequency, $N_B$ the number of bath oscillators, and $\xi$ the Kondo parameter. The two-level system is initially in the state $\ket{1}$ and the bath is initially in thermal equilibrium characterized by a thermal energy $k_B T=1/\beta$.

\begin{figure}[htbp]
     \begin{center}
     \includegraphics[width=0.45\textwidth]{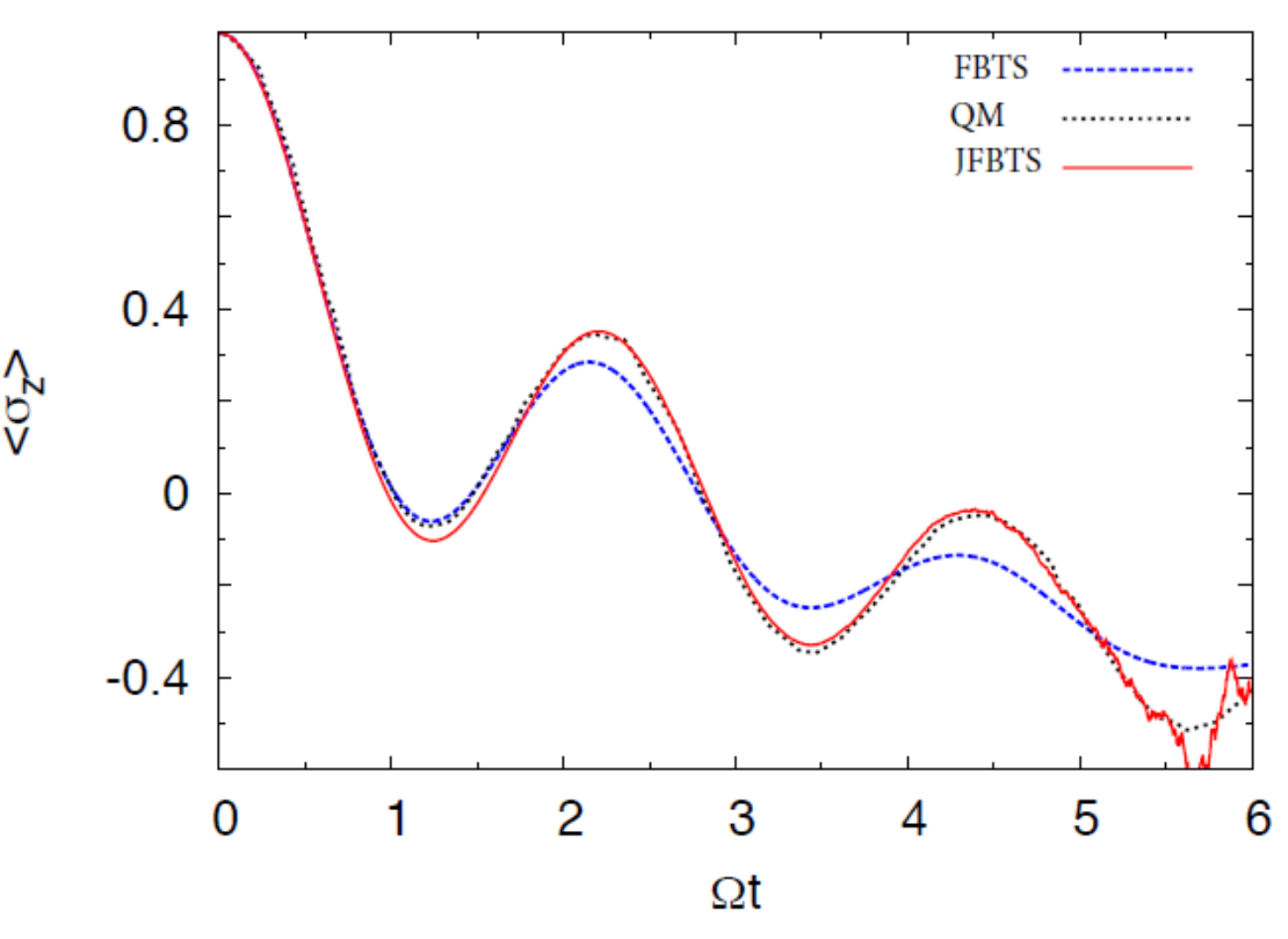}
     \end{center}
     \vspace{-17pt}
     \caption{Comparison of exact quantum~\cite{thompson99b}, forward-backward trajectory solution (FBTS) and jump forward-backward trajectory solution (JFBTS) results~\cite{hsieh13} for the asymmetric spin-boson model with parameters, $\epsilon = \Omega = 0.4$, $\xi = 0.13$, $\beta = 12.5$ and $\omega_c = 1.0$. }
   \label{fig:symmsb}
\end{figure}

Results for the symmetric spin-boson system with $\epsilon=0$ using the forward-backward trajectory solution are in quantitative agreement with exact quantum calculations~\cite{makarov94} for a wide range of parameter values~\cite{hsieh13} and will not be shown here. Instead, we prefer to focus on the asymmetric case ($\epsilon \ne 0$) where the forward-backward trajectory solution is not in quantitative agreement with the exact quantum results.
The introduction of a bias leads to significant differences between the symmetric and asymmetric spin-boson models~\cite{loss05}. Figure~\ref{fig:symmsb} compares the exact results for  $<\sigma_z(t)>$ for the asymmetric spin-boson model with simulations using the forward-backward trajectory solution and its jump analog. The forward-backward trajectory solution deviates from the exact results but this discrepancy can be corrected when jumps are included (results with 26 jumps are shown). The number of jumps needed to reproduce the exact result depends on factors such as the size of the time-step and the probability distribution chosen for the jumps~\cite{hsieh13}. While very few trajectories are needed to obtain converged results for the forward-backward trajectory solution, implementation of the jump forward-backward solution requires substantially more trajectories, depending on the number of jumps needed for a specific application.

Photosynthesis involves excitation energy transfer from antenna proteins to the reaction center.~\cite{fleming09,scholes10} The Fenna-Matthews-Olson (FMO) protein plays an important role in the excitation energy transfer process in green sulfur bacteria~\cite{fleming09}. The model Hamiltonian for this system comprises a seven-level quantum subsystem with each quantum level bilinearly coupled through a Debye spectral density to its own set of bath harmonic oscillators~\cite{fleming09a}. The quantum subsystem is initially in quantum state $\ket{1}$ and all bath oscillators are initially in thermal equilibrium. Numerically accurate quantum results are available~\cite{fleming09a, zhu11}, and simulations using the Poisson-bracket mapping equation~\cite{kelly-FMO} and partial linearized density matrix~\cite{huo11} algorithms have been carried out. (Also, the Poisson-bracket mapping equation was used to study the dynamics of a much more realistic model for FMO.~\cite{kelly-FMO2}) Since this model corresponds to a quantum subsystem bilinearly coupled to a harmonic bath, the QCLE is again exact.

\begin{figure}[htbp]
     \begin{center}
     \includegraphics[width=0.45\textwidth]{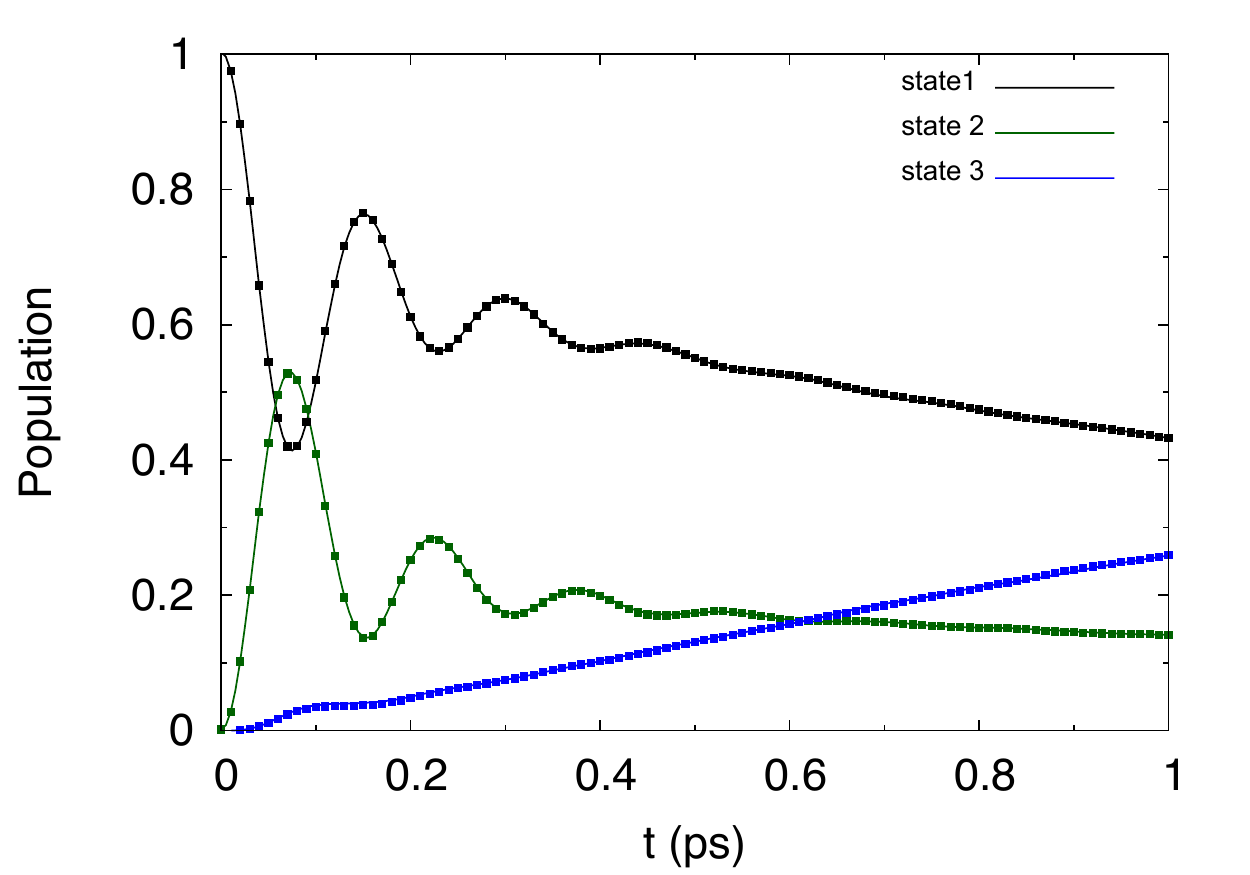}
     \end{center}
     \vspace{-20pt}
     \caption{Populations in states $\ket{1}$, $\ket{2}$, and $ \ket{3}$ of bacteriochlorophyll as function of time at a temperature of 77 K. The solid lines are the forward-backward trajectory solution results~\cite{hsieh13}, while the data points are extracted from numerically exact quantum results~\cite{zhu11}.}
   \label{fig:fmo}
\end{figure}
The populations in quantum states $\ket{1}$, $\ket{2}$, and $ \ket{3}$, computed using the forward-backward trajectory solution, are plotted versus time in Fig.~\ref{fig:fmo}.  Numerically exact full quantum results using the rescaled Hierarchical Coupled Reduced Master Equation algorithm~\cite{zhu11} are also shown for comparison. One can see that the two sets of results are indistinguishable on the scale of the figure. If the calculations are extended to very long times the population distributions obtained from the forward-backward trajectory solution closely approximate the thermal equilibrium distribution. This algorithm is able to accurately simulate dynamics of this multi-level system for long times in a computationally efficient manner since it only involves following Newtonian trajectories in an extended phase space.

\subsection*{Avoided Crossing and Conical Intersection Models}

Nonadiabatic dynamical events are especially important in systems where the adiabatic states are nearly degenerate at avoided crossings or at conical intersections where the adiabatic states cross. Plots of diabatic and adiabatic states for a two-level system as a function of a nuclear coordinate $R$ were shown in Fig.~\ref{fig:surf-hop} when surface-hopping dynamics was discussed. In the vicinity of an avoided crossing the nonadiabatic coupling matrix elements, $d_{\alpha \alpha'}(R)$, are large and, if a surface-hopping method is used to evolve the system, transitions between the two adiabatic states will occur with high probability.

A set of such avoided crossing models was constructed by Tully~\cite{tully90} and these have served as test cases for nonadiabatic methods. Figure~\ref{fig:surf-hop} is actually a sketch of the diabatic and adiabatic curves for Tully's single avoided crossing model. The Hamiltonian matrix in the diabatic representation is ${\bf H}_W=(P^2/2M) {\bf 1} + {\bf h}(R)$, where
\begin{eqnarray}\label{eq:tully2}
\hspace{-0.2in} {\bf h}(R) = \left[ \begin{array}{cc}
A[1 - e^{-B |R|}]\frac{R}{|R|} & Ce^{-DR^2} \\
Ce^{-DR^2} & A[1 - e^{-B |R|}]\frac{R}{|R|} \
\end{array}
\right].
\end{eqnarray}
The numerical values of parameters and all other details of this particular model are available in the literature.~\cite{tully90,nassimi10,hsieh13} Initially, the quantum subsystem is taken to be in the state $\ket{1}$ and the bath particle is modeled as a Gaussian wave packet centered at $R_0$ with initial bath momentum $P_0$ directed towards the interaction region. The forward-backward trajectory computations of the populations are in quantitative agreement with exact quantum results, except for very small initial momenta $P_0$ where small deviations are observed. Simulations using the jump forward-backward trajectory solution with 2 jumps converge to the exact quantum results~\cite{hsieh13}.

\begin{figure}[htbp]       \centering
\includegraphics[width=.9\columnwidth]{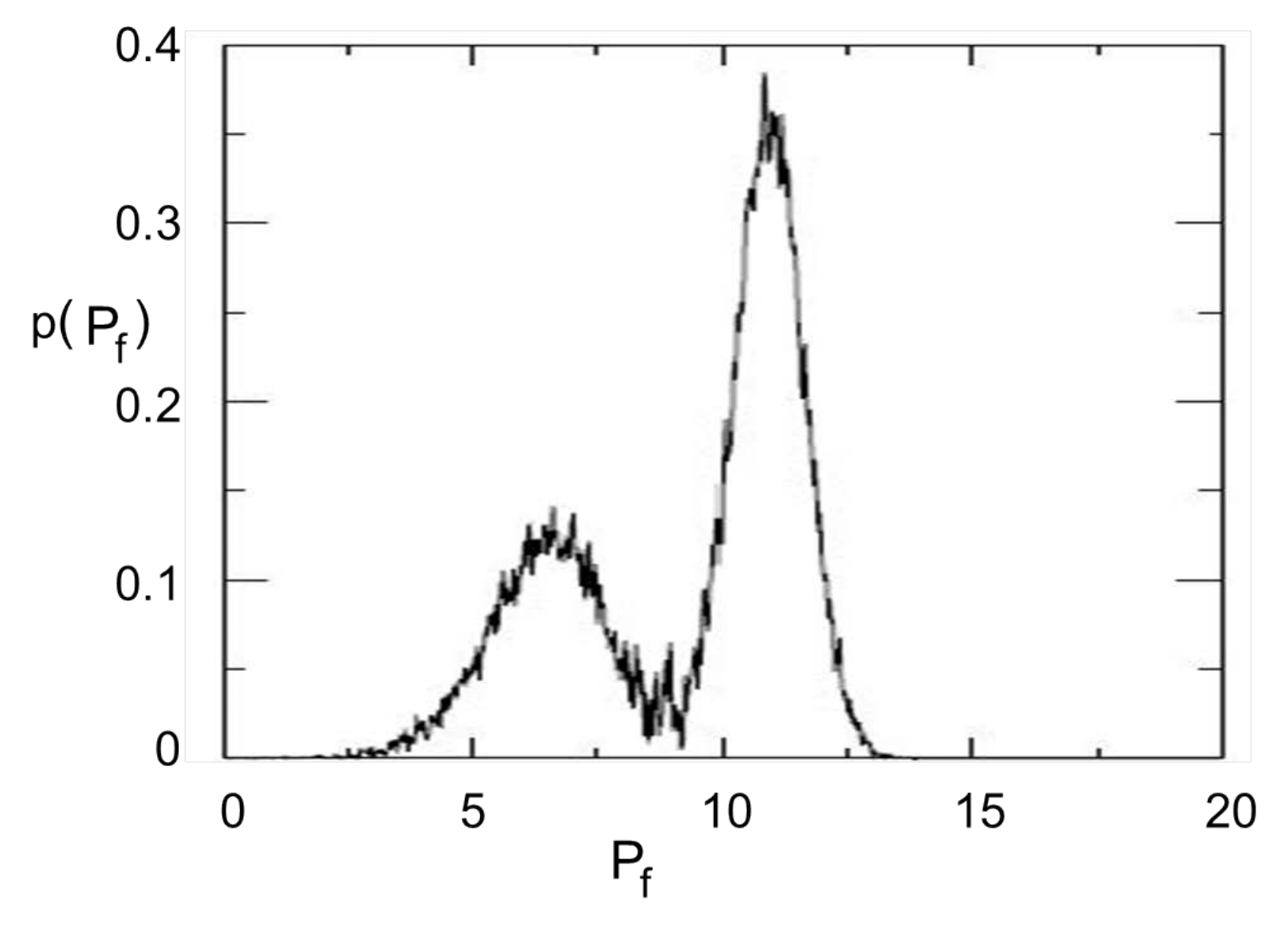}
    \caption{Plot of the momentum distribution $p(P_{{\rm f}})$ after passage through the avoided crossing obtained from Trotter-based surface-hopping solutions of the QCLE~\cite{kelly12}.  The parameter values are $A=0.01$, $B=1.6$, $C=0.005$ and $D=1$, and the initial momentum is $P_0=11$. All parameters are reported in atomic units.
}\label{fig:avoid-cross}
\end{figure}

The properties of the nuclear degrees of freedom in this model provide more stringent tests of simulation algorithms. Simulations based on the forward-backward initial-value representation yield a double-peak structure in accord with exact quantum results.~\cite{miller07} As the system passes through the avoided crossing and the coupling vanishes, the nuclear momenta have characteristically different values in the two asymptotic states. Consequently, the probability density of final nuclear momenta, $p(P_{{\rm f}})$, has a bimodal form. By contrast, computations using the forward-backward trajectory solution (and the Poisson-bracket mapping equation) yield a single-peak structure. The nuclear mean-field character of these solutions fails to capture this effect, although the quantum populations are described accurately. This is not a failure of the QCLE, but only of these specific algorithms. Both the Trotter-based surface-hopping and jump forward-backward trajectory solution algorithms are able to capture this nuclear quantum effect as can be seen from the plot of the momentum distribution in Fig.~\ref{fig:avoid-cross} obtained using the Trotter-based surface-hopping algorithm.~\cite{kelly12}

Conical intersections involve dynamical features that are different from those near avoided crossings, such as the appearance of a geometrical phase, and are believed to be responsible for the rapid population transfer observed in some systems.~\cite{migani04}
To examine such dynamics, we consider a two-level, two-mode quantum model for the coupled vibronic states of a linear ABA triatomic molecule constructed by Ferretti, Lami and Villiani.~\cite{ferretti_1,ferretti_2}  In this model, the nuclei are described by two vibrational degrees of freedom, $X$ and $Y$, the tuning and coupling coordinates.  The partially Wigner transformed Hamiltonian is
\begin{equation}\label{eq:flv_bath}
{\bf H}_W(R_s,P_s) = \left(\frac{P_X^2}{2M_X}+\frac{P_Y^2}{2M_Y}+\frac{1}{2}M_Y\omega_Y^2Y^2\right) {\bf 1} +{\bf h}(R_s),
\end{equation}
where the subsystem Hamiltonian is defined by the following matrix elements:
\begin{eqnarray}\label{eq:flv_qm}
h^{11}(R_s) & = & \frac{1}{2} M_X \omega_X^2 (X-X_1)^2, \nonumber \\
h^{22}(R_s) & = & \frac{1}{2} M_X \omega_X^2 (X-X_2)^2 + \Delta, \nonumber \\
h^{12}(R_s) & = & \gamma Y \exp\left(-\alpha(X-X_3)^2-\beta Y^2\right).
\end{eqnarray}
In these equations, $R_s=(X,Y)$, $P_s=(P_X,P_Y)$, $M_{X,Y}$ and $\omega_{X,Y}$ are the mass and frequency for the $X$ and $Y$ degrees of freedom, respectively. The quantum subsystem is initialized in the adiabatic ground state, while the vibronic $X$ and $Y$ initial states are taken to be Gaussian wave packets.
Further details of this model can be found in the literature.~\cite{kelly10,ferretti_1}

\begin{figure}[htbp]
     \begin{center}
     \includegraphics[width=0.49\textwidth]{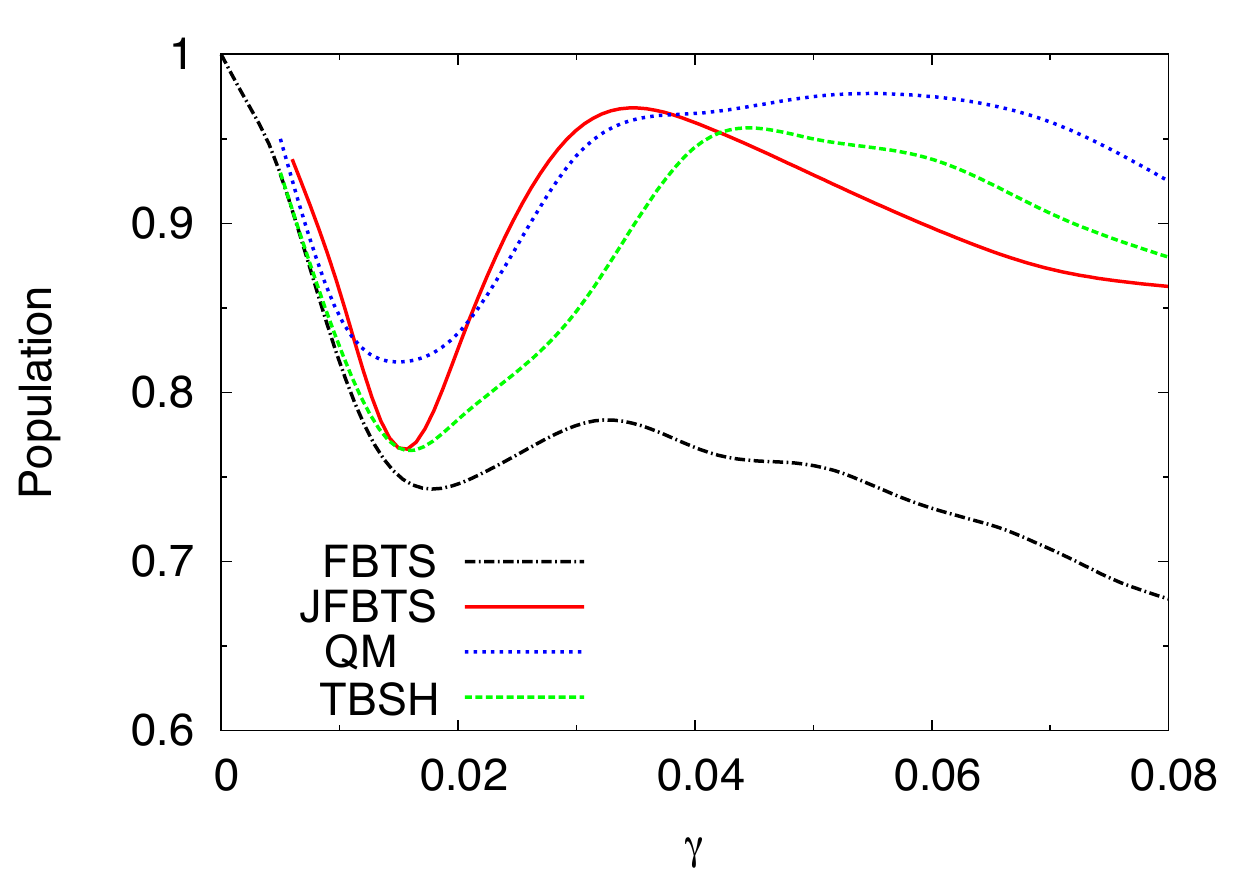}
     \end{center}
     \vspace{-17pt}
     \caption{Asymptotic adiabatic ground state population at 50~fs versus $\gamma$ compute using various simulation methods: forward-backward trajectory solution (FBTS), jump forward-backward trajectory solution (JFBTS), numerically exact quantum solution (QM) and Trotter-based surface hopping (TBSH)}
   \label{fig:flv_gamma}
\end{figure}
Figure~\ref{fig:flv_gamma} plots the adiabatic ground state population at $t=50$ fs as a function of the coupling strength $\gamma$.  We see that all results agree for small coupling strengths, somewhat less than $\gamma = 0.02$; however, the forward-backward trajectory solution results differ considerably from the jump forward-backward, Trotter-based and exact quantum results for larger coupling strengths.  At the higher coupling strengths, the errors introduced by the coherent state orthogonality approximation become significant but the 15-jump forward-backward solution and Trotter-based results shown in the figure are able to reproduce all major trends in population versus coupling strength curve. The description of dynamics in systems with a conical intersection places significant demands on simulation algorithms due to the strong nonadiabaticy that arises near the conical intersection point, especially for strong coupling. Both the jump forward-backward and Trotter-based solutions can account for this strong nonadiabaticity; however, many nonadiabatic events and more extensive statistical sampling are needed to improve the results further.

Systems with conical intersections also exhibit effects due to the existence of a geometric phase. These quantum effects manifest themselves in the nodal structure seen in the probability densities of the nuclear coordinates and provide a stringent test of the QCLE. Investigations of the geometric phase within the context of QCL dynamics have also been carried out for this model~\cite{kelly10}, as well as for a linear vibronic model that has a conical intersection~\cite{ryabinkin13,ryabinkin14}. The Trotter-based results on both of these models are able to capture nuclear nodal structure effects that are signatures of a geometric phase, again attesting to the accuracy of QCL dynamics.

\subsection*{Proton transfer in a polar solvent}

All of the above examples considered very simple models for both the subsystem and bath. Any of those model systems could have been (and were) solved using a full quantum description. The main motivation for developing quantum-classical dynamical models was to be able to simulate large complex many-body systems that are not amenable to a full quantum treatment. Our last example is of this type. We consider the system mentioned in the Introduction (see Fig.~\ref{fig:ptransfer}): proton transfer ($A$H-$B \rightleftharpoons A^{ - }$--H$^{ + }B$) in a phenol ($A$)-trimethylamine ($B$) complex solvated by polar methyl chloride molecules. We shall not consider the dynamics of this system by using first-principles, electronic-structure expressions for all of the interactions which are determined in the course of the dynamical evolution. Rather, we again appeal to a model for this system, but one that is far more realistic than those described above and  contains most of the elements needed to simulate quantum dynamics in complex environments. In this Azzouz-Borgis model~\cite{azzouz93} the phenol and trimethylamine molecular groups are treated as united-atom spheres, as are the methyl and chloride groups comprising the solvent molecules. The phenol-amine complex is solvated by condensed-phase methyl chloride molecules that interact among themselves, and with the proton, phenol and amine groups through intermolecular potentials. The proton is treated quantum mechanically and the Schr\"odinger equation is solved at each time step to determine the adiabatic energies and eigenfunctions that enter the quantum-classical dynamics. In addition to illustrating the effects of solvent polarization on proton transfer dynamics, this model has served as test case for quantum reactive dynamics in condensed phase systems and has been studied often using various approaches~\cite{hammesschiffer94,mcrae01,antoniou99,antoniou99a,kim03,yamamoto05,hanna05,hanna08}, where further details of the model and results can be found. The rate and mechanism of this quantum transfer reaction are of primary interest. This system provides a good example of how a quantum-classical description can be used to study a quantum rate process in a condensed phase environment that may be approximated by classical mechanics.

Following the strategy briefly described at the end of Sec.~\ref{sec:properties}, we may compute the rate constant from the reactive flux correlation function by sampling from quantum initial conditions and approximating the dynamics by evolution given by the QCLE. Expressions for reaction rate constants in this framework have been formulated.~\cite{sergi03,sergi04,kim05a,kim06,kim06c,hsieh14} The QCL expression for the time-dependent rate
coefficient, $k(t)$ is
\begin{equation}
k(t) =  - (\beta n_R^{eq})^{-1} {\rm Tr}_s{\int
{dX\;\hat{N}_P (X,t) (\frac{i}{\hbar}[\hat{N}_R ,\hat{\rho}_e] )_W}},
\label{eq:lrrate}
\end{equation}
where $\hat{N}_R$ and $\hat{N}_P$ are operators that characterize the reactant $R=(A$H-$B)$ and product $P=(A^{ - }$--H$^{ + }B)$ states and $n_R^{eq}$ is the equilibrium density of the reactant state. The quantum equilibrium canonical density is $\hat{\rho}_e$ and the dynamics of $\hat{N}_P (X,t)$ is given by the QCLE. If there is a significant time scale separation between the chemical and other relaxation processes, the plateau value of $k(t)$ yields the measured rate constant $k$ for the reaction $R \stackrel{k}{\rightarrow} P$.

\begin{figure}[h!]
     \begin{center}
     \includegraphics[width=0.45\textwidth]{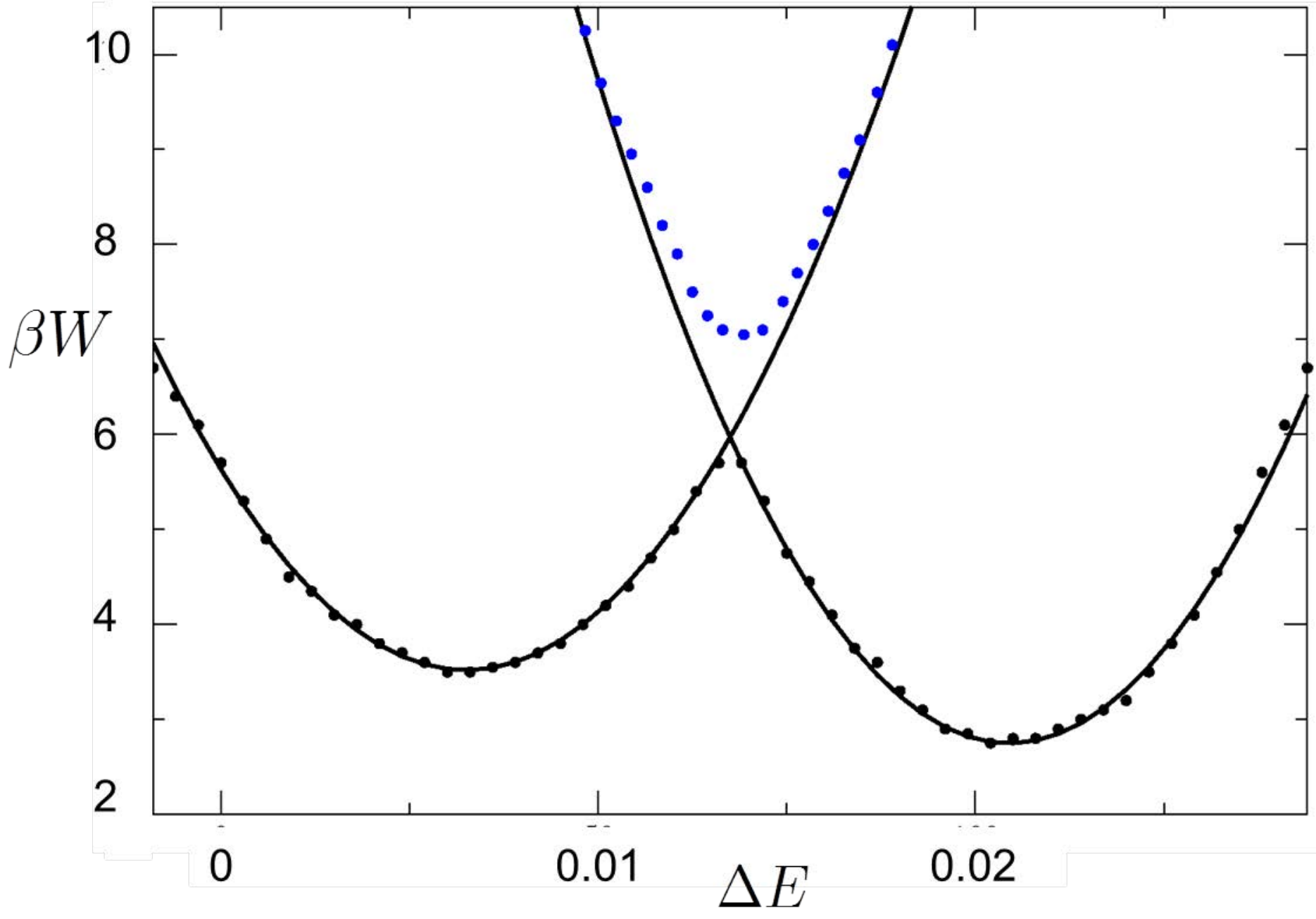}
     \end{center}
     \vspace{-17pt}
     \caption{Protonic free energy, $W$ (times $\beta=1/k_BT$), along the solvent polarization coordinate, $\Delta E({R})$, for the system in the ground (lower curve) and first excited (upper curve) adiabatic states. The solid lies are parabolic fits to the free energy in the left and right well regions.}
   \label{fig:proton-free}
\end{figure}

The solvent polarization, defined as the difference between the solvent electrical potentials at points $s$ and $s'$ within the complex, $\Delta E({R}) = \sum_{i,a} {z_a e ( {\vert \mbox{{R}} _i^a - s\vert^{-1} - \vert \mbox{{R}}_i^a - {s}'\vert }^{-1} )}$, can be used as a reaction coordinate to monitor the proton transfer reaction.~\cite{marcus85,warshel82} Here $z_a e$ is the charge on atom $a$, $s$ and $s'$ are two points within the complex, one at the center of mass and the other displaced from the center of mass, and the sums run over all solvent molecules $i$ and atoms $a$. The free energy along this reaction coordinate is plotted in Fig.~\ref{fig:proton-free} when the system is in the ground and excited protonic states.~\cite{hanna08}
This figure shows that when the system is in the ground state the values of $\Delta E({R})$ can be used to identify reactant R (left) and product P (right) species separated by a free energy barrier at $\Delta E^\ddag$. The free energy when the system is in the first excited state has a single minimum at $\Delta E({R})=\Delta E^\ddag$ and the avoided crossing with a small energy gap leads to strong nonadiabatic coupling in the vicinity of the barrier. Given this picture, we can choose $R$ and $P$ species as,
$\hat {N}_P =\theta (\Delta E(R) - \Delta E^\ddag)$ and $\hat {N}_R =\theta (\Delta E^\ddag-\Delta E(R))$. If the quantum equilibrium density is approximated by its adiabatic value, the expression for the rate coefficient can be written as
\begin{eqnarray}
k(t) &\approx&  \frac{1}{n_{R}^{eq}Z_{Q}}
\sum_{\alpha} \int {dX\; N_P^{\alpha \alpha} (X,t)}  \\ & \times
& \frac{P}{M}\cdot\nabla_{R}\Delta E({R})  \delta (\Delta E({R}) - \Delta E^\ddag
)e^{-\beta H_{\alpha}(X)},\nonumber \label{eq:ratef}
\end{eqnarray}
where $Z_{Q}$ is the partition function. This expression can be simulated using rare event sampling starting at the barrier top and the rate constant can be determined from relatively short-time QCL dynamics simulations using the Trotter-based surface-hopping algorithm.~\cite{hanna05,hanna08} The rate coefficient extracted from such simulations is $k=0.163$ ps$^{-1}$. The transmission coefficient $\kappa=k/k^{TST}$, defined to be the ratio of the rate coefficient to its transition state value, has the value $\kappa=0.65$. The reduction of the rate coefficient is due to dynamical recrossing of the barrier arising from both motion on the ground adiabatic surface and nonadiabatic effects involving transitions to the excited state surface. Other details concerning correlations between the dynamical evolution of the solvent polarization and the quantum mechanical average value of the proton position in the complex have also been extracted from such nonadiabatic QCL simulations and have served to elucidate the nature of the reaction mechanism.~\cite{hanna05}

\begin{figure}[h!]
     \begin{center}
     \includegraphics[width=0.49\textwidth]{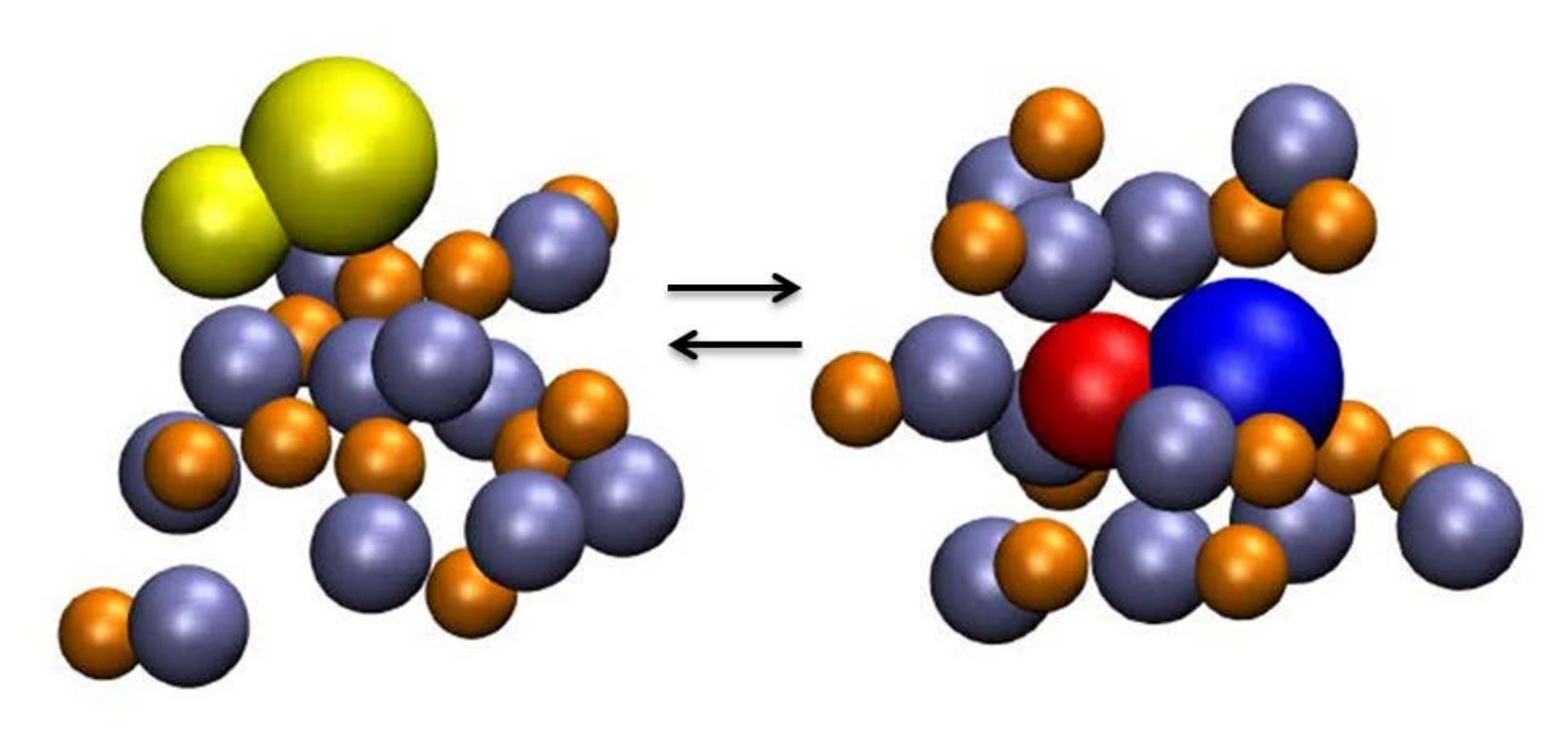}
     \end{center}
     \vspace{-17pt}
     \caption{Two configurations of the proton-phenol-amine complex in a polar molecule nanocluster.~\cite{kim-clus08} In the left configuration the complex is its covalent form (both spheres in the complex are yellow) and it tends to reside on the surface of the cluster. In the right configuration it is in its ionic form (red and blue spheres denote the phenol and amine groups in the complex with negative and positive charges) and is solvated in the interior of the cluster. The polar methyl chloride molecules are shown as linked spheres whose colors denote the partial charges (light purple - negative chloride and rust - positive methyl) in the molecule.}
   \label{fig:cluster}
\end{figure}

Proton transfer in the phenol-amine complex has also been investigated when the complex is solvated by a small cluster of methyl chloride molecules.~\cite{kim-clus06,kim-clus08} In addition to the computations of the rate coefficients for proton and deuteron transfer rates, the QCL simulations indicate that the cluster structure itself is changed as a result of the quantum particle transfer in the molecular complex. Figure~\ref{fig:cluster} shows two configurations of the cluster. In the left configuration the proton-phenol-amine complex is in the covalent form and the complex resides on the surface of the cluster. When a quantum proton transfer takes place in the complex to yield the ionic form, the complex moves to the interior of the cluster since this is the more favorable solvation state. Thus, the structure of the complex itself can also serve as a reaction coordinate.

The QCLE has been generalized to include the presence of a radiation field in order to be able to theoretically model multi-dimensional spectroscopy, which has then been applied to this proton transfer model, in both condensed phase and cluster environments.~\cite{hanna08b,hanna09,hanna10}

While various methods have been used to simulate quantum-classical Liouville dynamics~\cite{donoso98,horenko02,wan00,santer01,nassimi10,cecamchap}, we have chosen limit the above illustrative computational examples to the Trotter-based method, the forward-backward trajectory solution and its jump extension. The forward-backward solution is very simple to implement since it only requires the solution of a set of Hamiltonian equations in an extended phase space. The computational effort is comparable to that of Ehrenfest dynamics but it provides a much more accurate treatment of the dynamics. As discussed above, because of the coherent state orthogonality approximation that leads to a simple trajectory representation, it retains a mean-field character, albeit different from Ehrenfest dynamics. Its jump extension breaks this mean-field structure and yields a numerically exact solution of the quantum-classical Liouville equation, provided a sufficient number of ``jumps" are included in the simulation. This method can then be used to gauge the accuracy of the simple forward-backward scheme. The Trotter-based method makes use of the momentum-jump approximation that leads to a surface-hopping trajectory picture of the dynamics, which is different from  fewest-switching surface hopping since decoherence is taken into account. Like the jump forward-backward solution, it is also able to describe effects that lie outside mean-field-like descriptions and properly accounts for the back reaction of the quantum system on its environment.

\section{Summary and Conclusion} \label{sec:con}

There have been significant advances in the construction of quantum dynamical algorithms that are applicable for increasingly large systems~\cite{beck00,meyer90,alon07,marques04,burghardt13}. Nevertheless, at the present time it is still difficult to treat the full complexity of quantum dynamics in condensed-phase or large biochemical systems at a level of detail where the environment is not described in a highly idealized fashion. As a result, the study of quantum dynamics in open quantum-classical systems is a topic worth pursuing.

Quantum-classical Liouville dynamics, which was the focus of this review, is one of several quantum-classical dynamical schemes that are currently being developed and applied to study quantum dynamics. Problems are encountered when any mixed quantum-classical method is used. These problems center around the theoretical foundations of the mixed dynamics, and the way interactions between the quantum subsystem and its environment are treated. The manner in which theories account for (or do not account for) decoherence in the quantum subsystem is an important factor in the construction of quantum-classical dynamics. When approximations are made to quantum-classical Liouville dynamics it was shown that mean-field and surface-hoping descriptions of the dynamics could be obtained. In addition, it was also shown how decoherence is naturally accounted for in quantum-classical Liouville dynamics.

While the quantum-classical Liouville equation is able to provide an accurate description of many of the complex systems that are encountered in nature, it is not easy to simulate the dynamics prescribed by this equation. Several algorithms that yield either numerically exact or approximate solutions of this equation were presented, and the advantages and limitations of these algorithms were discussed. One of the important areas for future research on this topic is the development of more robust and generally accurate algorithms. In more a general context, it is also an interesting and challenging exercise to seek schemes that combine quantum and classical dynamics. The construction of mixed quantum and classical dynamical theories still presents many challenges and is fertile ground for future new developments.

\begin{acknowledgments}
This work was supported in part by a grant from the Natural Sciences and Engineering Council of Canada. A portion of the computational work was performed at the SciNet, which is funded by the Canada Foundation for Innovation under the auspices of Compute Canada, the Government of Ontario, Ontario Research Fund-Research Excellence, and the University of Toronto. Progress on this topic was possible only with the participation of many of my colleagues whose work is cited in this paper. I want to thank them for their collaboration in this research.
\end{acknowledgments}


%


\begin{thebibliography}{159}%
\makeatletter
\providecommand \@ifxundefined [1]{%
 \@ifx{#1\undefined}
}%
\providecommand \@ifnum [1]{%
 \ifnum #1\expandafter \@firstoftwo
 \else \expandafter \@secondoftwo
 \fi
}%
\providecommand \@ifx [1]{%
 \ifx #1\expandafter \@firstoftwo
 \else \expandafter \@secondoftwo
 \fi
}%
\providecommand \natexlab [1]{#1}%
\providecommand \enquote  [1]{``#1''}%
\providecommand \bibnamefont  [1]{#1}%
\providecommand \bibfnamefont [1]{#1}%
\providecommand \citenamefont [1]{#1}%
\providecommand \href@noop [0]{\@secondoftwo}%
\providecommand \href [0]{\begingroup \@sanitize@url \@href}%
\providecommand \@href[1]{\@@startlink{#1}\@@href}%
\providecommand \@@href[1]{\endgroup#1\@@endlink}%
\providecommand \@sanitize@url [0]{\catcode `\\12\catcode `\$12\catcode
  `\&12\catcode `\#12\catcode `\^12\catcode `\_12\catcode `\%12\relax}%
\providecommand \@@startlink[1]{}%
\providecommand \@@endlink[0]{}%
\providecommand \url  [0]{\begingroup\@sanitize@url \@url }%
\providecommand \@url [1]{\endgroup\@href {#1}{\urlprefix }}%
\providecommand \urlprefix  [0]{URL }%
\providecommand \Eprint [0]{\href }%
\providecommand \doibase [0]{http://dx.doi.org/}%
\providecommand \selectlanguage [0]{\@gobble}%
\providecommand \bibinfo  [0]{\@secondoftwo}%
\providecommand \bibfield  [0]{\@secondoftwo}%
\providecommand \translation [1]{[#1]}%
\providecommand \BibitemOpen [0]{}%
\providecommand \bibitemStop [0]{}%
\providecommand \bibitemNoStop [0]{.\EOS\space}%
\providecommand \EOS [0]{\spacefactor3000\relax}%
\providecommand \BibitemShut  [1]{\csname bibitem#1\endcsname}%
\let\auto@bib@innerbib\@empty
\bibitem [{\citenamefont {Redfield}(1965)}]{redfield65}%
  \BibitemOpen
  \bibfield  {author} {\bibinfo {author} {\bibfnamefont {A.}~\bibnamefont
  {Redfield}},\ }in\ \href@noop {} {\emph {\bibinfo {booktitle} {Advances in
  Magnetic Resonance}}},\ Vol.~\bibinfo {volume} {1},\ \bibinfo {editor}
  {edited by\ \bibinfo {editor} {\bibfnamefont {J.}~\bibnamefont {Waugh}}}\
  (\bibinfo  {publisher} {Academic Press},\ \bibinfo {address} {New York},\
  \bibinfo {year} {1965})\BibitemShut {NoStop}%
\bibitem [{\citenamefont {Lindblad}(1976)}]{lindblad76}%
  \BibitemOpen
  \bibfield  {author} {\bibinfo {author} {\bibfnamefont {G.}~\bibnamefont
  {Lindblad}},\ }\href@noop {} {\bibfield  {journal} {\bibinfo  {journal}
  {Commun. Math. Phys.}\ }\textbf {\bibinfo {volume} {48}},\ \bibinfo {pages}
  {119} (\bibinfo {year} {1976})}\BibitemShut {NoStop}%
\bibitem [{\citenamefont {Davies}(1976)}]{0book-davies}%
  \BibitemOpen
  \bibfield  {author} {\bibinfo {author} {\bibfnamefont {E.~B.}\ \bibnamefont
  {Davies}},\ }\href@noop {} {\emph {\bibinfo {title} {Quantum Theory of Open
  Systems}}}\ (\bibinfo  {publisher} {Academic Press},\ \bibinfo {address}
  {London},\ \bibinfo {year} {1976})\BibitemShut {NoStop}%
\bibitem [{\citenamefont {Weiss}(1999)}]{weiss99}%
  \BibitemOpen
  \bibinfo {editor} {\bibfnamefont {U.}~\bibnamefont {Weiss}},\ ed.,\
  \href@noop {} {\emph {\bibinfo {title} {Quantum Dissipative Systems}}}\
  (\bibinfo  {publisher} {World Scientific},\ \bibinfo {address} {Singapore},\
  \bibinfo {year} {1999})\BibitemShut {NoStop}%
\bibitem [{\citenamefont {Breuer}\ and\ \citenamefont
  {Petruccione}(2006)}]{breuer06}%
  \BibitemOpen
  \bibinfo {editor} {\bibfnamefont {H.-P.}\ \bibnamefont {Breuer}}\ and\
  \bibinfo {editor} {\bibfnamefont {F.}~\bibnamefont {Petruccione}},\ eds.,\
  \href@noop {} {\emph {\bibinfo {title} {The Theory of Open Quantum
  Systems}}}\ (\bibinfo  {publisher} {Oxford University Press},\ \bibinfo
  {address} {Oxforfd},\ \bibinfo {year} {2006})\BibitemShut {NoStop}%
\bibitem [{\citenamefont {Toutounji}(2005)}]{toutounji05}%
  \BibitemOpen
  \bibfield  {author} {\bibinfo {author} {\bibfnamefont {M.}~\bibnamefont
  {Toutounji}},\ }\href@noop {} {\bibfield  {journal} {\bibinfo  {journal} {J.
  Chem. Phys.}\ }\textbf {\bibinfo {volume} {123}},\ \bibinfo {pages} {244102}
  (\bibinfo {year} {2005})}\BibitemShut {NoStop}%
\bibitem [{\citenamefont {Toutounji}\ and\ \citenamefont
  {Kapral}(2001)}]{toutounji01}%
  \BibitemOpen
  \bibfield  {author} {\bibinfo {author} {\bibfnamefont {M.}~\bibnamefont
  {Toutounji}}\ and\ \bibinfo {author} {\bibfnamefont {R.}~\bibnamefont
  {Kapral}},\ }\href@noop {} {\bibfield  {journal} {\bibinfo  {journal} {Chem.
  Phys.}\ }\textbf {\bibinfo {volume} {268}},\ \bibinfo {pages} {79} (\bibinfo
  {year} {2001})}\BibitemShut {NoStop}%
\bibitem [{\citenamefont {Herman}(1994)}]{herman94}%
  \BibitemOpen
  \bibfield  {author} {\bibinfo {author} {\bibfnamefont {M.~F.}\ \bibnamefont
  {Herman}},\ }\href@noop {} {\bibfield  {journal} {\bibinfo  {journal} {Annu.
  Rev. Phys. Chem.}\ }\textbf {\bibinfo {volume} {45}},\ \bibinfo {pages} {83}
  (\bibinfo {year} {1994})}\BibitemShut {NoStop}%
\bibitem [{\citenamefont {Tully}(1998)}]{tully98}%
  \BibitemOpen
  \bibfield  {author} {\bibinfo {author} {\bibfnamefont {J.~C.}\ \bibnamefont
  {Tully}},\ }\href@noop {} {\bibfield  {journal} {\bibinfo  {journal} {Faraday
  Discuss.}\ }\textbf {\bibinfo {volume} {110}},\ \bibinfo {pages} {407}
  (\bibinfo {year} {1998})}\BibitemShut {NoStop}%
\bibitem [{\citenamefont {Ben-Nun}\ and\ \citenamefont
  {Marti\'nez}(1998)}]{martinez98}%
  \BibitemOpen
  \bibfield  {author} {\bibinfo {author} {\bibfnamefont {M.}~\bibnamefont
  {Ben-Nun}}\ and\ \bibinfo {author} {\bibfnamefont {T.~J.}\ \bibnamefont
  {Marti\'nez}},\ }\href@noop {} {\bibfield  {journal} {\bibinfo  {journal} {J.
  Chem. Phys.}\ }\textbf {\bibinfo {volume} {108}},\ \bibinfo {pages} {7244}
  (\bibinfo {year} {1998})}\BibitemShut {NoStop}%
\bibitem [{\citenamefont {Kapral}(2006)}]{kapral06_2}%
  \BibitemOpen
  \bibfield  {author} {\bibinfo {author} {\bibfnamefont {R.}~\bibnamefont
  {Kapral}},\ }\href@noop {} {\bibfield  {journal} {\bibinfo  {journal} {Annu.
  Rev. Phys. Chem.}\ }\textbf {\bibinfo {volume} {57}},\ \bibinfo {pages} {129}
  (\bibinfo {year} {2006})}\BibitemShut {NoStop}%
\bibitem [{\citenamefont {Tully}(2012)}]{tully12}%
  \BibitemOpen
  \bibfield  {author} {\bibinfo {author} {\bibfnamefont {J.~C.}\ \bibnamefont
  {Tully}},\ }\href@noop {} {\bibfield  {journal} {\bibinfo  {journal} {J.
  Chem. Phys.}\ }\textbf {\bibinfo {volume} {137}},\ \bibinfo {pages} {22A301}
  (\bibinfo {year} {2012})}\BibitemShut {NoStop}%
\bibitem [{\citenamefont {Agostini}\ \emph {et~al.}(2013)\citenamefont
  {Agostini}, \citenamefont {Abedi}, \citenamefont {Suzuki},\ and\
  \citenamefont {Gross}}]{agostini13}%
  \BibitemOpen
  \bibfield  {author} {\bibinfo {author} {\bibfnamefont {F.}~\bibnamefont
  {Agostini}}, \bibinfo {author} {\bibfnamefont {A.}~\bibnamefont {Abedi}},
  \bibinfo {author} {\bibfnamefont {Y.}~\bibnamefont {Suzuki}}, \ and\ \bibinfo
  {author} {\bibfnamefont {E.~K.~U.}\ \bibnamefont {Gross}},\ }\href@noop {}
  {\bibfield  {journal} {\bibinfo  {journal} {Mol. Phys.}\ }\textbf {\bibinfo
  {volume} {111}},\ \bibinfo {pages} {3625} (\bibinfo {year}
  {2013})}\BibitemShut {NoStop}%
\bibitem [{\citenamefont {Abedi}, \citenamefont {Maitra},\ and\ \citenamefont
  {Gross}(2010)}]{abedi10}%
  \BibitemOpen
  \bibfield  {author} {\bibinfo {author} {\bibfnamefont {A.}~\bibnamefont
  {Abedi}}, \bibinfo {author} {\bibfnamefont {N.}~\bibnamefont {Maitra}}, \
  and\ \bibinfo {author} {\bibfnamefont {E.~K.~U.}\ \bibnamefont {Gross}},\
  }\href@noop {} {\bibfield  {journal} {\bibinfo  {journal} {Phys. Rev. Lett.}\
  }\textbf {\bibinfo {volume} {105}},\ \bibinfo {pages} {123002} (\bibinfo
  {year} {2010})}\BibitemShut {NoStop}%
\bibitem [{\citenamefont {Sun}\ and\ \citenamefont {Miller}(1997)}]{sun97}%
  \BibitemOpen
  \bibfield  {author} {\bibinfo {author} {\bibfnamefont {X.}~\bibnamefont
  {Sun}}\ and\ \bibinfo {author} {\bibfnamefont {W.~H.}\ \bibnamefont
  {Miller}},\ }\href@noop {} {\bibfield  {journal} {\bibinfo  {journal} {J.
  Chem. Phys.}\ }\textbf {\bibinfo {volume} {106}},\ \bibinfo {pages} {916}
  (\bibinfo {year} {1997})}\BibitemShut {NoStop}%
\bibitem [{\citenamefont {Sun}, \citenamefont {Wang},\ and\ \citenamefont
  {Miller}(1998)}]{sun98}%
  \BibitemOpen
  \bibfield  {author} {\bibinfo {author} {\bibfnamefont {X.}~\bibnamefont
  {Sun}}, \bibinfo {author} {\bibfnamefont {H.~B.}\ \bibnamefont {Wang}}, \
  and\ \bibinfo {author} {\bibfnamefont {W.~H.}\ \bibnamefont {Miller}},\
  }\href@noop {} {\bibfield  {journal} {\bibinfo  {journal} {J. Chem. Phys.}\
  }\textbf {\bibinfo {volume} {109}},\ \bibinfo {pages} {7064} (\bibinfo {year}
  {1998})}\BibitemShut {NoStop}%
\bibitem [{\citenamefont {Makri}\ and\ \citenamefont
  {Thompson}(1998)}]{makri98}%
  \BibitemOpen
  \bibfield  {author} {\bibinfo {author} {\bibfnamefont {N.}~\bibnamefont
  {Makri}}\ and\ \bibinfo {author} {\bibfnamefont {K.}~\bibnamefont
  {Thompson}},\ }\href@noop {} {\bibfield  {journal} {\bibinfo  {journal}
  {Chem. Phys. Lett.}\ }\textbf {\bibinfo {volume} {291}},\ \bibinfo {pages}
  {101} (\bibinfo {year} {1998})}\BibitemShut {NoStop}%
\bibitem [{\citenamefont {Miller}(2009)}]{miller09}%
  \BibitemOpen
  \bibfield  {author} {\bibinfo {author} {\bibfnamefont {W.~H.}\ \bibnamefont
  {Miller}},\ }\href@noop {} {\bibfield  {journal} {\bibinfo  {journal} {J.
  Phys. Chem. A}\ }\textbf {\bibinfo {volume} {113}},\ \bibinfo {pages} {1406}
  (\bibinfo {year} {2009})}\BibitemShut {NoStop}%
\bibitem [{\citenamefont {Lambert}\ and\ \citenamefont
  {Makri}(2012)}]{makri12}%
  \BibitemOpen
  \bibfield  {author} {\bibinfo {author} {\bibfnamefont {R.}~\bibnamefont
  {Lambert}}\ and\ \bibinfo {author} {\bibfnamefont {N.}~\bibnamefont
  {Makri}},\ }\href@noop {} {\bibfield  {journal} {\bibinfo  {journal} {J.
  Chem. Phys.}\ }\textbf {\bibinfo {volume} {137}},\ \bibinfo {pages} {22A552}
  (\bibinfo {year} {2012})}\BibitemShut {NoStop}%
\bibitem [{\citenamefont {Habershon}\ \emph {et~al.}(2013)\citenamefont
  {Habershon}, \citenamefont {Manolopoulos}, \citenamefont {Markland},\ and\
  \citenamefont {Miller}}]{ring-polymer13}%
  \BibitemOpen
  \bibfield  {author} {\bibinfo {author} {\bibfnamefont {S.}~\bibnamefont
  {Habershon}}, \bibinfo {author} {\bibfnamefont {D.~E.}\ \bibnamefont
  {Manolopoulos}}, \bibinfo {author} {\bibfnamefont {T.~E.}\ \bibnamefont
  {Markland}}, \ and\ \bibinfo {author} {\bibfnamefont {T.~F.}\ \bibnamefont
  {Miller}},\ }\href@noop {} {\bibfield  {journal} {\bibinfo  {journal} {Annu.
  Rev. Phys. Chem.}\ }\textbf {\bibinfo {volume} {64}},\ \bibinfo {pages} {387}
  (\bibinfo {year} {2013})}\BibitemShut {NoStop}%
\bibitem [{\citenamefont {Joos}\ \emph {et~al.}(2003)\citenamefont {Joos},
  \citenamefont {Zeh}, \citenamefont {Kiefer}, \citenamefont {Giulini},
  \citenamefont {Kupsch},\ and\ \citenamefont {Stamatescu}}]{0book-joos03}%
  \BibitemOpen
  \bibfield  {author} {\bibinfo {author} {\bibfnamefont {E.}~\bibnamefont
  {Joos}}, \bibinfo {author} {\bibfnamefont {H.~D.}\ \bibnamefont {Zeh}},
  \bibinfo {author} {\bibfnamefont {C.}~\bibnamefont {Kiefer}}, \bibinfo
  {author} {\bibfnamefont {D.}~\bibnamefont {Giulini}}, \bibinfo {author}
  {\bibfnamefont {J.}~\bibnamefont {Kupsch}}, \ and\ \bibinfo {author}
  {\bibfnamefont {I.-O.}\ \bibnamefont {Stamatescu}},\ }\href@noop {} {\emph
  {\bibinfo {title} {Decoherence and the Appearance of a Classical World in
  Quantum Theory}}}\ (\bibinfo  {publisher} {Springer},\ \bibinfo {address}
  {New York},\ \bibinfo {year} {2003})\BibitemShut {NoStop}%
\bibitem [{\citenamefont {Zurek}(1991)}]{zurek91}%
  \BibitemOpen
  \bibfield  {author} {\bibinfo {author} {\bibfnamefont {W.~H.}\ \bibnamefont
  {Zurek}},\ }\href@noop {} {\bibfield  {journal} {\bibinfo  {journal} {Phys.
  Today}\ }\textbf {\bibinfo {volume} {44}},\ \bibinfo {pages} {36} (\bibinfo
  {year} {1991})}\BibitemShut {NoStop}%
\bibitem [{\citenamefont {Shiokawa}\ and\ \citenamefont
  {Kapral}(2002)}]{shiokawa02}%
  \BibitemOpen
  \bibfield  {author} {\bibinfo {author} {\bibfnamefont {K.}~\bibnamefont
  {Shiokawa}}\ and\ \bibinfo {author} {\bibfnamefont {R.}~\bibnamefont
  {Kapral}},\ }\href@noop {} {\bibfield  {journal} {\bibinfo  {journal} {J.
  Chem. Phys.}\ }\textbf {\bibinfo {volume} {117}},\ \bibinfo {pages} {7852}
  (\bibinfo {year} {2002})}\BibitemShut {NoStop}%
\bibitem [{\citenamefont {Wigner}(1932)}]{wigner32}%
  \BibitemOpen
  \bibfield  {author} {\bibinfo {author} {\bibfnamefont {E.}~\bibnamefont
  {Wigner}},\ }\href@noop {} {\bibfield  {journal} {\bibinfo  {journal} {Phys.
  Rev.}\ }\textbf {\bibinfo {volume} {40}},\ \bibinfo {pages} {749} (\bibinfo
  {year} {1932})}\BibitemShut {NoStop}%
\bibitem [{\citenamefont {Imre}\ \emph {et~al.}(1967)\citenamefont {Imre},
  \citenamefont {\"{O}zizmir}, \citenamefont {Rosenbaum},\ and\ \citenamefont
  {Zwiefel}}]{imre67}%
  \BibitemOpen
  \bibfield  {author} {\bibinfo {author} {\bibfnamefont {K.}~\bibnamefont
  {Imre}}, \bibinfo {author} {\bibfnamefont {E.}~\bibnamefont {\"{O}zizmir}},
  \bibinfo {author} {\bibfnamefont {M.}~\bibnamefont {Rosenbaum}}, \ and\
  \bibinfo {author} {\bibfnamefont {P.~F.}\ \bibnamefont {Zwiefel}},\
  }\href@noop {} {\bibfield  {journal} {\bibinfo  {journal} {J. Math. Phys.}\
  }\textbf {\bibinfo {volume} {5}},\ \bibinfo {pages} {1097} (\bibinfo {year}
  {1967})}\BibitemShut {NoStop}%
\bibitem [{\citenamefont {Aleksandrov}(1981)}]{alek81}%
  \BibitemOpen
  \bibfield  {author} {\bibinfo {author} {\bibfnamefont {I.~V.}\ \bibnamefont
  {Aleksandrov}},\ }\href@noop {} {\bibfield  {journal} {\bibinfo  {journal}
  {Z. Naturforsch.}\ }\textbf {\bibinfo {volume} {36}},\ \bibinfo {pages} {902}
  (\bibinfo {year} {1981})}\BibitemShut {NoStop}%
\bibitem [{\citenamefont {Gerasimenko}(1982)}]{geras82}%
  \BibitemOpen
  \bibfield  {author} {\bibinfo {author} {\bibfnamefont {V.~I.}\ \bibnamefont
  {Gerasimenko}},\ }\href@noop {} {\bibfield  {journal} {\bibinfo  {journal}
  {Theor. Math. Phys.}\ }\textbf {\bibinfo {volume} {50}},\ \bibinfo {pages}
  {77} (\bibinfo {year} {1982})}\BibitemShut {NoStop}%
\bibitem [{\citenamefont {Kapral}\ and\ \citenamefont
  {Ciccotti}(1999)}]{kapral99}%
  \BibitemOpen
  \bibfield  {author} {\bibinfo {author} {\bibfnamefont {R.}~\bibnamefont
  {Kapral}}\ and\ \bibinfo {author} {\bibfnamefont {G.}~\bibnamefont
  {Ciccotti}},\ }\href@noop {} {\bibfield  {journal} {\bibinfo  {journal} {J.
  Chem. Phys.}\ }\textbf {\bibinfo {volume} {110}},\ \bibinfo {pages} {8919}
  (\bibinfo {year} {1999})}\BibitemShut {NoStop}%
\bibitem [{\citenamefont {Mazur}\ and\ \citenamefont
  {Oppenheim}(1970)}]{mazur70}%
  \BibitemOpen
  \bibfield  {author} {\bibinfo {author} {\bibfnamefont {P.}~\bibnamefont
  {Mazur}}\ and\ \bibinfo {author} {\bibfnamefont {I.}~\bibnamefont
  {Oppenheim}},\ }\href@noop {} {\bibfield  {journal} {\bibinfo  {journal}
  {Physica}\ }\textbf {\bibinfo {volume} {50}},\ \bibinfo {pages} {141}
  (\bibinfo {year} {1970})}\BibitemShut {NoStop}%
\bibitem [{\citenamefont {Boucher}\ and\ \citenamefont
  {Traschen}(1988)}]{boucher88}%
  \BibitemOpen
  \bibfield  {author} {\bibinfo {author} {\bibfnamefont {W.}~\bibnamefont
  {Boucher}}\ and\ \bibinfo {author} {\bibfnamefont {J.}~\bibnamefont
  {Traschen}},\ }\href@noop {} {\bibfield  {journal} {\bibinfo  {journal}
  {Phys. Rev. D}\ }\textbf {\bibinfo {volume} {37}},\ \bibinfo {pages} {3522}
  (\bibinfo {year} {1988})}\BibitemShut {NoStop}%
\bibitem [{\citenamefont {Zhang}\ and\ \citenamefont
  {Balescu}(1988)}]{zhang88}%
  \BibitemOpen
  \bibfield  {author} {\bibinfo {author} {\bibfnamefont {W.~Y.}\ \bibnamefont
  {Zhang}}\ and\ \bibinfo {author} {\bibfnamefont {R.}~\bibnamefont
  {Balescu}},\ }\href@noop {} {\bibfield  {journal} {\bibinfo  {journal} {J.
  Plasma Phys.}\ }\textbf {\bibinfo {volume} {40}},\ \bibinfo {pages} {199}
  (\bibinfo {year} {1988})}\BibitemShut {NoStop}%
\bibitem [{\citenamefont {Donoso}\ and\ \citenamefont
  {Martens}(1998)}]{donoso98}%
  \BibitemOpen
  \bibfield  {author} {\bibinfo {author} {\bibfnamefont {A.}~\bibnamefont
  {Donoso}}\ and\ \bibinfo {author} {\bibfnamefont {C.~C.}\ \bibnamefont
  {Martens}},\ }\href@noop {} {\bibfield  {journal} {\bibinfo  {journal} {J.
  Phys. Chem. A}\ }\textbf {\bibinfo {volume} {102}},\ \bibinfo {pages} {4291}
  (\bibinfo {year} {1998})}\BibitemShut {NoStop}%
\bibitem [{\citenamefont {Horenko}\ \emph {et~al.}(2002)\citenamefont
  {Horenko}, \citenamefont {Salzmann}, \citenamefont {Schmidt},\ and\
  \citenamefont {Schutte}}]{horenko02}%
  \BibitemOpen
  \bibfield  {author} {\bibinfo {author} {\bibfnamefont {I.}~\bibnamefont
  {Horenko}}, \bibinfo {author} {\bibfnamefont {C.}~\bibnamefont {Salzmann}},
  \bibinfo {author} {\bibfnamefont {B.}~\bibnamefont {Schmidt}}, \ and\
  \bibinfo {author} {\bibfnamefont {C.}~\bibnamefont {Schutte}},\ }\href@noop
  {} {\bibfield  {journal} {\bibinfo  {journal} {J. Chem. Phys.}\ }\textbf
  {\bibinfo {volume} {117}},\ \bibinfo {pages} {11075} (\bibinfo {year}
  {2002})}\BibitemShut {NoStop}%
\bibitem [{\citenamefont {Shi}\ and\ \citenamefont {Geva}(2004)}]{shi04a}%
  \BibitemOpen
  \bibfield  {author} {\bibinfo {author} {\bibfnamefont {Q.}~\bibnamefont
  {Shi}}\ and\ \bibinfo {author} {\bibfnamefont {E.}~\bibnamefont {Geva}},\
  }\href@noop {} {\bibfield  {journal} {\bibinfo  {journal} {J. Chem. Phys.}\
  }\textbf {\bibinfo {volume} {121}},\ \bibinfo {pages} {3393} (\bibinfo {year}
  {2004})}\BibitemShut {NoStop}%
\bibitem [{\citenamefont {Thorndyke}\ and\ \citenamefont
  {Micha}(2005)}]{thorndyke05}%
  \BibitemOpen
  \bibfield  {author} {\bibinfo {author} {\bibfnamefont {B.}~\bibnamefont
  {Thorndyke}}\ and\ \bibinfo {author} {\bibfnamefont {D.~A.}\ \bibnamefont
  {Micha}},\ }\href@noop {} {\bibfield  {journal} {\bibinfo  {journal} {Chem.
  Phys. Lett.}\ }\textbf {\bibinfo {volume} {403}},\ \bibinfo {pages} {280}
  (\bibinfo {year} {2005})}\BibitemShut {NoStop}%
\bibitem [{\citenamefont {Bousquet}\ \emph {et~al.}(2011)\citenamefont
  {Bousquet}, \citenamefont {Hughes}, \citenamefont {Micha},\ and\
  \citenamefont {Burghardt}}]{burghardt11}%
  \BibitemOpen
  \bibfield  {author} {\bibinfo {author} {\bibfnamefont {D.}~\bibnamefont
  {Bousquet}}, \bibinfo {author} {\bibfnamefont {K.~H.}\ \bibnamefont
  {Hughes}}, \bibinfo {author} {\bibfnamefont {D.~A.}\ \bibnamefont {Micha}}, \
  and\ \bibinfo {author} {\bibfnamefont {I.}~\bibnamefont {Burghardt}},\
  }\href@noop {} {\bibfield  {journal} {\bibinfo  {journal} {J. Chem. Phys.}\
  }\textbf {\bibinfo {volume} {134}},\ \bibinfo {pages} {064116} (\bibinfo
  {year} {2011})}\BibitemShut {NoStop}%
\bibitem [{\citenamefont {Nielsen}, \citenamefont {Kapral},\ and\ \citenamefont
  {Ciccotti}(2001)}]{nielsen01}%
  \BibitemOpen
  \bibfield  {author} {\bibinfo {author} {\bibfnamefont {S.}~\bibnamefont
  {Nielsen}}, \bibinfo {author} {\bibfnamefont {R.}~\bibnamefont {Kapral}}, \
  and\ \bibinfo {author} {\bibfnamefont {G.}~\bibnamefont {Ciccotti}},\
  }\href@noop {} {\bibfield  {journal} {\bibinfo  {journal} {J. Chem. Phys.}\
  }\textbf {\bibinfo {volume} {115}},\ \bibinfo {pages} {5805} (\bibinfo {year}
  {2001})}\BibitemShut {NoStop}%
\bibitem [{\citenamefont {Bonella}, \citenamefont {Ciccotti},\ and\
  \citenamefont {Kapral.}(2010)}]{bonella10}%
  \BibitemOpen
  \bibfield  {author} {\bibinfo {author} {\bibfnamefont {S.}~\bibnamefont
  {Bonella}}, \bibinfo {author} {\bibfnamefont {G.}~\bibnamefont {Ciccotti}}, \
  and\ \bibinfo {author} {\bibfnamefont {R.}~\bibnamefont {Kapral.}},\
  }\href@noop {} {\bibfield  {journal} {\bibinfo  {journal} {Chem. Phys.
  Lett.}\ }\textbf {\bibinfo {volume} {484}},\ \bibinfo {pages} {399} (\bibinfo
  {year} {2010})}\BibitemShut {NoStop}%
\bibitem [{\citenamefont {Kapral}(2001)}]{kapral01}%
  \BibitemOpen
  \bibfield  {author} {\bibinfo {author} {\bibfnamefont {R.}~\bibnamefont
  {Kapral}},\ }\href@noop {} {\bibfield  {journal} {\bibinfo  {journal} {J.
  Phys. Chem. A}\ }\textbf {\bibinfo {volume} {105}},\ \bibinfo {pages} {2885}
  (\bibinfo {year} {2001})}\BibitemShut {NoStop}%
\bibitem [{\citenamefont {Nakajima}(1958)}]{nakajima58}%
  \BibitemOpen
  \bibfield  {author} {\bibinfo {author} {\bibfnamefont {S.}~\bibnamefont
  {Nakajima}},\ }\href@noop {} {\bibfield  {journal} {\bibinfo  {journal}
  {Prog. Theor. Phys.}\ }\textbf {\bibinfo {volume} {20}},\ \bibinfo {pages}
  {948} (\bibinfo {year} {1958})}\BibitemShut {NoStop}%
\bibitem [{\citenamefont {Zwanzig}(1961)}]{0chap-zwanzig61}%
  \BibitemOpen
  \bibfield  {author} {\bibinfo {author} {\bibfnamefont {R.}~\bibnamefont
  {Zwanzig}},\ }in\ \href@noop {} {\emph {\bibinfo {booktitle} {Lectures in
  Theoretical Physics}}},\ Vol.~\bibinfo {volume} {3},\ \bibinfo {editor}
  {edited by\ \bibinfo {editor} {\bibfnamefont {W.~E.}\ \bibnamefont
  {Brittin}}, \bibinfo {editor} {\bibfnamefont {B.}~\bibnamefont {Downs}}, \
  and\ \bibinfo {editor} {\bibfnamefont {J.}~\bibnamefont {Downs}}}\ (\bibinfo
  {publisher} {Interscience},\ \bibinfo {address} {New York},\ \bibinfo {year}
  {1961})\BibitemShut {NoStop}%
\bibitem [{\citenamefont {Tanimura}\ and\ \citenamefont
  {Mukamel}(1994)}]{mukamel94}%
  \BibitemOpen
  \bibfield  {author} {\bibinfo {author} {\bibfnamefont {Y.}~\bibnamefont
  {Tanimura}}\ and\ \bibinfo {author} {\bibfnamefont {S.}~\bibnamefont
  {Mukamel}},\ }\href@noop {} {\bibfield  {journal} {\bibinfo  {journal} {J.
  Chem. Phys.}\ }\textbf {\bibinfo {volume} {101}},\ \bibinfo {pages} {3049}
  (\bibinfo {year} {1994})}\BibitemShut {NoStop}%
\bibitem [{\citenamefont {Kapral}\ and\ \citenamefont
  {Ciccotti}(2002)}]{0chap-kapral02}%
  \BibitemOpen
  \bibfield  {author} {\bibinfo {author} {\bibfnamefont {R.}~\bibnamefont
  {Kapral}}\ and\ \bibinfo {author} {\bibfnamefont {G.}~\bibnamefont
  {Ciccotti}},\ }in\ \href@noop {} {\emph {\bibinfo {booktitle} {Bridging time
  scales: Molecular Simulations for the Next Decade}}},\ \bibinfo {editor}
  {edited by\ \bibinfo {editor} {\bibfnamefont {P.}~\bibnamefont {Nielaba}},
  \bibinfo {editor} {\bibfnamefont {M.}~\bibnamefont {Mareschal}}, \ and\
  \bibinfo {editor} {\bibfnamefont {G.}~\bibnamefont {Ciccotti}}}\ (\bibinfo
  {publisher} {Springer-Verlag},\ \bibinfo {address} {Berlin},\ \bibinfo {year}
  {2002})\ pp.\ \bibinfo {pages} {445--472}\BibitemShut {NoStop}%
\bibitem [{\citenamefont {Salcedo}(1996)}]{salcedo96}%
  \BibitemOpen
  \bibfield  {author} {\bibinfo {author} {\bibfnamefont {L.~L.}\ \bibnamefont
  {Salcedo}},\ }\href@noop {} {\bibfield  {journal} {\bibinfo  {journal} {Phys.
  Rev. A}\ }\textbf {\bibinfo {volume} {54}},\ \bibinfo {pages} {3657}
  (\bibinfo {year} {1996})}\BibitemShut {NoStop}%
\bibitem [{\citenamefont {Salcedo}(2012)}]{salcedo12}%
  \BibitemOpen
  \bibfield  {author} {\bibinfo {author} {\bibfnamefont {L.~L.}\ \bibnamefont
  {Salcedo}},\ }\href@noop {} {\bibfield  {journal} {\bibinfo  {journal} {Phys.
  Rev. A}\ }\textbf {\bibinfo {volume} {85}},\ \bibinfo {pages} {022127}
  (\bibinfo {year} {2012})}\BibitemShut {NoStop}%
\bibitem [{\citenamefont {Prezhdo}\ and\ \citenamefont
  {Kisil}(1997)}]{prezhdo97b}%
  \BibitemOpen
  \bibfield  {author} {\bibinfo {author} {\bibfnamefont {O.~V.}\ \bibnamefont
  {Prezhdo}}\ and\ \bibinfo {author} {\bibfnamefont {V.~V.}\ \bibnamefont
  {Kisil}},\ }\href@noop {} {\bibfield  {journal} {\bibinfo  {journal} {Phys.
  Rev. A}\ }\textbf {\bibinfo {volume} {56}},\ \bibinfo {pages} {162} (\bibinfo
  {year} {1997})}\BibitemShut {NoStop}%
\bibitem [{\citenamefont {Sergi}(2005)}]{sergi05}%
  \BibitemOpen
  \bibfield  {author} {\bibinfo {author} {\bibfnamefont {A.}~\bibnamefont
  {Sergi}},\ }\href@noop {} {\bibfield  {journal} {\bibinfo  {journal} {Phys.
  Rev. E}\ }\textbf {\bibinfo {volume} {72}},\ \bibinfo {pages} {066125}
  (\bibinfo {year} {2005})}\BibitemShut {NoStop}%
\bibitem [{\citenamefont {Prezhdo}(2006)}]{prezhdo06}%
  \BibitemOpen
  \bibfield  {author} {\bibinfo {author} {\bibfnamefont {O.~V.}\ \bibnamefont
  {Prezhdo}},\ }\href@noop {} {\bibfield  {journal} {\bibinfo  {journal} {J.
  Chem. Phys.}\ }\textbf {\bibinfo {volume} {124}},\ \bibinfo {pages} {201104}
  (\bibinfo {year} {2006})}\BibitemShut {NoStop}%
\bibitem [{\citenamefont {Salcedo}(2007)}]{salcedo07}%
  \BibitemOpen
  \bibfield  {author} {\bibinfo {author} {\bibfnamefont {L.~L.}\ \bibnamefont
  {Salcedo}},\ }\href@noop {} {\bibfield  {journal} {\bibinfo  {journal} {J.
  Chem. Phys.}\ }\textbf {\bibinfo {volume} {126}},\ \bibinfo {pages} {057101}
  (\bibinfo {year} {2007})}\BibitemShut {NoStop}%
\bibitem [{\citenamefont {F.~Agostini}\ and\ \citenamefont
  {Ciccotti}(2007)}]{agostini07}%
  \BibitemOpen
  \bibfield  {author} {\bibinfo {author} {\bibfnamefont {S.~C.}\ \bibnamefont
  {F.~Agostini}}\ and\ \bibinfo {author} {\bibfnamefont {G.}~\bibnamefont
  {Ciccotti}},\ }\href@noop {} {\bibfield  {journal} {\bibinfo  {journal}
  {EPL}\ }\textbf {\bibinfo {volume} {78}},\ \bibinfo {pages} {30001} (\bibinfo
  {year} {2007})}\BibitemShut {NoStop}%
\bibitem [{\citenamefont {Hall}\ and\ \citenamefont
  {Reginatto}(2005)}]{hall_reginatto05}%
  \BibitemOpen
  \bibfield  {author} {\bibinfo {author} {\bibfnamefont {M.~J.~W.}\
  \bibnamefont {Hall}}\ and\ \bibinfo {author} {\bibfnamefont {M.}~\bibnamefont
  {Reginatto}},\ }\href@noop {} {\bibfield  {journal} {\bibinfo  {journal}
  {Phys. Rev. A}\ }\textbf {\bibinfo {volume} {72}},\ \bibinfo {pages} {062109}
  (\bibinfo {year} {2005})}\BibitemShut {NoStop}%
\bibitem [{\citenamefont {Hall}(2008)}]{hall08}%
  \BibitemOpen
  \bibfield  {author} {\bibinfo {author} {\bibfnamefont {M.~J.~W.}\
  \bibnamefont {Hall}},\ }\href@noop {} {\bibfield  {journal} {\bibinfo
  {journal} {Phys. Rev. A}\ }\textbf {\bibinfo {volume} {78}},\ \bibinfo
  {pages} {042104} (\bibinfo {year} {2008})}\BibitemShut {NoStop}%
\bibitem [{\citenamefont {Sergi}\ and\ \citenamefont {Kapral}(2004)}]{sergi04}%
  \BibitemOpen
  \bibfield  {author} {\bibinfo {author} {\bibfnamefont {A.}~\bibnamefont
  {Sergi}}\ and\ \bibinfo {author} {\bibfnamefont {R.}~\bibnamefont {Kapral}},\
  }\href@noop {} {\bibfield  {journal} {\bibinfo  {journal} {J. Chem. Phys.}\
  }\textbf {\bibinfo {volume} {121}},\ \bibinfo {pages} {7565} (\bibinfo {year}
  {2004})}\BibitemShut {NoStop}%
\bibitem [{\citenamefont {Kim}\ and\ \citenamefont
  {Kapral}(2005{\natexlab{a}})}]{kim05}%
  \BibitemOpen
  \bibfield  {author} {\bibinfo {author} {\bibfnamefont {H.}~\bibnamefont
  {Kim}}\ and\ \bibinfo {author} {\bibfnamefont {R.}~\bibnamefont {Kapral}},\
  }\href@noop {} {\bibfield  {journal} {\bibinfo  {journal} {J. Chem. Phys.}\
  }\textbf {\bibinfo {volume} {122}},\ \bibinfo {pages} {214105} (\bibinfo
  {year} {2005}{\natexlab{a}})}\BibitemShut {NoStop}%
\bibitem [{\citenamefont {Kim}\ and\ \citenamefont
  {Kapral}(2005{\natexlab{b}})}]{kim05a}%
  \BibitemOpen
  \bibfield  {author} {\bibinfo {author} {\bibfnamefont {H.}~\bibnamefont
  {Kim}}\ and\ \bibinfo {author} {\bibfnamefont {R.}~\bibnamefont {Kapral}},\
  }\href@noop {} {\bibfield  {journal} {\bibinfo  {journal} {J. Chem. Phys.}\
  }\textbf {\bibinfo {volume} {123}},\ \bibinfo {pages} {194108} (\bibinfo
  {year} {2005}{\natexlab{b}})}\BibitemShut {NoStop}%
\bibitem [{\citenamefont {Hsieh}\ and\ \citenamefont {Kapral}(2014)}]{hsieh14}%
  \BibitemOpen
  \bibfield  {author} {\bibinfo {author} {\bibfnamefont {C.-Y.}\ \bibnamefont
  {Hsieh}}\ and\ \bibinfo {author} {\bibfnamefont {R.}~\bibnamefont {Kapral}},\
  }\href@noop {} {\bibfield  {journal} {\bibinfo  {journal} {Entropy}\ }\textbf
  {\bibinfo {volume} {16}},\ \bibinfo {pages} {200} (\bibinfo {year}
  {2014})}\BibitemShut {NoStop}%
\bibitem [{\citenamefont {Poulsen}, \citenamefont {Nyman},\ and\ \citenamefont
  {Rossky}(2003)}]{poulsen03}%
  \BibitemOpen
  \bibfield  {author} {\bibinfo {author} {\bibfnamefont {J.~A.}\ \bibnamefont
  {Poulsen}}, \bibinfo {author} {\bibfnamefont {G.}~\bibnamefont {Nyman}}, \
  and\ \bibinfo {author} {\bibfnamefont {P.~J.}\ \bibnamefont {Rossky}},\
  }\href@noop {} {\bibfield  {journal} {\bibinfo  {journal} {J. Chem. Phys.}\
  }\textbf {\bibinfo {volume} {119}},\ \bibinfo {pages} {12179} (\bibinfo
  {year} {2003})}\BibitemShut {NoStop}%
\bibitem [{\citenamefont {Ananth}\ and\ \citenamefont
  {Miller}(2010)}]{ananth10}%
  \BibitemOpen
  \bibfield  {author} {\bibinfo {author} {\bibfnamefont {N.}~\bibnamefont
  {Ananth}}\ and\ \bibinfo {author} {\bibfnamefont {T.~F.}\ \bibnamefont
  {Miller}},\ }\href@noop {} {\bibfield  {journal} {\bibinfo  {journal} {J.
  Chem. Phys.}\ }\textbf {\bibinfo {volume} {133}},\ \bibinfo {pages} {234103}
  (\bibinfo {year} {2010})}\BibitemShut {NoStop}%
\bibitem [{\citenamefont {Tully}(1990)}]{tully90}%
  \BibitemOpen
  \bibfield  {author} {\bibinfo {author} {\bibfnamefont {J.~C.}\ \bibnamefont
  {Tully}},\ }\href@noop {} {\bibfield  {journal} {\bibinfo  {journal} {J.
  Chem. Phys.}\ }\textbf {\bibinfo {volume} {93}},\ \bibinfo {pages} {1061}
  (\bibinfo {year} {1990})}\BibitemShut {NoStop}%
\bibitem [{\citenamefont {Tully}(1991)}]{tully91}%
  \BibitemOpen
  \bibfield  {author} {\bibinfo {author} {\bibfnamefont {J.~C.}\ \bibnamefont
  {Tully}},\ }\href@noop {} {\bibfield  {journal} {\bibinfo  {journal} {Int. J.
  Quantum Chem.}\ }\textbf {\bibinfo {volume} {25}},\ \bibinfo {pages} {299}
  (\bibinfo {year} {1991})}\BibitemShut {NoStop}%
\bibitem [{\citenamefont {Neria}\ and\ \citenamefont {Nitzan}(1993)}]{neria93}%
  \BibitemOpen
  \bibfield  {author} {\bibinfo {author} {\bibfnamefont {E.}~\bibnamefont
  {Neria}}\ and\ \bibinfo {author} {\bibfnamefont {A.}~\bibnamefont {Nitzan}},\
  }\href@noop {} {\bibfield  {journal} {\bibinfo  {journal} {J. Chem. Phys.}\
  }\textbf {\bibinfo {volume} {99}},\ \bibinfo {pages} {1109} (\bibinfo {year}
  {1993})}\BibitemShut {NoStop}%
\bibitem [{\citenamefont {Hammes-Schiffer}\ and\ \citenamefont
  {Tully}(1994)}]{hammesschiffer94}%
  \BibitemOpen
  \bibfield  {author} {\bibinfo {author} {\bibfnamefont {S.}~\bibnamefont
  {Hammes-Schiffer}}\ and\ \bibinfo {author} {\bibfnamefont {J.~C.}\
  \bibnamefont {Tully}},\ }\href@noop {} {\bibfield  {journal} {\bibinfo
  {journal} {J. Chem. Phys.}\ }\textbf {\bibinfo {volume} {101}},\ \bibinfo
  {pages} {4657} (\bibinfo {year} {1994})}\BibitemShut {NoStop}%
\bibitem [{\citenamefont {Bittner}\ and\ \citenamefont
  {Rossky}(1995)}]{bittner95}%
  \BibitemOpen
  \bibfield  {author} {\bibinfo {author} {\bibfnamefont {E.~R.}\ \bibnamefont
  {Bittner}}\ and\ \bibinfo {author} {\bibfnamefont {P.~J.}\ \bibnamefont
  {Rossky}},\ }\href@noop {} {\bibfield  {journal} {\bibinfo  {journal} {J.
  Chem. Phys.}\ }\textbf {\bibinfo {volume} {103}},\ \bibinfo {pages} {8130}
  (\bibinfo {year} {1995})}\BibitemShut {NoStop}%
\bibitem [{\citenamefont {Bittner}, \citenamefont {Schwartz},\ and\
  \citenamefont {Rossky}(1997)}]{bittner97}%
  \BibitemOpen
  \bibfield  {author} {\bibinfo {author} {\bibfnamefont {E.~R.}\ \bibnamefont
  {Bittner}}, \bibinfo {author} {\bibfnamefont {B.~J.}\ \bibnamefont
  {Schwartz}}, \ and\ \bibinfo {author} {\bibfnamefont {P.~J.}\ \bibnamefont
  {Rossky}},\ }\href@noop {} {\bibfield  {journal} {\bibinfo  {journal} {J.
  Mol. Struct.: THEOCHEM}\ }\textbf {\bibinfo {volume} {389}},\ \bibinfo
  {pages} {203} (\bibinfo {year} {1997})}\BibitemShut {NoStop}%
\bibitem [{\citenamefont {Schwartz}\ \emph {et~al.}(1996)\citenamefont
  {Schwartz}, \citenamefont {Bittner}, \citenamefont {Prezhdo},\ and\
  \citenamefont {Rossky}}]{schwartz96}%
  \BibitemOpen
  \bibfield  {author} {\bibinfo {author} {\bibfnamefont {B.~J.}\ \bibnamefont
  {Schwartz}}, \bibinfo {author} {\bibfnamefont {E.~R.}\ \bibnamefont
  {Bittner}}, \bibinfo {author} {\bibfnamefont {O.~V.}\ \bibnamefont
  {Prezhdo}}, \ and\ \bibinfo {author} {\bibfnamefont {P.~J.}\ \bibnamefont
  {Rossky}},\ }\href@noop {} {\bibfield  {journal} {\bibinfo  {journal} {J.
  Chem. Phys.}\ }\textbf {\bibinfo {volume} {104}},\ \bibinfo {pages} {5942}
  (\bibinfo {year} {1996})}\BibitemShut {NoStop}%
\bibitem [{\citenamefont {Bedard-Hearn}, \citenamefont {Larsen},\ and\
  \citenamefont {Schwartz}(2005)}]{bedard05}%
  \BibitemOpen
  \bibfield  {author} {\bibinfo {author} {\bibfnamefont {M.~J.}\ \bibnamefont
  {Bedard-Hearn}}, \bibinfo {author} {\bibfnamefont {R.~E.}\ \bibnamefont
  {Larsen}}, \ and\ \bibinfo {author} {\bibfnamefont {B.~J.}\ \bibnamefont
  {Schwartz}},\ }\href@noop {} {\bibfield  {journal} {\bibinfo  {journal} {J.
  Chem. Phys.}\ }\textbf {\bibinfo {volume} {123}},\ \bibinfo {pages} {234106}
  (\bibinfo {year} {2005})}\BibitemShut {NoStop}%
\bibitem [{\citenamefont {Subotnik}\ and\ \citenamefont
  {Shenvi}(2011)}]{subotnik11}%
  \BibitemOpen
  \bibfield  {author} {\bibinfo {author} {\bibfnamefont {J.~E.}\ \bibnamefont
  {Subotnik}}\ and\ \bibinfo {author} {\bibfnamefont {N.}~\bibnamefont
  {Shenvi}},\ }\href@noop {} {\bibfield  {journal} {\bibinfo  {journal} {J.
  Chem. Phys.}\ }\textbf {\bibinfo {volume} {134}},\ \bibinfo {pages} {024105}
  (\bibinfo {year} {2011})}\BibitemShut {NoStop}%
\bibitem [{\citenamefont {Shenvi}, \citenamefont {Subotnik},\ and\
  \citenamefont {Yang}(2011)}]{subotniki11a}%
  \BibitemOpen
  \bibfield  {author} {\bibinfo {author} {\bibfnamefont {N.}~\bibnamefont
  {Shenvi}}, \bibinfo {author} {\bibfnamefont {J.~E.}\ \bibnamefont
  {Subotnik}}, \ and\ \bibinfo {author} {\bibfnamefont {W.}~\bibnamefont
  {Yang}},\ }\href@noop {} {\bibfield  {journal} {\bibinfo  {journal} {J. Chem.
  Phys.}\ }\textbf {\bibinfo {volume} {134}},\ \bibinfo {pages} {144102}
  (\bibinfo {year} {2011})}\BibitemShut {NoStop}%
\bibitem [{\citenamefont {Landry}, \citenamefont {J.Falk},\ and\ \citenamefont
  {Subotnik}(2013)}]{subotnik11b}%
  \BibitemOpen
  \bibfield  {author} {\bibinfo {author} {\bibfnamefont {B.~R.}\ \bibnamefont
  {Landry}}, \bibinfo {author} {\bibfnamefont {M.}~\bibnamefont {J.Falk}}, \
  and\ \bibinfo {author} {\bibfnamefont {J.~E.}\ \bibnamefont {Subotnik}},\
  }\href@noop {} {\bibfield  {journal} {\bibinfo  {journal} {J. Chem. Phys.}\
  }\textbf {\bibinfo {volume} {139}},\ \bibinfo {pages} {211101} (\bibinfo
  {year} {2013})}\BibitemShut {NoStop}%
\bibitem [{\citenamefont {Subotnik}, \citenamefont {Ouyang},\ and\
  \citenamefont {Landry}(2013)}]{subotnik11c}%
  \BibitemOpen
  \bibfield  {author} {\bibinfo {author} {\bibfnamefont {J.~E.}\ \bibnamefont
  {Subotnik}}, \bibinfo {author} {\bibfnamefont {W.}~\bibnamefont {Ouyang}}, \
  and\ \bibinfo {author} {\bibfnamefont {B.~R.}\ \bibnamefont {Landry}},\
  }\href@noop {} {\bibfield  {journal} {\bibinfo  {journal} {J. Chem. Phys.}\
  }\textbf {\bibinfo {volume} {139}},\ \bibinfo {pages} {214107} (\bibinfo
  {year} {2013})}\BibitemShut {NoStop}%
\bibitem [{\citenamefont {Subotnik}(2011)}]{subotnik11d}%
  \BibitemOpen
  \bibfield  {author} {\bibinfo {author} {\bibfnamefont {J.~E.}\ \bibnamefont
  {Subotnik}},\ }\href@noop {} {\bibfield  {journal} {\bibinfo  {journal} {J.
  Phys. Chem. A}\ }\textbf {\bibinfo {volume} {115}},\ \bibinfo {pages} {12083}
  (\bibinfo {year} {2011})}\BibitemShut {NoStop}%
\bibitem [{\citenamefont {Jaeger}, \citenamefont {Fischer},\ and\ \citenamefont
  {Prezhdo}(2012)}]{prezhdo12}%
  \BibitemOpen
  \bibfield  {author} {\bibinfo {author} {\bibfnamefont {H.~M.}\ \bibnamefont
  {Jaeger}}, \bibinfo {author} {\bibfnamefont {S.}~\bibnamefont {Fischer}}, \
  and\ \bibinfo {author} {\bibfnamefont {O.~V.}\ \bibnamefont {Prezhdo}},\
  }\href@noop {} {\bibfield  {journal} {\bibinfo  {journal} {J. Chem. Phys.}\
  }\textbf {\bibinfo {volume} {137}},\ \bibinfo {pages} {22A545} (\bibinfo
  {year} {2012})}\BibitemShut {NoStop}%
\bibitem [{\citenamefont {Nielsen}, \citenamefont {Kapral},\ and\ \citenamefont
  {Ciccotti}(2000)}]{nielsen00b}%
  \BibitemOpen
  \bibfield  {author} {\bibinfo {author} {\bibfnamefont {S.}~\bibnamefont
  {Nielsen}}, \bibinfo {author} {\bibfnamefont {R.}~\bibnamefont {Kapral}}, \
  and\ \bibinfo {author} {\bibfnamefont {G.}~\bibnamefont {Ciccotti}},\
  }\href@noop {} {\bibfield  {journal} {\bibinfo  {journal} {J. Chem. Phys.}\
  }\textbf {\bibinfo {volume} {112}},\ \bibinfo {pages} {6543} (\bibinfo {year}
  {2000})}\BibitemShut {NoStop}%
\bibitem [{\citenamefont {MacKernan}, \citenamefont {Kapral},\ and\
  \citenamefont {Ciccotti}(2002)}]{mackernan02}%
  \BibitemOpen
  \bibfield  {author} {\bibinfo {author} {\bibfnamefont {D.}~\bibnamefont
  {MacKernan}}, \bibinfo {author} {\bibfnamefont {R.}~\bibnamefont {Kapral}}, \
  and\ \bibinfo {author} {\bibfnamefont {G.}~\bibnamefont {Ciccotti}},\
  }\href@noop {} {\bibfield  {journal} {\bibinfo  {journal} {J. Phys.: Condes.
  Matter}\ }\textbf {\bibinfo {volume} {14}},\ \bibinfo {pages} {9069}
  (\bibinfo {year} {2002})}\BibitemShut {NoStop}%
\bibitem [{\citenamefont {Sergi}\ \emph {et~al.}(2003)\citenamefont {Sergi},
  \citenamefont {MacKernan}, \citenamefont {Ciccotti},\ and\ \citenamefont
  {Kapral}}]{sergi03b}%
  \BibitemOpen
  \bibfield  {author} {\bibinfo {author} {\bibfnamefont {A.}~\bibnamefont
  {Sergi}}, \bibinfo {author} {\bibfnamefont {D.}~\bibnamefont {MacKernan}},
  \bibinfo {author} {\bibfnamefont {G.}~\bibnamefont {Ciccotti}}, \ and\
  \bibinfo {author} {\bibfnamefont {R.}~\bibnamefont {Kapral}},\ }\href@noop {}
  {\bibfield  {journal} {\bibinfo  {journal} {Theor. Chem. Acc.}\ }\textbf
  {\bibinfo {volume} {110}},\ \bibinfo {pages} {49} (\bibinfo {year}
  {2003})}\BibitemShut {NoStop}%
\bibitem [{\citenamefont {MacKernan}, \citenamefont {Ciccotti},\ and\
  \citenamefont {Kapral.}(2008)}]{mackernan08}%
  \BibitemOpen
  \bibfield  {author} {\bibinfo {author} {\bibfnamefont {D.}~\bibnamefont
  {MacKernan}}, \bibinfo {author} {\bibfnamefont {G.}~\bibnamefont {Ciccotti}},
  \ and\ \bibinfo {author} {\bibfnamefont {R.}~\bibnamefont {Kapral.}},\
  }\href@noop {} {\bibfield  {journal} {\bibinfo  {journal} {J. Phys. Chem. B}\
  }\textbf {\bibinfo {volume} {112}},\ \bibinfo {pages} {424} (\bibinfo {year}
  {2008})}\BibitemShut {NoStop}%
\bibitem [{\citenamefont {Hanna}\ and\ \citenamefont {Kapral}(2005)}]{hanna05}%
  \BibitemOpen
  \bibfield  {author} {\bibinfo {author} {\bibfnamefont {G.}~\bibnamefont
  {Hanna}}\ and\ \bibinfo {author} {\bibfnamefont {R.}~\bibnamefont {Kapral}},\
  }\href@noop {} {\bibfield  {journal} {\bibinfo  {journal} {J. Chem. Phys.}\
  }\textbf {\bibinfo {volume} {122}},\ \bibinfo {pages} {244505} (\bibinfo
  {year} {2005})}\BibitemShut {NoStop}%
\bibitem [{\citenamefont {Uken}, \citenamefont {Sergi},\ and\ \citenamefont
  {Petruccione}(2013)}]{sergi13}%
  \BibitemOpen
  \bibfield  {author} {\bibinfo {author} {\bibfnamefont {D.~A.}\ \bibnamefont
  {Uken}}, \bibinfo {author} {\bibfnamefont {A.}~\bibnamefont {Sergi}}, \ and\
  \bibinfo {author} {\bibfnamefont {F.}~\bibnamefont {Petruccione}},\
  }\href@noop {} {\bibfield  {journal} {\bibinfo  {journal} {Phys. Rev. E}\
  }\textbf {\bibinfo {volume} {88}},\ \bibinfo {pages} {033301} (\bibinfo
  {year} {2013})}\BibitemShut {NoStop}%
\bibitem [{\citenamefont {Wan}\ and\ \citenamefont {Schofield}(2000)}]{wan00}%
  \BibitemOpen
  \bibfield  {author} {\bibinfo {author} {\bibfnamefont {C.}~\bibnamefont
  {Wan}}\ and\ \bibinfo {author} {\bibfnamefont {J.}~\bibnamefont
  {Schofield}},\ }\href@noop {} {\bibfield  {journal} {\bibinfo  {journal} {J.
  Chem. Phys.}\ }\textbf {\bibinfo {volume} {113}},\ \bibinfo {pages} {7047}
  (\bibinfo {year} {2000})}\BibitemShut {NoStop}%
\bibitem [{\citenamefont {Santer}, \citenamefont {Manthe},\ and\ \citenamefont
  {Stock}(2001)}]{santer01}%
  \BibitemOpen
  \bibfield  {author} {\bibinfo {author} {\bibfnamefont {M.}~\bibnamefont
  {Santer}}, \bibinfo {author} {\bibfnamefont {U.}~\bibnamefont {Manthe}}, \
  and\ \bibinfo {author} {\bibfnamefont {G.}~\bibnamefont {Stock}},\
  }\href@noop {} {\bibfield  {journal} {\bibinfo  {journal} {J. Chem. Phys.}\
  }\textbf {\bibinfo {volume} {114}},\ \bibinfo {pages} {2001} (\bibinfo {year}
  {2001})}\BibitemShut {NoStop}%
\bibitem [{\citenamefont {Wan}\ and\ \citenamefont {Schofield}(2002)}]{wan02}%
  \BibitemOpen
  \bibfield  {author} {\bibinfo {author} {\bibfnamefont {C.}~\bibnamefont
  {Wan}}\ and\ \bibinfo {author} {\bibfnamefont {J.}~\bibnamefont
  {Schofield}},\ }\href@noop {} {\bibfield  {journal} {\bibinfo  {journal} {J.
  Chem. Phys.}\ }\textbf {\bibinfo {volume} {116}},\ \bibinfo {pages} {494}
  (\bibinfo {year} {2002})}\BibitemShut {NoStop}%
\bibitem [{\citenamefont {Grunwald}, \citenamefont {Kelly},\ and\ \citenamefont
  {Kapral}(2009)}]{cecamchap}%
  \BibitemOpen
  \bibfield  {author} {\bibinfo {author} {\bibfnamefont {R.}~\bibnamefont
  {Grunwald}}, \bibinfo {author} {\bibfnamefont {A.}~\bibnamefont {Kelly}}, \
  and\ \bibinfo {author} {\bibfnamefont {R.}~\bibnamefont {Kapral}},\ }in\
  \href@noop {} {\emph {\bibinfo {booktitle} {Energy Transfer Dynamics in
  Biomaterial Systems}}},\ \bibinfo {editor} {edited by\ \bibinfo {editor}
  {\bibfnamefont {I.}~\bibnamefont {Burghardt}}}\ (\bibinfo  {publisher}
  {Springer},\ \bibinfo {address} {Berlin},\ \bibinfo {year} {2009})\ pp.\
  \bibinfo {pages} {383--413}\BibitemShut {NoStop}%
\bibitem [{\citenamefont {Ehrenfest}(1927)}]{ehrenfest27}%
  \BibitemOpen
  \bibfield  {author} {\bibinfo {author} {\bibfnamefont {P.}~\bibnamefont
  {Ehrenfest}},\ }\href@noop {} {\bibfield  {journal} {\bibinfo  {journal} {Z.
  Phys.}\ }\textbf {\bibinfo {volume} {45}},\ \bibinfo {pages} {455} (\bibinfo
  {year} {1927})}\BibitemShut {NoStop}%
\bibitem [{\citenamefont {Dirac}(1930)}]{dirac30}%
  \BibitemOpen
  \bibfield  {author} {\bibinfo {author} {\bibfnamefont {P.~A.~M.}\
  \bibnamefont {Dirac}},\ }\href@noop {} {\bibfield  {journal} {\bibinfo
  {journal} {Prc. Camb. Philos. Soc.}\ }\textbf {\bibinfo {volume} {26}},\
  \bibinfo {pages} {376} (\bibinfo {year} {1930})}\BibitemShut {NoStop}%
\bibitem [{\citenamefont {McLachlan}(1964)}]{mclachlan64}%
  \BibitemOpen
  \bibfield  {author} {\bibinfo {author} {\bibfnamefont {A.~D.}\ \bibnamefont
  {McLachlan}},\ }\href@noop {} {\bibfield  {journal} {\bibinfo  {journal}
  {Mol. Phys.}\ }\textbf {\bibinfo {volume} {8}},\ \bibinfo {pages} {39}
  (\bibinfo {year} {1964})}\BibitemShut {NoStop}%
\bibitem [{\citenamefont {Prezhdo}\ and\ \citenamefont
  {Rossky}(1997)}]{prezhdo97}%
  \BibitemOpen
  \bibfield  {author} {\bibinfo {author} {\bibfnamefont {O.~V.}\ \bibnamefont
  {Prezhdo}}\ and\ \bibinfo {author} {\bibfnamefont {P.~J.}\ \bibnamefont
  {Rossky}},\ }\href@noop {} {\bibfield  {journal} {\bibinfo  {journal} {J.
  Chem. Phys.}\ }\textbf {\bibinfo {volume} {107}},\ \bibinfo {pages} {825}
  (\bibinfo {year} {1997})}\BibitemShut {NoStop}%
\bibitem [{\citenamefont {Subotnik}(2010)}]{subotnik10}%
  \BibitemOpen
  \bibfield  {author} {\bibinfo {author} {\bibfnamefont {J.~E.}\ \bibnamefont
  {Subotnik}},\ }\href@noop {} {\bibfield  {journal} {\bibinfo  {journal} {J.
  Chem. Phys.}\ }\textbf {\bibinfo {volume} {132}},\ \bibinfo {pages} {134112}
  (\bibinfo {year} {2010})}\BibitemShut {NoStop}%
\bibitem [{\citenamefont {Zhu}\ \emph {et~al.}(2004)\citenamefont {Zhu},
  \citenamefont {Nangia}, \citenamefont {Jasper},\ and\ \citenamefont
  {Truhlar}}]{truhlar04}%
  \BibitemOpen
  \bibfield  {author} {\bibinfo {author} {\bibfnamefont {C.}~\bibnamefont
  {Zhu}}, \bibinfo {author} {\bibfnamefont {S.}~\bibnamefont {Nangia}},
  \bibinfo {author} {\bibfnamefont {A.~W.}\ \bibnamefont {Jasper}}, \ and\
  \bibinfo {author} {\bibfnamefont {D.~G.}\ \bibnamefont {Truhlar}},\
  }\href@noop {} {\bibfield  {journal} {\bibinfo  {journal} {J. Chem. Phys.}\
  }\textbf {\bibinfo {volume} {121}},\ \bibinfo {pages} {7658} (\bibinfo {year}
  {2004})}\BibitemShut {NoStop}%
\bibitem [{\citenamefont {Jasper}\ \emph {et~al.}(2004)\citenamefont {Jasper},
  \citenamefont {Zhu}, \citenamefont {Nangia},\ and\ \citenamefont
  {Truhlar}}]{truhlar04b}%
  \BibitemOpen
  \bibfield  {author} {\bibinfo {author} {\bibfnamefont {A.~W.}\ \bibnamefont
  {Jasper}}, \bibinfo {author} {\bibfnamefont {C.}~\bibnamefont {Zhu}},
  \bibinfo {author} {\bibfnamefont {S.}~\bibnamefont {Nangia}}, \ and\ \bibinfo
  {author} {\bibfnamefont {D.~G.}\ \bibnamefont {Truhlar}},\ }\href@noop {}
  {\bibfield  {journal} {\bibinfo  {journal} {Faraday Discuss.}\ }\textbf
  {\bibinfo {volume} {127}},\ \bibinfo {pages} {1} (\bibinfo {year}
  {2004})}\BibitemShut {NoStop}%
\bibitem [{\citenamefont {Akimov}, \citenamefont {Long},\ and\ \citenamefont
  {Prezhdo}(2014)}]{prezhdo14}%
  \BibitemOpen
  \bibfield  {author} {\bibinfo {author} {\bibfnamefont {A.~V.}\ \bibnamefont
  {Akimov}}, \bibinfo {author} {\bibfnamefont {R.}~\bibnamefont {Long}}, \ and\
  \bibinfo {author} {\bibfnamefont {O.~V.}\ \bibnamefont {Prezhdo}},\
  }\href@noop {} {\bibfield  {journal} {\bibinfo  {journal} {J. Chem. Phys.}\
  }\textbf {\bibinfo {volume} {140}},\ \bibinfo {pages} {194107} (\bibinfo
  {year} {2014})}\BibitemShut {NoStop}%
\bibitem [{\citenamefont {Miller}\ and\ \citenamefont
  {McCurdy}(1978)}]{miller78}%
  \BibitemOpen
  \bibfield  {author} {\bibinfo {author} {\bibfnamefont {W.~H.}\ \bibnamefont
  {Miller}}\ and\ \bibinfo {author} {\bibfnamefont {C.~W.}\ \bibnamefont
  {McCurdy}},\ }\href@noop {} {\bibfield  {journal} {\bibinfo  {journal} {J.
  Chem. Phys.}\ }\textbf {\bibinfo {volume} {69}},\ \bibinfo {pages} {5163}
  (\bibinfo {year} {1978})}\BibitemShut {NoStop}%
\bibitem [{\citenamefont {Stock}\ and\ \citenamefont {Thoss}(1997)}]{stock97}%
  \BibitemOpen
  \bibfield  {author} {\bibinfo {author} {\bibfnamefont {G.}~\bibnamefont
  {Stock}}\ and\ \bibinfo {author} {\bibfnamefont {M.}~\bibnamefont {Thoss}},\
  }\href@noop {} {\bibfield  {journal} {\bibinfo  {journal} {Phys. Rev. Lett.}\
  }\textbf {\bibinfo {volume} {78}},\ \bibinfo {pages} {578} (\bibinfo {year}
  {1997})}\BibitemShut {NoStop}%
\bibitem [{\citenamefont {Thoss}\ and\ \citenamefont {Stock}(1999)}]{thoss99}%
  \BibitemOpen
  \bibfield  {author} {\bibinfo {author} {\bibfnamefont {M.}~\bibnamefont
  {Thoss}}\ and\ \bibinfo {author} {\bibfnamefont {G.}~\bibnamefont {Stock}},\
  }\href@noop {} {\bibfield  {journal} {\bibinfo  {journal} {Phys. Rev. A}\
  }\textbf {\bibinfo {volume} {59}},\ \bibinfo {pages} {64} (\bibinfo {year}
  {1999})}\BibitemShut {NoStop}%
\bibitem [{\citenamefont {Miller}(2001)}]{miller01}%
  \BibitemOpen
  \bibfield  {author} {\bibinfo {author} {\bibfnamefont {W.~H.}\ \bibnamefont
  {Miller}},\ }\href@noop {} {\bibfield  {journal} {\bibinfo  {journal} {J.
  Phys. Chem. A}\ }\textbf {\bibinfo {volume} {105}},\ \bibinfo {pages} {2942}
  (\bibinfo {year} {2001})}\BibitemShut {NoStop}%
\bibitem [{\citenamefont {Bonella}\ and\ \citenamefont
  {Coker}(2003)}]{bonella03}%
  \BibitemOpen
  \bibfield  {author} {\bibinfo {author} {\bibfnamefont {S.}~\bibnamefont
  {Bonella}}\ and\ \bibinfo {author} {\bibfnamefont {D.~F.}\ \bibnamefont
  {Coker}},\ }\href@noop {} {\bibfield  {journal} {\bibinfo  {journal} {J.
  Chem. Phys.}\ }\textbf {\bibinfo {volume} {118}},\ \bibinfo {pages} {4370}
  (\bibinfo {year} {2003})}\BibitemShut {NoStop}%
\bibitem [{\citenamefont {Bonella}\ and\ \citenamefont
  {Coker}(2005)}]{bonella05}%
  \BibitemOpen
  \bibfield  {author} {\bibinfo {author} {\bibfnamefont {S.}~\bibnamefont
  {Bonella}}\ and\ \bibinfo {author} {\bibfnamefont {D.~F.}\ \bibnamefont
  {Coker}},\ }\href@noop {} {\bibfield  {journal} {\bibinfo  {journal} {J.
  Chem. Phys.}\ }\textbf {\bibinfo {volume} {122}},\ \bibinfo {pages} {194102}
  (\bibinfo {year} {2005})}\BibitemShut {NoStop}%
\bibitem [{\citenamefont {Dunkel}, \citenamefont {Bonella},\ and\ \citenamefont
  {Coker}(2008)}]{dunkel08}%
  \BibitemOpen
  \bibfield  {author} {\bibinfo {author} {\bibfnamefont {E.}~\bibnamefont
  {Dunkel}}, \bibinfo {author} {\bibfnamefont {S.}~\bibnamefont {Bonella}}, \
  and\ \bibinfo {author} {\bibfnamefont {D.~F.}\ \bibnamefont {Coker}},\
  }\href@noop {} {\bibfield  {journal} {\bibinfo  {journal} {J. Chem. Phys.}\
  }\textbf {\bibinfo {volume} {129}},\ \bibinfo {pages} {114106} (\bibinfo
  {year} {2008})}\BibitemShut {NoStop}%
\bibitem [{\citenamefont {Schwinger}(1965)}]{chap-schwinger65}%
  \BibitemOpen
  \bibfield  {author} {\bibinfo {author} {\bibfnamefont {J.}~\bibnamefont
  {Schwinger}},\ }in\ \href@noop {} {\emph {\bibinfo {booktitle} {Quantum
  Theory of Angular Momentum}}},\ \bibinfo {editor} {edited by\ \bibinfo
  {editor} {\bibfnamefont {L.~C.}\ \bibnamefont {Biedenharn}}\ and\ \bibinfo
  {editor} {\bibfnamefont {H.~V.}\ \bibnamefont {Dam}}}\ (\bibinfo  {publisher}
  {Academic Press},\ \bibinfo {address} {New York},\ \bibinfo {year} {1965})\
  p.\ \bibinfo {pages} {229}\BibitemShut {NoStop}%
\bibitem [{\citenamefont {Meyer}\ and\ \citenamefont {Miller}(1979)}]{meyer79}%
  \BibitemOpen
  \bibfield  {author} {\bibinfo {author} {\bibfnamefont {H.~D.}\ \bibnamefont
  {Meyer}}\ and\ \bibinfo {author} {\bibfnamefont {W.~H.}\ \bibnamefont
  {Miller}},\ }\href@noop {} {\bibfield  {journal} {\bibinfo  {journal} {J.
  Chem. Phys.}\ }\textbf {\bibinfo {volume} {70}},\ \bibinfo {pages} {3214}
  (\bibinfo {year} {1979})}\BibitemShut {NoStop}%
\bibitem [{\citenamefont {Stock}\ and\ \citenamefont {Thoss.}(2005)}]{stock05}%
  \BibitemOpen
  \bibfield  {author} {\bibinfo {author} {\bibfnamefont {G.}~\bibnamefont
  {Stock}}\ and\ \bibinfo {author} {\bibfnamefont {M.}~\bibnamefont {Thoss.}},\
  }\href@noop {} {\bibfield  {journal} {\bibinfo  {journal} {Adv. Chem. Phys.}\
  }\textbf {\bibinfo {volume} {131}},\ \bibinfo {pages} {243} (\bibinfo {year}
  {2005})}\BibitemShut {NoStop}%
\bibitem [{\citenamefont {Kim}, \citenamefont {Nassimi},\ and\ \citenamefont
  {Kapral.}(2008)}]{kim-map08}%
  \BibitemOpen
  \bibfield  {author} {\bibinfo {author} {\bibfnamefont {H.}~\bibnamefont
  {Kim}}, \bibinfo {author} {\bibfnamefont {A.}~\bibnamefont {Nassimi}}, \ and\
  \bibinfo {author} {\bibfnamefont {R.}~\bibnamefont {Kapral.}},\ }\href@noop
  {} {\bibfield  {journal} {\bibinfo  {journal} {J. Chem. Phys.}\ }\textbf
  {\bibinfo {volume} {129}},\ \bibinfo {pages} {084102} (\bibinfo {year}
  {2008})}\BibitemShut {NoStop}%
\bibitem [{\citenamefont {Nassimi}, \citenamefont {Bonella},\ and\
  \citenamefont {Kapral.}(2010)}]{nassimi10}%
  \BibitemOpen
  \bibfield  {author} {\bibinfo {author} {\bibfnamefont {A.}~\bibnamefont
  {Nassimi}}, \bibinfo {author} {\bibfnamefont {S.}~\bibnamefont {Bonella}}, \
  and\ \bibinfo {author} {\bibfnamefont {R.}~\bibnamefont {Kapral.}},\
  }\href@noop {} {\bibfield  {journal} {\bibinfo  {journal} {J. Chem. Phys.}\
  }\textbf {\bibinfo {volume} {133}},\ \bibinfo {pages} {134115} (\bibinfo
  {year} {2010})}\BibitemShut {NoStop}%
\bibitem [{\citenamefont {Kelly}\ \emph {et~al.}(2012)\citenamefont {Kelly},
  \citenamefont {van Zon}, \citenamefont {Schofield},\ and\ \citenamefont
  {Kapral}}]{kelly12}%
  \BibitemOpen
  \bibfield  {author} {\bibinfo {author} {\bibfnamefont {A.}~\bibnamefont
  {Kelly}}, \bibinfo {author} {\bibfnamefont {R.}~\bibnamefont {van Zon}},
  \bibinfo {author} {\bibfnamefont {J.~M.}\ \bibnamefont {Schofield}}, \ and\
  \bibinfo {author} {\bibfnamefont {R.}~\bibnamefont {Kapral}},\ }\href@noop {}
  {\bibfield  {journal} {\bibinfo  {journal} {J. Chem. Phys.}\ }\textbf
  {\bibinfo {volume} {136}},\ \bibinfo {pages} {084101} (\bibinfo {year}
  {2012})}\BibitemShut {NoStop}%
\bibitem [{\citenamefont {Donoso}\ and\ \citenamefont
  {Martens}(2001)}]{donoso01}%
  \BibitemOpen
  \bibfield  {author} {\bibinfo {author} {\bibfnamefont {A.}~\bibnamefont
  {Donoso}}\ and\ \bibinfo {author} {\bibfnamefont {C.~C.}\ \bibnamefont
  {Martens}},\ }\href@noop {} {\bibfield  {journal} {\bibinfo  {journal} {Phys.
  Rev. Lett.}\ }\textbf {\bibinfo {volume} {87}},\ \bibinfo {pages} {223202}
  (\bibinfo {year} {2001})}\BibitemShut {NoStop}%
\bibitem [{\citenamefont {Donoso}, \citenamefont {Zheng},\ and\ \citenamefont
  {Martens}(2003)}]{donoso03}%
  \BibitemOpen
  \bibfield  {author} {\bibinfo {author} {\bibfnamefont {A.}~\bibnamefont
  {Donoso}}, \bibinfo {author} {\bibfnamefont {Y.}~\bibnamefont {Zheng}}, \
  and\ \bibinfo {author} {\bibfnamefont {C.~C.}\ \bibnamefont {Martens}},\
  }\href@noop {} {\bibfield  {journal} {\bibinfo  {journal} {J. Chem. Phys.}\
  }\textbf {\bibinfo {volume} {119}},\ \bibinfo {pages} {5010} (\bibinfo {year}
  {2003})}\BibitemShut {NoStop}%
\bibitem [{\citenamefont {Ananth}, \citenamefont {Venkataraman},\ and\
  \citenamefont {Miller}(2007)}]{miller07}%
  \BibitemOpen
  \bibfield  {author} {\bibinfo {author} {\bibfnamefont {N.}~\bibnamefont
  {Ananth}}, \bibinfo {author} {\bibfnamefont {C.}~\bibnamefont
  {Venkataraman}}, \ and\ \bibinfo {author} {\bibfnamefont {W.~H.}\
  \bibnamefont {Miller}},\ }\href@noop {} {\bibfield  {journal} {\bibinfo
  {journal} {J. Chem. Phys.}\ }\textbf {\bibinfo {volume} {127}},\ \bibinfo
  {pages} {084114} (\bibinfo {year} {2007})}\BibitemShut {NoStop}%
\bibitem [{\citenamefont {Rekik}\ \emph {et~al.}(2013)\citenamefont {Rekik},
  \citenamefont {Hsieh}, \citenamefont {Freedman},\ and\ \citenamefont
  {Hanna}}]{rekik13}%
  \BibitemOpen
  \bibfield  {author} {\bibinfo {author} {\bibfnamefont {N.}~\bibnamefont
  {Rekik}}, \bibinfo {author} {\bibfnamefont {C.-Y.}\ \bibnamefont {Hsieh}},
  \bibinfo {author} {\bibfnamefont {H.}~\bibnamefont {Freedman}}, \ and\
  \bibinfo {author} {\bibfnamefont {G.}~\bibnamefont {Hanna}},\ }\href@noop {}
  {\bibfield  {journal} {\bibinfo  {journal} {J. Chem. Phys.}\ }\textbf
  {\bibinfo {volume} {138}},\ \bibinfo {pages} {144106} (\bibinfo {year}
  {2013})}\BibitemShut {NoStop}%
\bibitem [{\citenamefont {Bonella}\ and\ \citenamefont
  {Coker}(2001)}]{bonella01}%
  \BibitemOpen
  \bibfield  {author} {\bibinfo {author} {\bibfnamefont {S.}~\bibnamefont
  {Bonella}}\ and\ \bibinfo {author} {\bibfnamefont {D.~F.}\ \bibnamefont
  {Coker}},\ }\href@noop {} {\bibfield  {journal} {\bibinfo  {journal} {Chem.
  Phys.}\ }\textbf {\bibinfo {volume} {268}},\ \bibinfo {pages} {323} (\bibinfo
  {year} {2001})}\BibitemShut {NoStop}%
\bibitem [{\citenamefont {Kelly}\ and\ \citenamefont
  {Markland}(2013)}]{kelly-markland13}%
  \BibitemOpen
  \bibfield  {author} {\bibinfo {author} {\bibfnamefont {A.}~\bibnamefont
  {Kelly}}\ and\ \bibinfo {author} {\bibfnamefont {T.~E.}\ \bibnamefont
  {Markland}},\ }\href@noop {} {\bibfield  {journal} {\bibinfo  {journal} {J.
  Chem. Phys.}\ }\textbf {\bibinfo {volume} {139}},\ \bibinfo {pages} {014104}
  (\bibinfo {year} {2013})}\BibitemShut {NoStop}%
\bibitem [{\citenamefont {Kim}\ and\ \citenamefont {Rhee}(2014)}]{rhee14}%
  \BibitemOpen
  \bibfield  {author} {\bibinfo {author} {\bibfnamefont {H.~W.}\ \bibnamefont
  {Kim}}\ and\ \bibinfo {author} {\bibfnamefont {Y.~M.}\ \bibnamefont {Rhee}},\
  }\href@noop {} {\bibfield  {journal} {\bibinfo  {journal} {J. Chem.Phys.}\
  }\textbf {\bibinfo {volume} {140}},\ \bibinfo {pages} {184106} (\bibinfo
  {year} {2014})}\BibitemShut {NoStop}%
\bibitem [{\citenamefont {Sun}\ and\ \citenamefont {Miller}(1999)}]{sun99}%
  \BibitemOpen
  \bibfield  {author} {\bibinfo {author} {\bibfnamefont {X.}~\bibnamefont
  {Sun}}\ and\ \bibinfo {author} {\bibfnamefont {W.~H.}\ \bibnamefont
  {Miller}},\ }\href@noop {} {\bibfield  {journal} {\bibinfo  {journal} {J.
  Chem. Phys.}\ }\textbf {\bibinfo {volume} {110}},\ \bibinfo {pages} {6635}
  (\bibinfo {year} {1999})}\BibitemShut {NoStop}%
\bibitem [{\citenamefont {Thompson}\ and\ \citenamefont
  {Makri}(1999{\natexlab{a}})}]{thompson99}%
  \BibitemOpen
  \bibfield  {author} {\bibinfo {author} {\bibfnamefont {K.}~\bibnamefont
  {Thompson}}\ and\ \bibinfo {author} {\bibfnamefont {N.}~\bibnamefont
  {Makri}},\ }\href@noop {} {\bibfield  {journal} {\bibinfo  {journal} {J.
  Chem. Phys.}\ }\textbf {\bibinfo {volume} {110}},\ \bibinfo {pages} {1343}
  (\bibinfo {year} {1999}{\natexlab{a}})}\BibitemShut {NoStop}%
\bibitem [{\citenamefont {Wang}, \citenamefont {Thoss},\ and\ \citenamefont
  {Miller}(2000)}]{wang00}%
  \BibitemOpen
  \bibfield  {author} {\bibinfo {author} {\bibfnamefont {H.~B.}\ \bibnamefont
  {Wang}}, \bibinfo {author} {\bibfnamefont {M.}~\bibnamefont {Thoss}}, \ and\
  \bibinfo {author} {\bibfnamefont {W.~H.}\ \bibnamefont {Miller}},\
  }\href@noop {} {\bibfield  {journal} {\bibinfo  {journal} {J. Chem. Phys.}\
  }\textbf {\bibinfo {volume} {112}},\ \bibinfo {pages} {47} (\bibinfo {year}
  {2000})}\BibitemShut {NoStop}%
\bibitem [{\citenamefont {Thoss}, \citenamefont {Wang},\ and\ \citenamefont
  {Miller}(2001)}]{thoss01}%
  \BibitemOpen
  \bibfield  {author} {\bibinfo {author} {\bibfnamefont {M.}~\bibnamefont
  {Thoss}}, \bibinfo {author} {\bibfnamefont {H.~B.}\ \bibnamefont {Wang}}, \
  and\ \bibinfo {author} {\bibfnamefont {W.~H.}\ \bibnamefont {Miller}},\
  }\href@noop {} {\bibfield  {journal} {\bibinfo  {journal} {J. Chem. Phys.}\
  }\textbf {\bibinfo {volume} {114}},\ \bibinfo {pages} {9220} (\bibinfo {year}
  {2001})}\BibitemShut {NoStop}%
\bibitem [{\citenamefont {Bukhman}\ and\ \citenamefont
  {Makri}(2009)}]{bukhman09}%
  \BibitemOpen
  \bibfield  {author} {\bibinfo {author} {\bibfnamefont {E.}~\bibnamefont
  {Bukhman}}\ and\ \bibinfo {author} {\bibfnamefont {N.}~\bibnamefont
  {Makri}},\ }\href@noop {} {\bibfield  {journal} {\bibinfo  {journal} {J.
  Phys. Chem. A}\ }\textbf {\bibinfo {volume} {113}},\ \bibinfo {pages} {7183}
  (\bibinfo {year} {2009})}\BibitemShut {NoStop}%
\bibitem [{\citenamefont {Huo}\ and\ \citenamefont {Coker}(2011)}]{huo11}%
  \BibitemOpen
  \bibfield  {author} {\bibinfo {author} {\bibfnamefont {P.}~\bibnamefont
  {Huo}}\ and\ \bibinfo {author} {\bibfnamefont {D.~F.}\ \bibnamefont
  {Coker}},\ }\href@noop {} {\bibfield  {journal} {\bibinfo  {journal} {J.
  Chem. Phys.}\ }\textbf {\bibinfo {volume} {135}},\ \bibinfo {pages} {201101}
  (\bibinfo {year} {2011})}\BibitemShut {NoStop}%
\bibitem [{\citenamefont {Huo}\ and\ \citenamefont {Coker}(2012)}]{huo12_jcp}%
  \BibitemOpen
  \bibfield  {author} {\bibinfo {author} {\bibfnamefont {P.}~\bibnamefont
  {Huo}}\ and\ \bibinfo {author} {\bibfnamefont {D.~F.}\ \bibnamefont
  {Coker}},\ }\href@noop {} {\bibfield  {journal} {\bibinfo  {journal} {J.
  Chem. Phys.}\ }\textbf {\bibinfo {volume} {137}},\ \bibinfo {pages} {22A535}
  (\bibinfo {year} {2012})}\BibitemShut {NoStop}%
\bibitem [{\citenamefont {Hsieh}\ and\ \citenamefont {Kapral}(2012)}]{hsieh12}%
  \BibitemOpen
  \bibfield  {author} {\bibinfo {author} {\bibfnamefont {C.-Y.}\ \bibnamefont
  {Hsieh}}\ and\ \bibinfo {author} {\bibfnamefont {R.}~\bibnamefont {Kapral}},\
  }\href@noop {} {\bibfield  {journal} {\bibinfo  {journal} {J. Chem. Phys.}\
  }\textbf {\bibinfo {volume} {137}},\ \bibinfo {pages} {22A507} (\bibinfo
  {year} {2012})}\BibitemShut {NoStop}%
\bibitem [{\citenamefont {Hsieh}\ and\ \citenamefont {Kapral}(2013)}]{hsieh13}%
  \BibitemOpen
  \bibfield  {author} {\bibinfo {author} {\bibfnamefont {C.-Y.}\ \bibnamefont
  {Hsieh}}\ and\ \bibinfo {author} {\bibfnamefont {R.}~\bibnamefont {Kapral}},\
  }\href@noop {} {\bibfield  {journal} {\bibinfo  {journal} {J. Chem. Phys.}\
  }\textbf {\bibinfo {volume} {138}},\ \bibinfo {pages} {134110} (\bibinfo
  {year} {2013})}\BibitemShut {NoStop}%
\bibitem [{\citenamefont {Hsieh}, \citenamefont {Sscofield},\ and\
  \citenamefont {Kapral}(2013)}]{hsieh13b}%
  \BibitemOpen
  \bibfield  {author} {\bibinfo {author} {\bibfnamefont {C.-Y.}\ \bibnamefont
  {Hsieh}}, \bibinfo {author} {\bibfnamefont {J.~M.}\ \bibnamefont
  {Sscofield}}, \ and\ \bibinfo {author} {\bibfnamefont {R.}~\bibnamefont
  {Kapral}},\ }\href@noop {} {\bibfield  {journal} {\bibinfo  {journal} {Mol.
  Phys.}\ }\textbf {\bibinfo {volume} {111}},\ \bibinfo {pages} {3546}
  (\bibinfo {year} {2013})}\BibitemShut {NoStop}%
\bibitem [{\citenamefont {MacKernan}, \citenamefont {Ciccotti},\ and\
  \citenamefont {Kapral}(2002)}]{mackernan02a}%
  \BibitemOpen
  \bibfield  {author} {\bibinfo {author} {\bibfnamefont {D.}~\bibnamefont
  {MacKernan}}, \bibinfo {author} {\bibfnamefont {G.}~\bibnamefont {Ciccotti}},
  \ and\ \bibinfo {author} {\bibfnamefont {R.}~\bibnamefont {Kapral}},\
  }\href@noop {} {\bibfield  {journal} {\bibinfo  {journal} {J. Chem. Phys.}\
  }\textbf {\bibinfo {volume} {116}},\ \bibinfo {pages} {2346} (\bibinfo {year}
  {2002})}\BibitemShut {NoStop}%
\bibitem [{\citenamefont {Bonella}\ \emph {et~al.}(2009)\citenamefont
  {Bonella}, \citenamefont {Coker}, \citenamefont {Kernan}, \citenamefont
  {Kapral},\ and\ \citenamefont {Ciccotti}}]{bonella09}%
  \BibitemOpen
  \bibfield  {author} {\bibinfo {author} {\bibfnamefont {S.}~\bibnamefont
  {Bonella}}, \bibinfo {author} {\bibfnamefont {D.~F.}\ \bibnamefont {Coker}},
  \bibinfo {author} {\bibfnamefont {D.~M.}\ \bibnamefont {Kernan}}, \bibinfo
  {author} {\bibfnamefont {R.}~\bibnamefont {Kapral}}, \ and\ \bibinfo {author}
  {\bibfnamefont {G.}~\bibnamefont {Ciccotti}},\ }in\ \href@noop {} {\emph
  {\bibinfo {booktitle} {Energy Transfer Dynamics in Biomaterial Systems}}},\
  \bibinfo {series} {Springer Series in Chemical Physics}, Vol.~\bibinfo
  {volume} {93},\ \bibinfo {editor} {edited by\ \bibinfo {editor}
  {\bibfnamefont {I.}~\bibnamefont {Burghardt}}, \bibinfo {editor}
  {\bibfnamefont {V.}~\bibnamefont {May}}, \bibinfo {editor} {\bibfnamefont
  {D.~A.}\ \bibnamefont {Micha}}, \ and\ \bibinfo {editor} {\bibfnamefont
  {E.~R.}\ \bibnamefont {Bittner}}}\ (\bibinfo  {publisher} {Springer Berlin
  Heidelberg},\ \bibinfo {address} {Berlin, Heidelberg},\ \bibinfo {year}
  {2009})\BibitemShut {NoStop}%
\bibitem [{\citenamefont {Makarov}\ and\ \citenamefont
  {Makri}(1994)}]{makarov94}%
  \BibitemOpen
  \bibfield  {author} {\bibinfo {author} {\bibfnamefont {D.~E.}\ \bibnamefont
  {Makarov}}\ and\ \bibinfo {author} {\bibfnamefont {N.}~\bibnamefont
  {Makri}},\ }\href@noop {} {\bibfield  {journal} {\bibinfo  {journal} {Chem.
  Phys. Lett}\ }\textbf {\bibinfo {volume} {221}},\ \bibinfo {pages} {482}
  (\bibinfo {year} {1994})}\BibitemShut {NoStop}%
\bibitem [{\citenamefont {DiVincenzo}\ and\ \citenamefont
  {Loss}(2005)}]{loss05}%
  \BibitemOpen
  \bibfield  {author} {\bibinfo {author} {\bibfnamefont {D.~P.}\ \bibnamefont
  {DiVincenzo}}\ and\ \bibinfo {author} {\bibfnamefont {D.}~\bibnamefont
  {Loss}},\ }\href {\doibase 10.1103/PhysRevB.71.035318} {\bibfield  {journal}
  {\bibinfo  {journal} {Phys. Rev. B}\ }\textbf {\bibinfo {volume} {71}},\
  \bibinfo {pages} {035318} (\bibinfo {year} {2005})}\BibitemShut {NoStop}%
\bibitem [{\citenamefont {Thompson}\ and\ \citenamefont
  {Makri}(1999{\natexlab{b}})}]{thompson99b}%
  \BibitemOpen
  \bibfield  {author} {\bibinfo {author} {\bibfnamefont {K.}~\bibnamefont
  {Thompson}}\ and\ \bibinfo {author} {\bibfnamefont {N.}~\bibnamefont
  {Makri}},\ }\href@noop {} {\bibfield  {journal} {\bibinfo  {journal} {Phys.
  Rev. E}\ }\textbf {\bibinfo {volume} {59}},\ \bibinfo {pages} {R4729}
  (\bibinfo {year} {1999}{\natexlab{b}})}\BibitemShut {NoStop}%
\bibitem [{\citenamefont {Cheng}\ and\ \citenamefont
  {Fleming}(2009)}]{fleming09}%
  \BibitemOpen
  \bibfield  {author} {\bibinfo {author} {\bibfnamefont {Y.-C.}\ \bibnamefont
  {Cheng}}\ and\ \bibinfo {author} {\bibfnamefont {G.~R.}\ \bibnamefont
  {Fleming}},\ }\href@noop {} {\bibfield  {journal} {\bibinfo  {journal} {Annu.
  Rev. Phys. Chem.}\ }\textbf {\bibinfo {volume} {60}},\ \bibinfo {pages} {241}
  (\bibinfo {year} {2009})}\BibitemShut {NoStop}%
\bibitem [{\citenamefont {Scholes}(2010)}]{scholes10}%
  \BibitemOpen
  \bibfield  {author} {\bibinfo {author} {\bibfnamefont {G.~D.}\ \bibnamefont
  {Scholes}},\ }\href@noop {} {\bibfield  {journal} {\bibinfo  {journal} {J.
  Phys. Chem. Lett.}\ }\textbf {\bibinfo {volume} {1}},\ \bibinfo {pages} {2}
  (\bibinfo {year} {2010})}\BibitemShut {NoStop}%
\bibitem [{\citenamefont {Fleming}\ and\ \citenamefont
  {Ishizaki}(2009)}]{fleming09a}%
  \BibitemOpen
  \bibfield  {author} {\bibinfo {author} {\bibfnamefont {G.~R.}\ \bibnamefont
  {Fleming}}\ and\ \bibinfo {author} {\bibfnamefont {A.}~\bibnamefont
  {Ishizaki}},\ }\href@noop {} {\bibfield  {journal} {\bibinfo  {journal}
  {PNAS}\ }\textbf {\bibinfo {volume} {106}},\ \bibinfo {pages} {17255}
  (\bibinfo {year} {2009})}\BibitemShut {NoStop}%
\bibitem [{\citenamefont {Zhu}\ \emph {et~al.}(2011)\citenamefont {Zhu},
  \citenamefont {Kais}, \citenamefont {Rebentrost},\ and\ \citenamefont
  {Aspuru-Guzik}}]{zhu11}%
  \BibitemOpen
  \bibfield  {author} {\bibinfo {author} {\bibfnamefont {J.}~\bibnamefont
  {Zhu}}, \bibinfo {author} {\bibfnamefont {S.}~\bibnamefont {Kais}}, \bibinfo
  {author} {\bibfnamefont {P.}~\bibnamefont {Rebentrost}}, \ and\ \bibinfo
  {author} {\bibfnamefont {A.}~\bibnamefont {Aspuru-Guzik}},\ }\href@noop {}
  {\bibfield  {journal} {\bibinfo  {journal} {J. Phys. Chem. B}\ }\textbf
  {\bibinfo {volume} {115}},\ \bibinfo {pages} {1531} (\bibinfo {year}
  {2011})}\BibitemShut {NoStop}%
\bibitem [{\citenamefont {Kelly}\ and\ \citenamefont {Rhee}(2011)}]{kelly-FMO}%
  \BibitemOpen
  \bibfield  {author} {\bibinfo {author} {\bibfnamefont {A.}~\bibnamefont
  {Kelly}}\ and\ \bibinfo {author} {\bibfnamefont {Y.~M.}\ \bibnamefont
  {Rhee}},\ }\href@noop {} {\bibfield  {journal} {\bibinfo  {journal} {J. Phys.
  Chem.Lett.}\ }\textbf {\bibinfo {volume} {2}},\ \bibinfo {pages} {808}
  (\bibinfo {year} {2011})}\BibitemShut {NoStop}%
\bibitem [{\citenamefont {Kim}\ \emph {et~al.}(2012)\citenamefont {Kim},
  \citenamefont {Kelly}, \citenamefont {Park},\ and\ \citenamefont
  {Rhee}}]{kelly-FMO2}%
  \BibitemOpen
  \bibfield  {author} {\bibinfo {author} {\bibfnamefont {H.~W.}\ \bibnamefont
  {Kim}}, \bibinfo {author} {\bibfnamefont {A.}~\bibnamefont {Kelly}}, \bibinfo
  {author} {\bibfnamefont {J.~W.}\ \bibnamefont {Park}}, \ and\ \bibinfo
  {author} {\bibfnamefont {Y.~M.}\ \bibnamefont {Rhee}},\ }\href@noop {}
  {\bibfield  {journal} {\bibinfo  {journal} {J. Am. Chem. Soc.}\ }\textbf
  {\bibinfo {volume} {134}},\ \bibinfo {pages} {11640} (\bibinfo {year}
  {2012})}\BibitemShut {NoStop}%
\bibitem [{\citenamefont {Migani}\ and\ \citenamefont
  {Olivucci}(2004)}]{migani04}%
  \BibitemOpen
  \bibfield  {author} {\bibinfo {author} {\bibfnamefont {A.}~\bibnamefont
  {Migani}}\ and\ \bibinfo {author} {\bibfnamefont {M.}~\bibnamefont
  {Olivucci}},\ }in\ \href@noop {} {\emph {\bibinfo {booktitle} {Conical
  Intersection Electronic Structure, Dynamics and Spectroscopy}}},\ \bibinfo
  {editor} {edited by\ \bibinfo {editor} {\bibfnamefont {W.}~\bibnamefont
  {Domcke}}, \bibinfo {editor} {\bibfnamefont {D.~R.}\ \bibnamefont {Yarkony}},
  \ and\ \bibinfo {editor} {\bibfnamefont {H.}~\bibnamefont {Köppel}}}\
  (\bibinfo  {publisher} {World Scientific},\ \bibinfo {address} {Singapore},\
  \bibinfo {year} {2004})\ p.\ \bibinfo {pages} {271}\BibitemShut {NoStop}%
\bibitem [{\citenamefont {Ferretti}\ \emph {et~al.}(1996)\citenamefont
  {Ferretti}, \citenamefont {Granucci}, \citenamefont {Lami}, \citenamefont
  {Persico},\ and\ \citenamefont {Villani}}]{ferretti_1}%
  \BibitemOpen
  \bibfield  {author} {\bibinfo {author} {\bibfnamefont {A.}~\bibnamefont
  {Ferretti}}, \bibinfo {author} {\bibfnamefont {G.}~\bibnamefont {Granucci}},
  \bibinfo {author} {\bibfnamefont {A.}~\bibnamefont {Lami}}, \bibinfo {author}
  {\bibfnamefont {M.}~\bibnamefont {Persico}}, \ and\ \bibinfo {author}
  {\bibfnamefont {G.}~\bibnamefont {Villani}},\ }\href@noop {} {\bibfield
  {journal} {\bibinfo  {journal} {J. Chem. Phys.}\ }\textbf {\bibinfo {volume}
  {104}},\ \bibinfo {pages} {5517} (\bibinfo {year} {1996})}\BibitemShut
  {NoStop}%
\bibitem [{\citenamefont {Ferretti}, \citenamefont {Lami},\ and\ \citenamefont
  {Villani}(1996)}]{ferretti_2}%
  \BibitemOpen
  \bibfield  {author} {\bibinfo {author} {\bibfnamefont {A.}~\bibnamefont
  {Ferretti}}, \bibinfo {author} {\bibfnamefont {A.}~\bibnamefont {Lami}}, \
  and\ \bibinfo {author} {\bibfnamefont {G.}~\bibnamefont {Villani}},\
  }\href@noop {} {\bibfield  {journal} {\bibinfo  {journal} {J. Chem. Phys.}\
  }\textbf {\bibinfo {volume} {106}},\ \bibinfo {pages} {934} (\bibinfo {year}
  {1996})}\BibitemShut {NoStop}%
\bibitem [{\citenamefont {Kelly}\ and\ \citenamefont {Kapral}(2010)}]{kelly10}%
  \BibitemOpen
  \bibfield  {author} {\bibinfo {author} {\bibfnamefont {A.}~\bibnamefont
  {Kelly}}\ and\ \bibinfo {author} {\bibfnamefont {R.}~\bibnamefont {Kapral}},\
  }\href@noop {} {\bibfield  {journal} {\bibinfo  {journal} {J. Chem. Phys.}\
  }\textbf {\bibinfo {volume} {133}},\ \bibinfo {pages} {084502} (\bibinfo
  {year} {2010})}\BibitemShut {NoStop}%
\bibitem [{\citenamefont {Ryabinkin}\ and\ \citenamefont
  {Izmaylov}(2013)}]{ryabinkin13}%
  \BibitemOpen
  \bibfield  {author} {\bibinfo {author} {\bibfnamefont {G.}~\bibnamefont
  {Ryabinkin}}\ and\ \bibinfo {author} {\bibfnamefont {A.~F.}\ \bibnamefont
  {Izmaylov}},\ }\href@noop {} {\bibfield  {journal} {\bibinfo  {journal}
  {Phys. Rev. Lett.}\ }\textbf {\bibinfo {volume} {111}},\ \bibinfo {pages}
  {220406} (\bibinfo {year} {2013})}\BibitemShut {NoStop}%
\bibitem [{\citenamefont {Ryabinkin}\ \emph {et~al.}(2014)\citenamefont
  {Ryabinkin}, \citenamefont {Hsieh}, \citenamefont {Kapral},\ and\
  \citenamefont {Izmaylov}}]{ryabinkin14}%
  \BibitemOpen
  \bibfield  {author} {\bibinfo {author} {\bibfnamefont {G.}~\bibnamefont
  {Ryabinkin}}, \bibinfo {author} {\bibfnamefont {C.-Y.}\ \bibnamefont
  {Hsieh}}, \bibinfo {author} {\bibfnamefont {R.}~\bibnamefont {Kapral}}, \
  and\ \bibinfo {author} {\bibfnamefont {A.~F.}\ \bibnamefont {Izmaylov}},\
  }\href@noop {} {\bibfield  {journal} {\bibinfo  {journal} {J. Chem. Phys.}\
  }\textbf {\bibinfo {volume} {140}},\ \bibinfo {pages} {084104} (\bibinfo
  {year} {2014})}\BibitemShut {NoStop}%
\bibitem [{\citenamefont {Azzouz}\ and\ \citenamefont
  {Borgis}(1993)}]{azzouz93}%
  \BibitemOpen
  \bibfield  {author} {\bibinfo {author} {\bibfnamefont {H.}~\bibnamefont
  {Azzouz}}\ and\ \bibinfo {author} {\bibfnamefont {D.}~\bibnamefont
  {Borgis}},\ }\href@noop {} {\bibfield  {journal} {\bibinfo  {journal} {J.
  Chem. Phys.}\ }\textbf {\bibinfo {volume} {98}},\ \bibinfo {pages} {7361}
  (\bibinfo {year} {1993})}\BibitemShut {NoStop}%
\bibitem [{\citenamefont {McRae}\ \emph {et~al.}(2001)\citenamefont {McRae},
  \citenamefont {Schenter}, \citenamefont {Garrett}, \citenamefont
  {Svetlicic},\ and\ \citenamefont {Truhlar}}]{mcrae01}%
  \BibitemOpen
  \bibfield  {author} {\bibinfo {author} {\bibfnamefont {R.~P.}\ \bibnamefont
  {McRae}}, \bibinfo {author} {\bibfnamefont {G.~K.}\ \bibnamefont {Schenter}},
  \bibinfo {author} {\bibfnamefont {B.~C.}\ \bibnamefont {Garrett}}, \bibinfo
  {author} {\bibfnamefont {Z.}~\bibnamefont {Svetlicic}}, \ and\ \bibinfo
  {author} {\bibfnamefont {D.~G.}\ \bibnamefont {Truhlar}},\ }\href@noop {}
  {\bibfield  {journal} {\bibinfo  {journal} {J. Chem. Phys.}\ }\textbf
  {\bibinfo {volume} {115}},\ \bibinfo {pages} {8460} (\bibinfo {year}
  {2001})}\BibitemShut {NoStop}%
\bibitem [{\citenamefont {Antoniou}\ and\ \citenamefont
  {Schwartz}(1999{\natexlab{a}})}]{antoniou99}%
  \BibitemOpen
  \bibfield  {author} {\bibinfo {author} {\bibfnamefont {D.}~\bibnamefont
  {Antoniou}}\ and\ \bibinfo {author} {\bibfnamefont {S.~D.}\ \bibnamefont
  {Schwartz}},\ }\href@noop {} {\bibfield  {journal} {\bibinfo  {journal} {J.
  Chem. Phys.}\ }\textbf {\bibinfo {volume} {110}},\ \bibinfo {pages} {465}
  (\bibinfo {year} {1999}{\natexlab{a}})}\BibitemShut {NoStop}%
\bibitem [{\citenamefont {Antoniou}\ and\ \citenamefont
  {Schwartz}(1999{\natexlab{b}})}]{antoniou99a}%
  \BibitemOpen
  \bibfield  {author} {\bibinfo {author} {\bibfnamefont {D.}~\bibnamefont
  {Antoniou}}\ and\ \bibinfo {author} {\bibfnamefont {S.~D.}\ \bibnamefont
  {Schwartz}},\ }\href@noop {} {\bibfield  {journal} {\bibinfo  {journal} {J.
  Chem. Phys.}\ }\textbf {\bibinfo {volume} {110}},\ \bibinfo {pages} {7359}
  (\bibinfo {year} {1999}{\natexlab{b}})}\BibitemShut {NoStop}%
\bibitem [{\citenamefont {Kim}\ and\ \citenamefont
  {Hammes-Schiffer}(2003)}]{kim03}%
  \BibitemOpen
  \bibfield  {author} {\bibinfo {author} {\bibfnamefont {S.~Y.}\ \bibnamefont
  {Kim}}\ and\ \bibinfo {author} {\bibfnamefont {S.}~\bibnamefont
  {Hammes-Schiffer}},\ }\href@noop {} {\bibfield  {journal} {\bibinfo
  {journal} {J. Chem. Phys.}\ }\textbf {\bibinfo {volume} {119}},\ \bibinfo
  {pages} {4389} (\bibinfo {year} {2003})}\BibitemShut {NoStop}%
\bibitem [{\citenamefont {Yamamoto}\ and\ \citenamefont
  {Miller}(2005)}]{yamamoto05}%
  \BibitemOpen
  \bibfield  {author} {\bibinfo {author} {\bibfnamefont {T.}~\bibnamefont
  {Yamamoto}}\ and\ \bibinfo {author} {\bibfnamefont {W.~H.}\ \bibnamefont
  {Miller}},\ }\href@noop {} {\bibfield  {journal} {\bibinfo  {journal} {J.
  Chem. Phys.}\ }\textbf {\bibinfo {volume} {122}},\ \bibinfo {pages} {044106}
  (\bibinfo {year} {2005})}\BibitemShut {NoStop}%
\bibitem [{\citenamefont {Hanna}\ and\ \citenamefont {Kapral}(2008)}]{hanna08}%
  \BibitemOpen
  \bibfield  {author} {\bibinfo {author} {\bibfnamefont {G.}~\bibnamefont
  {Hanna}}\ and\ \bibinfo {author} {\bibfnamefont {R.}~\bibnamefont {Kapral}},\
  }\href@noop {} {\bibfield  {journal} {\bibinfo  {journal} {J. Chem. Phys.}\
  }\textbf {\bibinfo {volume} {128}},\ \bibinfo {pages} {164520} (\bibinfo
  {year} {2008})}\BibitemShut {NoStop}%
\bibitem [{\citenamefont {Sergi}\ and\ \citenamefont {Kapral}(2003)}]{sergi03}%
  \BibitemOpen
  \bibfield  {author} {\bibinfo {author} {\bibfnamefont {A.}~\bibnamefont
  {Sergi}}\ and\ \bibinfo {author} {\bibfnamefont {R.}~\bibnamefont {Kapral}},\
  }\href@noop {} {\bibfield  {journal} {\bibinfo  {journal} {J. Chem. Phys.}\
  }\textbf {\bibinfo {volume} {118}},\ \bibinfo {pages} {8566} (\bibinfo {year}
  {2003})}\BibitemShut {NoStop}%
\bibitem [{\citenamefont {Kim}\ and\ \citenamefont {Kapral}(ress)}]{kim06}%
  \BibitemOpen
  \bibfield  {author} {\bibinfo {author} {\bibfnamefont {H.}~\bibnamefont
  {Kim}}\ and\ \bibinfo {author} {\bibfnamefont {R.}~\bibnamefont {Kapral}},\
  }\href@noop {} {\bibfield  {journal} {\bibinfo  {journal} {Chem. Phys.
  Lett.}\ } (\bibinfo {year} {in press})}\BibitemShut {NoStop}%
\bibitem [{\citenamefont {Kim}, \citenamefont {Hanna},\ and\ \citenamefont
  {Kapral}(2006)}]{kim06c}%
  \BibitemOpen
  \bibfield  {author} {\bibinfo {author} {\bibfnamefont {H.}~\bibnamefont
  {Kim}}, \bibinfo {author} {\bibfnamefont {G.}~\bibnamefont {Hanna}}, \ and\
  \bibinfo {author} {\bibfnamefont {R.}~\bibnamefont {Kapral}},\ }\href@noop {}
  {\bibfield  {journal} {\bibinfo  {journal} {J. Chem. Phys.}\ }\textbf
  {\bibinfo {volume} {125}},\ \bibinfo {pages} {084509} (\bibinfo {year}
  {2006})}\BibitemShut {NoStop}%
\bibitem [{\citenamefont {Marcus}\ and\ \citenamefont
  {Sutin}(1985)}]{marcus85}%
  \BibitemOpen
  \bibfield  {author} {\bibinfo {author} {\bibfnamefont {R.~A.}\ \bibnamefont
  {Marcus}}\ and\ \bibinfo {author} {\bibfnamefont {N.}~\bibnamefont {Sutin}},\
  }\href@noop {} {\bibfield  {journal} {\bibinfo  {journal} {Biochim. Biophys.
  Acta}\ }\textbf {\bibinfo {volume} {811}},\ \bibinfo {pages} {265} (\bibinfo
  {year} {1985})}\BibitemShut {NoStop}%
\bibitem [{\citenamefont {Warshel}(1982)}]{warshel82}%
  \BibitemOpen
  \bibfield  {author} {\bibinfo {author} {\bibfnamefont {A.}~\bibnamefont
  {Warshel}},\ }\href@noop {} {\bibfield  {journal} {\bibinfo  {journal} {J.
  Phys. Chem.}\ }\textbf {\bibinfo {volume} {86}},\ \bibinfo {pages} {2218}
  (\bibinfo {year} {1982})}\BibitemShut {NoStop}%
\bibitem [{\citenamefont {Kim}\ and\ \citenamefont
  {Kapral}(2006)}]{kim-clus06}%
  \BibitemOpen
  \bibfield  {author} {\bibinfo {author} {\bibfnamefont {H.}~\bibnamefont
  {Kim}}\ and\ \bibinfo {author} {\bibfnamefont {R.}~\bibnamefont {Kapral}},\
  }\href@noop {} {\bibfield  {journal} {\bibinfo  {journal} {J. Chem. Phys.}\
  }\textbf {\bibinfo {volume} {125}},\ \bibinfo {pages} {234309} (\bibinfo
  {year} {2006})}\BibitemShut {NoStop}%
\bibitem [{\citenamefont {Kim}\ and\ \citenamefont
  {Kapral}(2008)}]{kim-clus08}%
  \BibitemOpen
  \bibfield  {author} {\bibinfo {author} {\bibfnamefont {H.}~\bibnamefont
  {Kim}}\ and\ \bibinfo {author} {\bibfnamefont {R.}~\bibnamefont {Kapral}},\
  }\href@noop {} {\bibfield  {journal} {\bibinfo  {journal} {ChemPhysChem}\
  }\textbf {\bibinfo {volume} {9}},\ \bibinfo {pages} {470} (\bibinfo {year}
  {2008})}\BibitemShut {NoStop}%
\bibitem [{\citenamefont {Hanna}\ and\ \citenamefont {Geva}(2008)}]{hanna08b}%
  \BibitemOpen
  \bibfield  {author} {\bibinfo {author} {\bibfnamefont {G.}~\bibnamefont
  {Hanna}}\ and\ \bibinfo {author} {\bibfnamefont {E.}~\bibnamefont {Geva}},\
  }\href@noop {} {\bibfield  {journal} {\bibinfo  {journal} {J. Phys. Chem. B}\
  }\textbf {\bibinfo {volume} {112}},\ \bibinfo {pages} {4048} (\bibinfo {year}
  {2008})}\BibitemShut {NoStop}%
\bibitem [{\citenamefont {Hanna}\ and\ \citenamefont {Geva}(2009)}]{hanna09}%
  \BibitemOpen
  \bibfield  {author} {\bibinfo {author} {\bibfnamefont {G.}~\bibnamefont
  {Hanna}}\ and\ \bibinfo {author} {\bibfnamefont {E.}~\bibnamefont {Geva}},\
  }\href@noop {} {\bibfield  {journal} {\bibinfo  {journal} {J. Phys. Chem. B}\
  }\textbf {\bibinfo {volume} {113}},\ \bibinfo {pages} {9278} (\bibinfo {year}
  {2009})}\BibitemShut {NoStop}%
\bibitem [{\citenamefont {Hanna}\ and\ \citenamefont {Geva}(2010)}]{hanna10}%
  \BibitemOpen
  \bibfield  {author} {\bibinfo {author} {\bibfnamefont {G.}~\bibnamefont
  {Hanna}}\ and\ \bibinfo {author} {\bibfnamefont {E.}~\bibnamefont {Geva}},\
  }\href@noop {} {\bibfield  {journal} {\bibinfo  {journal} {Chem. Phys.}\
  }\textbf {\bibinfo {volume} {370}},\ \bibinfo {pages} {201} (\bibinfo {year}
  {2010})}\BibitemShut {NoStop}%
\bibitem [{\citenamefont {Beck}\ \emph {et~al.}(2000)\citenamefont {Beck},
  \citenamefont {Jäckle}, \citenamefont {Worth},\ and\ \citenamefont
  {Meyer}}]{beck00}%
  \BibitemOpen
  \bibfield  {author} {\bibinfo {author} {\bibfnamefont {M.}~\bibnamefont
  {Beck}}, \bibinfo {author} {\bibfnamefont {A.}~\bibnamefont {Jäckle}},
  \bibinfo {author} {\bibfnamefont {G.}~\bibnamefont {Worth}}, \ and\ \bibinfo
  {author} {\bibfnamefont {H.-D.}\ \bibnamefont {Meyer}},\ }\href@noop {}
  {\bibfield  {journal} {\bibinfo  {journal} {Phys. Rep.}\ }\textbf {\bibinfo
  {volume} {324}},\ \bibinfo {pages} {1 } (\bibinfo {year} {2000})}\BibitemShut
  {NoStop}%
\bibitem [{\citenamefont {Meyer}, \citenamefont {Manthe},\ and\ \citenamefont
  {Cederbaum}(1990)}]{meyer90}%
  \BibitemOpen
  \bibfield  {author} {\bibinfo {author} {\bibfnamefont {H.-D.}\ \bibnamefont
  {Meyer}}, \bibinfo {author} {\bibfnamefont {U.}~\bibnamefont {Manthe}}, \
  and\ \bibinfo {author} {\bibfnamefont {L.}~\bibnamefont {Cederbaum}},\
  }\href@noop {} {\bibfield  {journal} {\bibinfo  {journal} {Chem. Phys.
  Lett.}\ }\textbf {\bibinfo {volume} {165}},\ \bibinfo {pages} {73 } (\bibinfo
  {year} {1990})}\BibitemShut {NoStop}%
\bibitem [{\citenamefont {Alon}, \citenamefont {Streltsov},\ and\ \citenamefont
  {Cederbaum}(2007)}]{alon07}%
  \BibitemOpen
  \bibfield  {author} {\bibinfo {author} {\bibfnamefont {O.~E.}\ \bibnamefont
  {Alon}}, \bibinfo {author} {\bibfnamefont {A.~I.}\ \bibnamefont {Streltsov}},
  \ and\ \bibinfo {author} {\bibfnamefont {L.~S.}\ \bibnamefont {Cederbaum}},\
  }\href@noop {} {\bibfield  {journal} {\bibinfo  {journal} {J. Chem. Phys.}\
  }\textbf {\bibinfo {volume} {127}},\ \bibinfo {eid} {154103} (\bibinfo {year}
  {2007})}\BibitemShut {NoStop}%
\bibitem [{\citenamefont {Marques}\ and\ \citenamefont
  {Gross}(2004)}]{marques04}%
  \BibitemOpen
  \bibfield  {author} {\bibinfo {author} {\bibfnamefont {M.}~\bibnamefont
  {Marques}}\ and\ \bibinfo {author} {\bibfnamefont {E.}~\bibnamefont
  {Gross}},\ }\href@noop {} {\bibfield  {journal} {\bibinfo  {journal} {Annu.
  Rev. Phys. Chem.}\ }\textbf {\bibinfo {volume} {55}},\ \bibinfo {pages} {427}
  (\bibinfo {year} {2004})}\BibitemShut {NoStop}%
\bibitem [{\citenamefont {Roemer}, \citenamefont {Ruckenbauer},\ and\
  \citenamefont {Burghardt}(2013)}]{burghardt13}%
  \BibitemOpen
  \bibfield  {author} {\bibinfo {author} {\bibfnamefont {S.}~\bibnamefont
  {Roemer}}, \bibinfo {author} {\bibfnamefont {M.}~\bibnamefont {Ruckenbauer}},
  \ and\ \bibinfo {author} {\bibfnamefont {I.}~\bibnamefont {Burghardt}},\
  }\href@noop {} {\bibfield  {journal} {\bibinfo  {journal} {J. Chem. Phys.}\
  }\textbf {\bibinfo {volume} {138}},\ \bibinfo {pages} {064106} (\bibinfo
  {year} {2013})}\BibitemShut {NoStop}%
\end{thebibliography}
\end{document}